\documentclass[aip,jap,amsmath,amssymb,reprint,groupedaddress]{revtex4-1}

\usepackage{graphicx}
\usepackage{dcolumn}
\usepackage{bm}
\usepackage{array}
\usepackage[utf8]{inputenc}
\usepackage[T1]{fontenc}
\usepackage{mathptmx}

\usepackage{booktabs}
\usepackage{placeins}
\usepackage{etoolbox}
\usepackage{url}
\usepackage[hidelinks]{hyperref}

\makeatletter
\def\@email#1#2{%
 \endgroup
 \patchcmd{\titleblock@produce}
  {\frontmatter@RRAPformat}
  {\frontmatter@RRAPformat{\produce@RRAP{*#1\href{mailto:#2}{#2}}}\frontmatter@RRAPformat}
  {}{}
}%
\makeatother

\begin{document}

\title{Free-vortex-wake dynamics of airborne wind energy kites in crosswind flight}

\author{Yinong Tian}
\author{Shan Wang$^{*}$}
\email[Author to whom correspondence should be addressed: ]{shan.wang@centec.tecnico.ulisboa.pt}
\affiliation{Centre for Marine Technology and Ocean Engineering (CENTEC), Instituto Superior T\'ecnico, Universidade de Lisboa, Av. Rovisco Pais 1, 1049-001 Lisboa, Portugal}

\date{\today}

\begin{abstract}
The wake of an airborne wind energy kite is shaped by its crosswind path and, for on-board generation, by the slipstreams of its rotors, and in farms it reaches downstream units. It is computed here by a vortex step method with a time-marching free vortex wake, implemented by the authors in their open-source code OpenFAWE with a lumped-mass tether, actuator-disc rotors releasing vortex rings and several kites. It reproduces an independent vortex step code to $10^{-9}$ on the rigid-wing Makani M600 and within 0.3~\% in lift and 0.9~\% in drag on the TU Delft V3 leading-edge inflatable kite, matching its Reynolds-averaged Navier--Stokes lift within 3.1~\%. For fixed kites the lift deficit of a finite wake is the far field of its starting vortex, which fixes the wake length before a run, and the vortex core must stay below the narrowest panel width. With these settings the helical tip-vortex pair of the M600 on its loop lowers its lift by about 0.5~\% from 5 to 13~m/s. Free rotor rings do not converge with the time step, whereas an undeformed slipstream tube converges once its rings are refined at the wing and raises the lift by 1.9~\%. A slipstream deflected by the wing lowers it by 3.5~\%, so the treatment near the wing sets the sign of the rotor effect. A second unit one loop diameter downstream loses 0.5~\% of its lift and 2.2~\% of its aerodynamic power, with or without rotors.
\end{abstract}

\maketitle

\section{Introduction}\label{sec:i}

Airborne wind energy (AWE) systems replace the tower of a wind turbine by a tether and harvest the wind with a kite or an aircraft that flies crosswind \cite{Vermillion2021}. The kite flies at an airspeed of about its lift-to-drag ratio times the wind speed, and the power is generated either on the ground through the tether or on board through rotors carried by the kite \cite{Loyd1980,Cherubini2015,Fagiano2022}. By Loyd's crosswind formula \cite{Loyd1980} the power is proportional to the cube of the lift coefficient and inversely proportional to the square of the drag coefficient, so that an error of 1~\% in a coefficient shifts the power estimate by 2~\% to 3~\%. The formula neglects the induction by which the kite slows the oncoming flow \cite{Loyd1980}, and in a vortex description of the flow this induction comes from the wake that the kite leaves behind \cite{Katz2001,Trevisi2023a}.

The wake of a kite differs from the wake of a wind turbine in what sets its shape. A turbine wake is made by blades that rotate about a fixed axis, and its helical vortices, their induction and their stability have been studied as such \cite{Hardin1982,Fukumoto2005,Sherry2013}, and the break-down of its tip vortices and the recovery of the wake depend on the turbulence of the inflow \cite{Hodgson2023}. A kite wake is made by one wing that follows a closed path on the sphere of its tether, so that a circle produces an annular helical wake with one turn per loop and a figure-eight produces a sheet that crosses itself once per cycle. The periodic motion of a lifting body changes its wake, as the stability analysis of the wakes of floating wind turbines shows \cite{Kleine2022}, and in a kite this motion is the whole source of the wake. A kite that generates power on board, as the Makani M600 does with eight rotors on its wing \cite{Echeverri2020}, adds the slipstreams of its rotors to the wake of its wing. A vortex particle computation of an M600-like kite \cite{Mehr2025} found that these slipstreams lower the lift of the wing by up to about 7~\%, whereas a design model of an airborne wind energy aircraft with on-board turbines represents each turbine wake as a vortex cylinder and switches its interaction with the wing off for turbines on pylons \cite{Trevisi2026}. In a farm the units downstream fly in the wakes of those upstream \cite{Haas2022}, as wind turbines in a group do, for which a vortex filament wake model resolves the interference in line, side by side and staggered unit by unit \cite{Liu2023} and computations of two floating turbines show the dependence on the layout \cite{Huang2023}. Unsteady Reynolds-averaged Navier--Stokes (URANS) computations show a velocity deficit of 10~\% at ten orbit radii downstream of a kite \cite{Kheiri2022}, and large-eddy simulations of farms of 25 units with the kites represented by actuator sectors report wake losses that grow from row to row and reach 17~\% in the last row for ground generation and 25~\% to 45~\% for on-board generation, depending on the density of the layout \cite{Haas2022}.

System models of AWE keep the aerodynamics of the kite simple, in that energy assessments use constant coefficients or analytical polars \cite{Argatov2009,Argatov2010,Argatov2016,Archer2014,Bechtle2019} and the others take their coefficients from flight identification \cite{vanderVlugt2019,Fechner2015} or from steady computations \cite{Malz2019,Eijkelhof2022}. In these models the wake is ignored, enters as the induced drag of a straight wake \cite{Trevisi2023b} or is approximated by momentum theory \cite{DeLellis2018}. An analytical vortex model of the wake shows that ground-generation systems fly closer to their own wake because of the reel-out and have a lower power potential than a geometrically identical fly-generation system \cite{Trevisi2023a,Trevisi2023b}. Two analytical wake models, one from the conservation of mass and one from the conservation of mass and momentum, give the speed and the shape of the wake of a crosswind kite and were checked against computational results \cite{Karakouzian2022}, and in the optimal control of a dual-kite system the optimum without a wake differs by 145~\% from the reference with a wake \cite{DeSchutter2026}. These models leave the wake out or prescribe its shape, which leaves open how the wake builds up along the path and what of it the kite and a downstream unit feel.

Vortex methods that resolve the load distribution over the lifting surface fall into two classes by their treatment of the wake \cite{Tian2026}. With a straight wake, the vortex step method determines the circulation from the polar at the three-quarter-chord point and trails semi-infinite straight vortices along the oncoming flow \cite{Damiani2019}. It computed the loads of the M600 in KiteFAST \cite{Damiani2019,Jonkman2021}, entered an aero-structural model of leading-edge inflatable kites \cite{Cayon2023} and reached mean absolute errors of 9~\% in lift and 13~\% in drag against wind-tunnel data of the V3 kite \cite{Poland2026pre}. With a free wake that moves with the local velocity, kites have been computed with the lifting-line code AWSM \cite{Vimalakanthan2018} and the unsteady panel code UnPaM \cite{CastroFernandez2021}, and the lifting-line code QBlade and the vortex-particle code DUST served as references for engineering wake models \cite{Trevisi2023b,DeSchutter2026}. Free-vortex filament methods are established for rotors and wind turbines \cite{Leishman2002,Bhagwat2001,Sebastian2012,Shaler2020,Tian2026}, a wind turbine wake can also be built from annular vortex elements released by the blades \cite{Tian2024}, and their results depend on the regularisation of the filaments and on the wake length \cite{Sebastian2012,Branlard2022,Tian2025}. For kites the convergence of the loads with wake length, time step and core radius had not been reported in the form that is standard for grid-refinement studies \cite{Roache1994,Celik2008}. Nor had it been settled how a free vortex wake should represent the slipstreams of rotors that pass within a fraction of a chord of a lifting surface, with which parameters and with what effect on the wing.

This paper formulates the vortex step method with a time-marching free vortex wake for kites on prescribed crosswind motions, with a lumped-mass tether, actuator-disc rotors that release vortex rings and several kites in one wake. The method is implemented in the aerodynamic module of OpenFAWE, an open-source code developed by the authors \cite{OpenFAWE}, with which all computations of this paper were made. Section II gives the method. Section III describes the models of the TU Delft V3 and the Makani M600, verifies the implementation against an independent vortex step code, wind-tunnel data and RANS, and derives from the fixed kites the rules for the wake length, the time step and the core radius. Section IV applies the method to the rigid-wing kite on the crosswind loop of its controller, first alone, then with its eight rotors and finally as one of two units in one wake, and Sec.~V concludes.

\section{Method}\label{sec:ii}

\subsection{Lifting surfaces and vortex system}

The flow is incompressible, with free stream $\mathbf{U}_\infty$ and density $\rho$, and a kite is represented by lifting surfaces that carry bound vortices and by the lattice of vortex rings that its trailing and shed vorticity forms behind it. A lifting surface is given by $N+1$ leading-edge nodes $\mathbf{x}^{\mathrm{le}}_k$ and trailing-edge nodes $\mathbf{x}^{\mathrm{te}}_k$. The bound vortex runs through the quarter-chord nodes $\mathbf{b}_k = \mathbf{x}^{\mathrm{le}}_k + \tfrac14\mathbf{c}_k$, with the chord vectors $\mathbf{c}_k = \mathbf{x}^{\mathrm{te}}_k - \mathbf{x}^{\mathrm{le}}_k$, and panel $j$ between the nodes $j$ and $j+1$ has the bound segment $\mathrm{d}\mathbf{l}_j = \mathbf{b}_{j+1}-\mathbf{b}_j$ of width $w_j = |\mathrm{d}\mathbf{l}_j|$. The spanwise station $s^*_j$ of its evaluation points follows the full-cosine approximation of van Garrel \cite{vanGarrel2003} that OLAF uses \cite{Shaler2020}, which is the mid-width for uniform panels. With $\mathbf{b}(s)$ and $\mathbf{c}(s)$ interpolated linearly along the span, the load point and the closure point of the panel lie at the quarter and the three-quarter chord,

\begin{equation}
\mathbf{x}^{l}_j = \mathbf{b}(s^*_j), \qquad \mathbf{x}^{c}_j = \mathbf{x}^{l}_j + \tfrac12\,\mathbf{c}(s^*_j).
\label{eq:1}
\end{equation}

The panel carries the unit chord tangent $\mathbf{t}_j$, the normal $\mathbf{n}_j$, the span direction $\mathbf{s}_j = \mathrm{d}\mathbf{l}_j/w_j$, the chord $c_j$ and a section polar $C_l(\alpha)$, $C_d(\alpha)$, $C_m(\alpha)$ that is interpolated linearly between the tabulated angles and never extrapolated. The lift and drag coefficients $C_L$ and $C_D$ of a kite are the force components normal and parallel to $\mathbf{U}_\infty$ and $C_{M_y}$ is its pitching moment, made dimensionless with $\tfrac12\rho U_\infty^2$, the reference area $S$ and the reference chord $c$.

The method of this paper is the vortex step method with a free vortex wake (VSM-F). Figure~1 and Table~I set it against the two other bound-vortex models with which Sec.~III compares it, each with a straight wake (S) or a free wake (F) on the same geometry, section data and oncoming flow. The lifting line is the nonlinear formulation of OLAF \cite{Shaler2020,Branlard2022,vanGarrel2003,Phillips2000} and places its closure points on the bound vortex [Fig.~1(a), 1(d)]. The lifting surface is a vortex-lattice method \cite{Katz2001} with eight cosine-spaced vortex rings along the chord of every panel and a closure point at the three-quarter chord of every ring [Fig.~1(c), 1(f)], inviscid in the version LSi and matched to the polar through a virtual incidence of the panel normal in the version LSp. The vortex step method keeps one bound segment and one unknown circulation per panel, as the lifting line does, and satisfies the closure at the three-quarter chord, as the method of Weissinger on which the vortex step method of KiteFAST is based \cite{Damiani2019}, with the load point on the bound vortex [Fig.~1(b), 1(e)]. The trailing vortices in Fig.~1 are cut 0.8 chords behind the trailing edge. In the straight wake the VSM uses the semi-infinite horseshoe vortices and the core treatment of its reference implementation \cite{Cayon2023,VSMCode}, the straight wakes of the other two models are one lattice cell per panel extending 1000~m downstream, and a straight wake has a single steady solution.

\begin{figure*}[!t]
\centering
\includegraphics[width=0.85\textwidth]{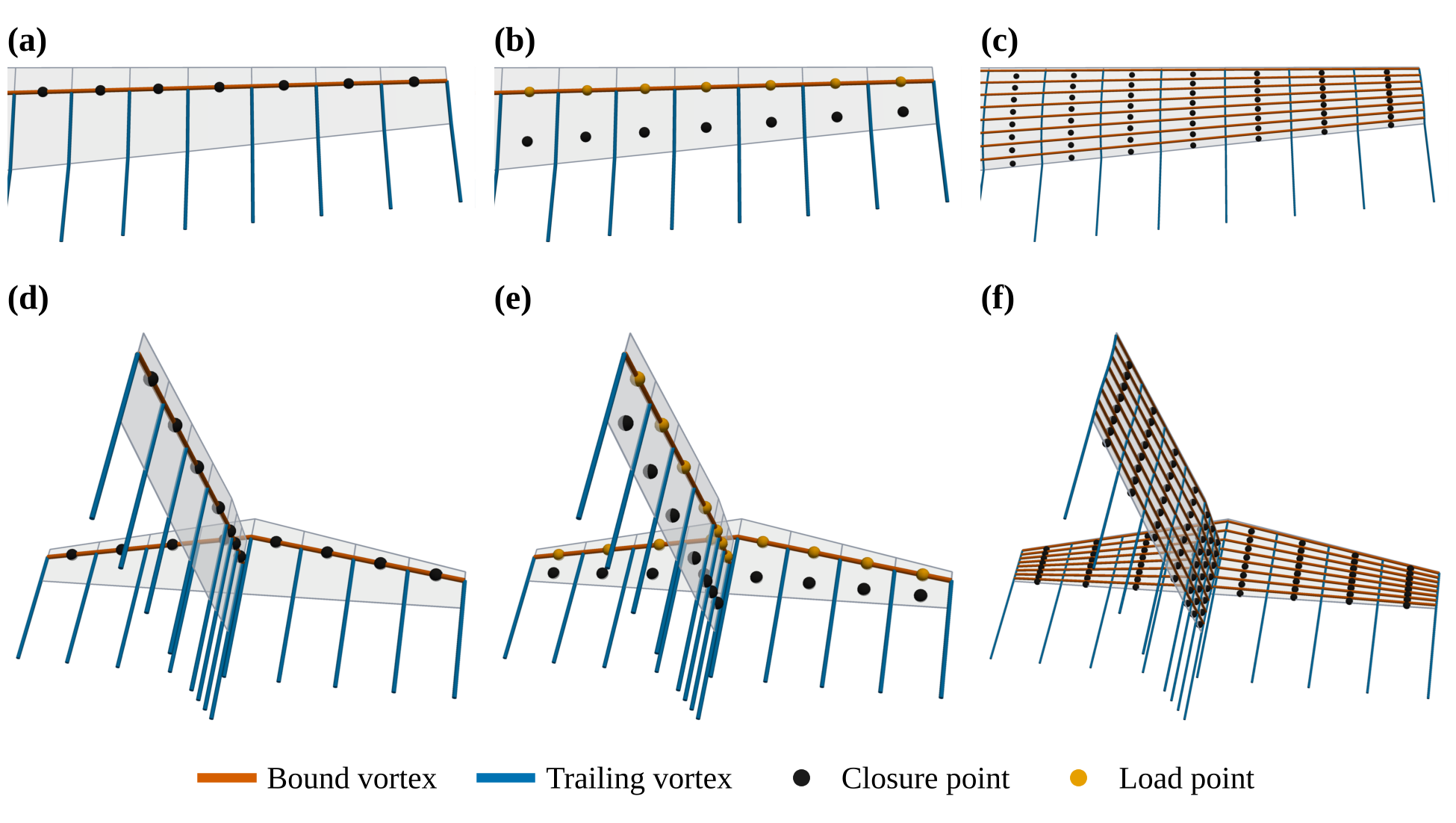}
\caption{\label{fig:1}Bound-vortex models on the M600. (a) to (c) Starboard outer wing, (d) to (f) horizontal and vertical tail, with the lifting line, the vortex step method and the lifting surface.}
\end{figure*}

\begin{table*}[!t]
\caption{\label{tab:1}The three bound-vortex models compared in Sec.~III.}
\small\setlength{\tabcolsep}{5pt}\renewcommand{\arraystretch}{1.12}
\begin{tabular}{@{}>{\raggedright\arraybackslash}p{0.13\textwidth}>{\raggedright\arraybackslash}p{0.25\textwidth}>{\raggedright\arraybackslash}p{0.27\textwidth}>{\raggedright\arraybackslash}p{0.25\textwidth}@{}}
\toprule
 & Lifting line (LL) & Vortex step method (VSM) & Lifting surface (LSi, LSp) \\
\midrule
Bound vortex & one segment per panel at $\tfrac14$ chord & one segment per panel at $\tfrac14$ chord & eight vortex rings per panel on the mean surface \\
Closure point & $\tfrac14$ chord, on the bound vortex & $\tfrac34$ chord, Eq.~(\ref{eq:1}) & $\tfrac34$ chord of every vortex ring \\
Closure & section lift from the polar at the $\tfrac14$-chord flow & section lift from the polar at the $\tfrac34$-chord flow, Eq.~(\ref{eq:8}) & flow tangency, with a virtual incidence matched to the polar in LSp \\
Unknowns per panel & 1 & 1 & 8 \\
Loads & Kutta--Joukowski force and profile drag at $\tfrac14$ chord & magnitude from $\tfrac34$ chord, direction from $\tfrac14$ chord, Eq.~(\ref{eq:10}) & Kutta--Joukowski force on all vortex-ring segments \\
Source & OLAF \cite{Shaler2020,vanGarrel2003} & Cayon \emph{et al.} \cite{Cayon2023,VSMCode} & vortex-lattice method \cite{Katz2001} \\
\bottomrule
\end{tabular}
\end{table*}

The vortex system of VSM-F is a lattice of quadrilateral vortex rings whose rows are the bound-vortex line $\mathbf{b}_k$, the trailing edge $\mathbf{x}^{\mathrm{te}}_k$ and the wake rows $m = 1,\dots,n$, numbered from the newest to the oldest. The bound vortex ring of panel $j$, between the bound row and the trailing edge, carries the bound circulation $\Gamma_j$, and the wake vortex ring of panel $j$ between rows $m-1$ and $m$ carries $\Gamma_j^{(m)}$. A filament shared by two vortex rings carries the difference of their strengths, so that a trailing filament between panels $j$ and $j+1$ carries $\Gamma_j^{(m)} - \Gamma_{j+1}^{(m)}$ and a shed filament on row $m$ carries $\Gamma_j^{(m+1)} - \Gamma_j^{(m)}$. The state at time $t_n$ consists of the wake node positions and the vortex-ring strengths, and several lifting surfaces, of one kite or of several kites, share one circulation system and one wake. The velocity induced at a point $\mathbf{x}$ is the sum over all straight filaments,

\begin{equation}
\begin{aligned} \mathbf{u}(\mathbf{x}) &= \sum_k F_\nu\!\left(\frac{h_k}{r_{c,k}}\right)\frac{\Gamma_k}{4\pi}\,\mathbf{K}_k, \\ \mathbf{K}_k &= \frac{(r_1+r_2)(\mathbf{r}_1\times\mathbf{r}_2)}{r_1 r_2(r_1 r_2+\mathbf{r}_1\cdot\mathbf{r}_2)}, \\ F_\nu(\eta) &= 1-\exp(-\alpha_{\mathrm{O}}\,\eta^2), \qquad \alpha_{\mathrm{O}} = 1.25643, \end{aligned}
\label{eq:2}
\end{equation}

\noindent with $\mathbf{r}_1$ and $\mathbf{r}_2$ the vectors from the ends of filament $k$ to $\mathbf{x}$, $r_1$ and $r_2$ their lengths, $h_k$ the distance of $\mathbf{x}$ from its line and $r_{c,k}$ its core radius. $F_\nu$ is the Lamb--Oseen function \cite{Lamb1932} as implemented in OLAF \cite{Shaler2020,OpenFAST}, where it is set to one for $h > 2.82\,r_c$. At a wake node the filaments that end at this node are excluded from the sum, and any other evaluation point that coincides with a filament stops the run. The core radius is a numerical regularisation and a model of the viscous core \cite{Tian2026,Shaler2020}. In the baseline it is constant in space and time, $r_c = 0.15$ m. A core that grows with the age of the filament by the diffusion law of OLAF \cite{Shaler2020}, $r_c^2 = r_{c,0}^2 + 4\alpha_{\mathrm{O}}\delta\nu\tau$ with the kinematic viscosity $\nu$, the eddy-viscosity factor $\delta$ and the age $\tau$, is used in Sec.~III to test the effect of viscous diffusion.

\subsection{Time marching}

The wing starts impulsively at $t = 0$ with its bound vortex rings alone, and the trailing-edge nodes and all wake nodes are material points. Between two releases their positions $\mathbf{X}$ obey

\begin{equation}
\frac{\mathrm{d}\mathbf{X}}{\mathrm{d}t} = \mathbf{U}_\infty + \mathbf{u}\!\left(\mathbf{X};\,\boldsymbol{\Gamma}\right), \qquad \mathbf{R}(\boldsymbol{\Gamma};\,\mathbf{X}) = \mathbf{0}
\label{eq:3}
\end{equation}

\noindent where $\mathbf{u}$ is the velocity of Eq.~(2) induced by the complete vortex system and $\mathbf{R} = \mathbf{0}$ is the closure that fixes the bound circulation for a given wake. While the Jacobian $\partial\mathbf{R}/\partial\boldsymbol{\Gamma}$ is non-singular, the closure defines $\boldsymbol{\Gamma}(\mathbf{X})$ by the implicit-function theorem, and Eq.~(3) is a semi-explicit differential-algebraic system of index one \cite{Hairer1996}, and Sec.~III examines when this condition is lost. One step from $t_n$ to $t_{n+1} = t_n + \Delta t$ consists of four operations (Fig.~2). The node positions are first convected with the circulation frozen at $t_n$,

\begin{equation}
\mathbf{X}^{n+1} = \mathbf{X}^{n} + \Delta t\left[\mathbf{U}_\infty + \mathbf{u}\big(\mathbf{X}^n;\ \boldsymbol{\Gamma}^n\big)\right]
\label{eq:4}
\end{equation}

\noindent the explicit Euler step of the computations of this paper, so that the coupled scheme is the half-explicit Euler method \cite{Hairer1996} for Eq.~(3). The code also offers Heun and classical fourth-order Runge--Kutta stages, and the scheme stays first-order accurate in $\Delta t$ with any of them because the circulation is held at $t_n$ during the step. The stability, consistency and convergence of such time-marching free-vortex schemes were analysed for rotors by Bhagwat and Leishman \cite{Bhagwat2001}. The time step is also the streamwise resolution of the wake, because consecutive rows are released $U_\infty\Delta t$ apart. The convected copy of the trailing-edge row then becomes wake row 1, every older row moves one position downstream, and the wake vortex rings keep their strengths,

\begin{equation}
\Gamma_j^{(1)} = \Gamma_j^{n+1}, \qquad \Gamma_j^{(m+1)}\big|_{t_{n+1}} = \Gamma_j^{(m)}\big|_{t_n}, \quad m \ge 1
\label{eq:5}
\end{equation}

\noindent so that the vortex ring released during the step carries the new bound circulation. The net circulation of the filament along the trailing edge is then zero, which is the Kutta condition in the form used by OLAF \cite{Shaler2020}, and the shed vorticity $\Gamma_j^{n+1} - \Gamma_j^{n}$ lies on the spanwise filament of wake row 1. Every wake vortex ring keeps its released strength for the whole run, with no dissipation, truncation, merging or decay, so that after $n$ steps the wake extends $n\,U_\infty\Delta t$ downstream. The circulation from the closure and the loads complete the step. Every later step places the lifting surfaces at the new pose, convects the wake, sheds a row of wake vortex rings, solves the closure on the new geometry and advances the tether (Fig.~2), and a run that misses the acceptance criterion of the closure stops and is recorded as failed.

\begin{figure}[!htb]
\centering
\includegraphics[width=0.72\linewidth]{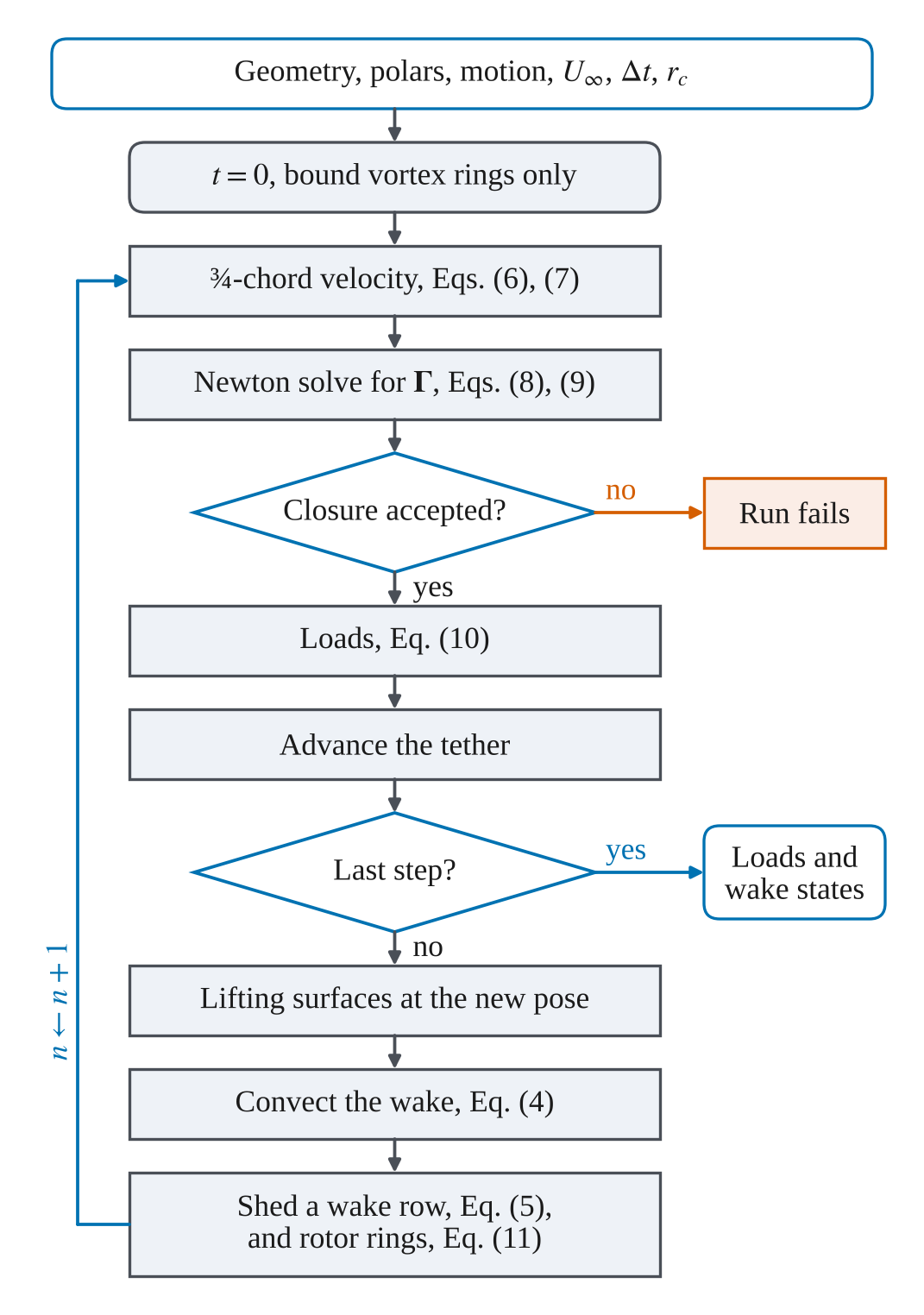}
\caption{\label{fig:2}One time step of the vortex step method with a free vortex wake along a prescribed motion.}
\end{figure}

\subsection{Closure and loads}

The vortex step closure \cite{Damiani2019,Cayon2023} and the lattice free wake \cite{Shaler2020,Branlard2022,Leishman2002} exist separately in the reference codes, and their combination is formulated here. At the new time the velocity at the closure point of panel $i$ is split into three parts,

\begin{equation}
\mathbf{u}_i = \mathbf{U}_{\text{rel},i} + \sum_j \mathbf{M}_{ij}\,\Gamma_j + \mathbf{u}^{\text{hist}}_i , \qquad \mathbf{M}_{ij} = \mathbf{A}_{ij} - \delta_{ij}\,\mathbf{a}^{2D}_i
\label{eq:6}
\end{equation}

\noindent with $\mathbf{U}_{\text{rel},i}$ the free stream relative to the panel, which for a moving kite includes the velocity of the panel. $\mathbf{A}_{ij}$ is the velocity induced by unit circulation on the two vortex rings of panel $j$ that carry the unknown circulation, the bound vortex ring and the newest wake vortex ring, and $\mathbf{u}^{\text{hist}}_i$ is the known velocity induced by all older wake vortex rings with their retained strengths. Because the section polar already contains the two-dimensional effect of the panel's own bound vortex, the velocity of an infinite straight vortex along the bound segment is subtracted on the diagonal \cite{Damiani2019},

\begin{equation}
\mathbf{a}^{2D}_i = \frac{|\mathrm{d}\mathbf{l}_i|\;\mathrm{d}\mathbf{l}_i\times\mathbf{r}_i}{2\pi\,|\mathrm{d}\mathbf{l}_i\times\mathbf{r}_i|^2}
\label{eq:7}
\end{equation}

\noindent with $\mathbf{r}_i$ the vector from the midpoint of the bound segment to the closure point. This term is inviscid and carries no core, whereas every filament in $\mathbf{A}$ and $\mathbf{u}^{\text{hist}}$ is regularised by Eq.~(2). The circulation follows from the section lift,

\begin{equation}
\begin{aligned} \Gamma_i &= G_i(\boldsymbol{\Gamma}) = \frac{c_i\, C_l(\alpha_i)\, |\mathbf{u}_i\times\mathbf{s}_i|^2}{2\,|\mathbf{U}_{\text{rel},i}\times\mathbf{s}_i|}, \\ \alpha_i &= \operatorname{atan2}(\mathbf{u}_i\cdot\mathbf{n}_i,\ \mathbf{u}_i\cdot\mathbf{t}_i), \end{aligned}
\label{eq:8}
\end{equation}

\noindent which is the closure of Cayon et al. \cite{Cayon2023,VSMCode} with the velocity of Eq.~(6). A state with reversed chordwise flow, $\mathbf{u}_i\cdot\mathbf{t}_i \le 0$, or with an angle outside the polar table is outside the domain of $\mathbf{G}$. The residual $\mathbf{R}(\boldsymbol{\Gamma}) = \mathbf{G}(\boldsymbol{\Gamma}) - \boldsymbol{\Gamma}$ is piecewise smooth because the polars are piecewise linear, and inside a polar segment its Jacobian follows from Eqs.~(6) and (8) with $v_t = \mathbf{u}_i\cdot\mathbf{t}_i$ and $v_n = \mathbf{u}_i\cdot\mathbf{n}_i$,

\begin{equation}
\begin{aligned} J_{ij} = {} & \frac{c_i}{2\,|\mathbf{U}_{\text{rel},i}\times\mathbf{s}_i|}\Big[C_l'(\alpha_i)\,|\mathbf{u}_i\times\mathbf{s}_i|^2 \\ & \times\frac{v_t\,\mathbf{M}_{ij}\cdot\mathbf{n}_i - v_n\,\mathbf{M}_{ij}\cdot\mathbf{t}_i}{v_t^2+v_n^2} \\ & + 2\,C_l(\alpha_i)\,(\mathbf{u}_i\times\mathbf{s}_i)\cdot(\mathbf{M}_{ij}\times\mathbf{s}_i)\Big] - \delta_{ij}. \end{aligned}
\label{eq:9}
\end{equation}

The first term is the change of the section lift through the angle of attack, the second through the transverse speed. The diagonal $\beta_i = J_{ii}+1 = \partial G_i/\partial\Gamma_i$ is the self-coefficient of panel $i$, which is negative when an increase of its own circulation lowers its own angle of attack, as the downwash of its trailing filaments does.

The root of $\mathbf{R}$ is found by a Newton iteration with a backtracking line search \cite{Nocedal2006}, which halves the step from 1 down to $2^{-31}$ until the squared residual falls by the factor $1 - 10^{-4}\lambda$ for the step length $\lambda$ and the trial state lies inside the domain. A trial state outside the domain is rejected, never clipped. The Jacobian is formed by forward differences with the increment $\epsilon^{1/3}\max(1,|\Gamma_j|)$, $\epsilon$ the machine precision, and by backward differences or a halved increment when the forward state leaves the domain. At converged steady solutions with a 100-chord straight wake the differences agree with Eq.~(9) to $7\times10^{-7}$ (V3) and $5\times10^{-6}$ (M600) of the largest Jacobian entry. The iteration starts from the circulation of the previous time step, and from zero at the start, is limited to 60 updates and accepts a solution with $\max_i |R_i| \le 10^{-10}\ \mathrm{m^2/s} + 10^{-8}\,\overline{|\Gamma|}$. The straight-wake VSM keeps the damped fixed-point iteration of the reference code \cite{VSMCode,Damiani2019} with the same acceptance criterion. The magnitude of the section force follows from the three-quarter-chord closure and its direction from the flow angle $\alpha^{l}_i$ at the load point, the aerodynamic-centre option of the reference code \cite{Cayon2023,VSMCode},

\begin{equation}
\begin{aligned} \mathbf{F}_i &= \tfrac12\rho\,|\mathbf{u}_i\times\mathbf{s}_i|^2\,c_i\,|\mathrm{d}\mathbf{l}_i|\left[C_l(\alpha_i)\,\hat{\mathbf{l}}_i + C_d(\alpha_i)\,\hat{\mathbf{d}}_i\right], \\ \hat{\mathbf{l}}_i &= \frac{\hat{\mathbf{v}}_i\times\mathbf{s}_i}{|\hat{\mathbf{v}}_i\times\mathbf{s}_i|}, \qquad \hat{\mathbf{d}}_i = \mathbf{s}_i\times\hat{\mathbf{l}}_i, \end{aligned}
\label{eq:10}
\end{equation}

\noindent with $\hat{\mathbf{v}}_i = \cos\alpha^{l}_i\,\mathbf{t}_i + \sin\alpha^{l}_i\,\mathbf{n}_i$. The velocity at the load point is split like Eq.~(6), without the bound spanwise filaments and without the term of Eq.~(7), and the section moment $\tfrac12\rho\,|\mathbf{u}_i\times\mathbf{s}_i|^2 c_i^2\,|\mathrm{d}\mathbf{l}_i|\,C_m(\alpha_i)$ acts about $\mathbf{s}_i$. The loads are quasi-steady, without an added-mass term. For a moving kite the force and the moment about the body origin are also expressed in body axes, and the rate of work of the aerodynamic load on the body, $\mathbf{F}\cdot\mathbf{v}_0 + \mathbf{M}_0\cdot\boldsymbol{\omega}$ with $\mathbf{v}_0$ and $\boldsymbol{\omega}$ the velocity of the origin and the angular velocity, is the aerodynamic power. With a long straight wake without cores the method reduces to the straight-wake VSM, and on an eight-panel elliptic wing with a wake of 1000~m the influence coefficients differ from the semi-infinite ones by less than $5\times10^{-8}\ \mathrm{m^{-1}}$.

\subsection{Motion, tether and rotors}

A kite is a rigid body with origin $\mathbf{x}_0(t)$, attitude $\mathbf{R}(t)$ (body to inertial axes, with body $x$ chordwise towards the trailing edge, $y$ along the span and $z$ towards the lift side), velocity $\mathbf{v}_0(t)$ and angular velocity $\boldsymbol{\omega}(t)$ in an inertial frame whose $X$ axis points along the wind and whose $Z$ axis points up. The lifting surfaces are placed at every time step, $\mathbf{x} = \mathbf{R}\,\mathbf{x}_{\mathrm{body}} + \mathbf{x}_0$, and a panel point $\mathbf{p}$ moves with $\mathbf{v}_0 + \boldsymbol{\omega}\times(\mathbf{p}-\mathbf{x}_0)$, which enters Eq.~(6) as the velocity of the panel. The wake stays in the inertial frame, where Eq.~(4) convects it, while the bound rows and the trailing-edge row follow the body. The motion is given as samples at the wake time step, and the consistency of its velocity and angular velocity with its position and attitude is checked before a run by fourth-order central differences of the samples (Sec.~III B). The relative wind at the body origin, $\mathbf{R}^{\mathsf{T}}(\mathbf{U}_\infty - \mathbf{v}_0)$, defines the incidence and the sideslip of the kite as in a flight-mechanics model, and its dynamic pressure forms the body-axis force and moment coefficients of Sec.~IV.

The M600 flies the nominal crosswind loop of its controller, transcribed from the open-source Makani release \cite{MakaniCode}. The body origin runs on the commanded circle, whose radius, elevation and azimuth offset come from the land playbook at the wind speed, with the kite speed that gives the commanded airspeed in the uniform wind and with the commanded attitude at every loop angle, namely the incidence and sideslip schedules and the tether roll from the curvature balance with the feed-forward term and its saturation band. The tables of the playbook are read through periodic cubic splines, so that the motion is twice continuously differentiable, and the circle is placed on the sphere that the tether and the bridle reach.

The tether is a lumped-mass line transcribed from the Makani simulator \cite{MakaniCode} with $n_t$ interior nodes of mass $\mu L/n_t$ between a fixed ground point and the bridle point of the kite, with a tension-only longitudinal spring $T = EA\,\varepsilon$ and a longitudinal damping along each segment, a bending moment $EI\,\kappa$ from the curvature of the circle through three consecutive nodes with its damping, the normal component of the drag of a cylinder on each node, $\mathbf{F} = -\tfrac12\rho C_{d,t} d\,\ell_{\mathrm{cyl}}\,|\mathbf{v}_n|\,\mathbf{v}_n$, and gravity. The line is integrated with the classical fourth-order Runge--Kutta method in sub-steps of the wake time step below the stability bound of its longitudinal modes, with the body pose interpolated between the motion samples. The bridle point is the Makani bridle model \cite{MakaniCode}, in which the two bridle legs meet the tether at a point that stays at a fixed distance from the bridle pivot of the kite and swings about the span axis towards the last tether node. The tether sees the ambient wind only, and the force of its last segment acts on the body at the bridle point. It is recorded with the aerodynamic loads at every step and does not alter the prescribed motion. The line is first run alone along two periods of the motion so that the coupled run starts from a periodic line state, and the period-to-period change of its tension is reported.

The eight rotors of the M600 are actuator discs of 2.3~m diameter at the hub positions of the Makani configuration, with their axes 3$^\circ$ from the body $x$ axis \cite{MakaniCode}. They turn with the body at a prescribed angular speed $\Omega$. The thrust $T$ and the shaft power of a disc are read from the rotor table of the open-source Makani code \cite{MakaniCode} as functions of $\Omega$ and of the axial freestream $V$, the component along the disc axis of the ambient wind plus the velocity induced by the wake of the lifting surfaces and by the rings of the other discs, minus the velocity of the hub. The rings of the disc itself are left out because the table contains the rotor's own induction. The disc acts on the flow through the vortex tube of a uniformly loaded actuator disc, whose sheet strength $\gamma$ follows from momentum theory \cite{Leishman2006}. The tube is released as one vortex ring per time step at the rim of the disc, a closed polygon of 16~straight filaments with the circulation

\begin{equation}
T = \rho A \left(V + \tfrac12\gamma\right)\gamma, \qquad \Gamma_r = \gamma\left(V_\infty + \tfrac12\gamma\right)\Delta t
\label{eq:11}
\end{equation}

\noindent with the disc area $A$ and the axial component $V_\infty$ of the ambient wind relative to the hub. The ring is the sheet strength times the length of tube released in one step, as in the free vortex-ring actuator disc \cite{Yu2017} and in the annular vortex elements of a wind turbine wake \cite{Tian2024}. From its release the ring drifts with the ambient wind plus $\gamma/2$ along the axis of its disc at release, so that the rings lie $(V_\infty + \gamma/2)\Delta t$ apart, the tube wall advances at the mean of the axial velocities outside and inside, and the tube carries exactly the strength $\gamma$. The rings drift without the induced velocity, so the tube passes the wing undeformed. Section IV B derives this treatment, tube T3, and compares it with a model in which the wing deflects the slipstream. The rings induce velocities at the closure and load points of the lifting surfaces, in the wake and at the other hubs from the step after their release on. At the lifting-surface points their filaments use a surface core of 0.1~m, and every ring near these points is replaced there by sub-rings at most half this core apart, Eq.~(20) of Sec.~IV B. Elsewhere they use a core equal to their node spacing, 0.45~m, and rings older than 2~s are removed. In generation the thrust is negative, the rings slow the axial flow inside the tube, and their sense is opposite to that of a propeller slipstream. Momentum theory has no solution beyond a thrust coefficient of one in the turbine state \cite{Leishman2006}, where the computation stops. The gyroscopic moment of the rotors is not included. The angular speed is held at 97.4~rad/s, the speed at which the table gives the largest shaft power at the mean commanded airspeed of the 9~m/s loop, 46.5~m/s, and read along the loop at the commanded airspeed the table then gives a mean shaft power of 544~kW and a mean thrust of $-$14.2~kN for the eight rotors. The Makani controller schedules the rotor speed \cite{MakaniCode}, whereas the computations hold it at this one operating point of the table.

\subsection{Several kites and evaluation of the induced velocity}

For the two-unit computations of Sec.~IV C the lifting surfaces of every kite are placed at its own pose and joined into one assembly, so that all panels share one circulation system, Eq.~(6), and one wake, Eq.~(3). The wake of each kite induces velocities on every other kite, the wakes convect together, and with rotors the discs of all kites release their rings into one set of rings. The structural velocity of a panel point is that of the kite it belongs to, identified by the nearest body origin, which every step checks to be unambiguous while the kites stay farther apart than their own size. Each kite has its own tether and ground point. The single-kite computations of Sec.~IV A keep the complete free wake. In the computations with rotors and with two kites and in their references without rotors (Secs.~IV B and IV C), whose number of filaments grows several times faster, a wake row older than one loop keeps its circulation and its induction on every other element but is convected with the ambient velocity at its own position instead of Eq.~(4). This is the buffer zone of OLAF, whose markers ``convect at the average ambient velocity'' between the free and the truncated wake \cite{Shaler2020}, without the truncation. The frozen far wake leaves the first loop of a run unchanged, and Sec.~III measures its effect on the loads against the complete free wake for one unit, while Sec.~IV C compares it with a run in which two loops convect freely for two units.

The cost of a step is dominated by the convection, Eq.~(4), and grows with the square of the number of wake segments. Three interchangeable evaluations of Eq.~(2) are implemented, a NumPy reference \cite{Harris2020}, a direct sum compiled with Numba \cite{Lam2015} with the same summation order and the targets distributed over threads, and a tree code. The tree code is of Barnes--Hut type \cite{Barnes1986} with Cartesian multipole expansions as used for vortex sheets by Lindsay and Krasny \cite{Lindsay2001}. Unlike the tree option of OLAF, which represents the segments by particles \cite{Shaler2020}, it keeps the segments and expands the line integrals of their strength vectors $\Gamma_k\mathbf{L}_k$ about the centre of a node up to second order, with the exact second moment $L_jL_l/12$ of a straight segment, so that, as in other multipole methods \cite{Greengard1987,Lindsay2001}, the relative truncation error of a node decreases like $\theta^{3}$ with the opening parameter $\theta$. A node of radius $R$ at distance $D$ is evaluated by its expansion only when $R < \theta D$, the target lies at least three core radii outside the sphere that contains the node and no segment incident to the target lies in the node. Otherwise the children of the node are visited, the leaves are summed exactly with Eq.~(2), so that excluded pairs and pairs within three core radii of a node are always summed exactly, and the far field omits the core factor. The tree is applied to the wake-on-wake sums of the convection only, whereas $\mathbf{A}$ and $\mathbf{u}^{\text{hist}}$ are summed directly, so that the tree affects the loads through the convected wake shape alone. The settings used throughout are $\theta = 0.2$, second-order moments and at most 32~segments per leaf.

\section{Models and verification}\label{sec:iii}

\subsection{Kite models}

Two public configurations are computed (Fig.~3, Table~II), an arched soft kite of low aspect ratio without a tail, flown on a pumping (ground-generation) system, and a rigid wing of high aspect ratio whose tail operates in the wake of the wing, built for on-board (fly) generation. In the parameter study of this section both are held fixed in a uniform stream, and from Sec.~IV on the M600 flies its crosswind loop on its tether. The aerodynamic models consist of 36~spanwise panels for the V3 and of 34, 8 and 7 panels on the wing, the horizontal tail and the vertical tail of the M600. Figures~3(a) and 3(c) draw the bridle, a shortened tether and the ground station schematically and the fuselage of the M600 for orientation only. Figures~3(b) and 3(d) show the free wakes 20 chords (V3 at 5.41$^\circ$) and 40 chords (M600 at 0$^\circ$) after the impulsive start, with all trailing filaments and every third (V3) or fourth (M600) row of shed filaments coloured by the magnitude of their circulation, and the starting vortex at the downstream end exceeds the colour range.

\begin{figure*}[!t]
\centering
\includegraphics[width=0.75\textwidth]{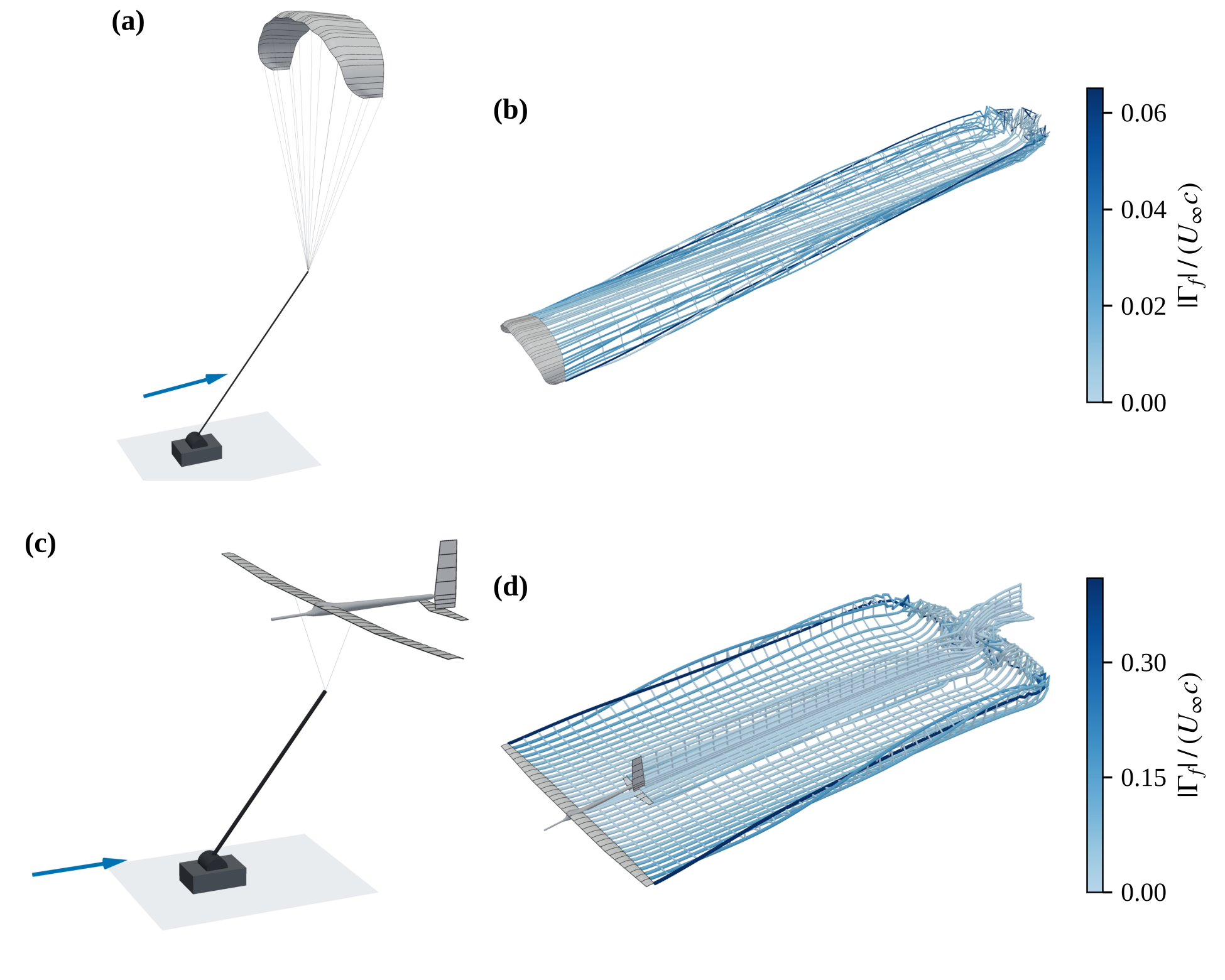}
\caption{\label{fig:3}The V3 kite (a, b) and the M600 (c, d) with their ground units and their free wakes after the impulsive start.}
\end{figure*}

The V3 is a 25~m$^2$ leading-edge inflatable kite of TU Delft, flown on a 20~kW pumping system whose test flights are documented with the elevation, azimuth and tether force of the traction phase \cite{vanderVlugt2019,Oehler2019}. Poland et al. measured the loads of a rigid model at a scale of 1 to 6.5 in an open-jet wind tunnel and reported them with 99~\% confidence intervals \cite{Poland2026load}, and the model reproduces the undeformed design geometry of the RANS simulations of Viré et al. \cite{Vire2020,Vire2022}. The computational model is built from the public geometry \cite{V3Mesh,Poland2026load} with section polars from two-dimensional RANS simulations at $Re = 5\times10^5$, distributed with the open-source vortex step code \cite{VSMCode}, whose configuration file states $10^6$, a value the code does not read. The coefficients use the reference quantities of the wind-tunnel study scaled to full size, the free-stream speed follows from $Re = 5\times10^5$, the three angles of attack are corrected angles of the wind-tunnel data set \cite{Poland2026load,V3LoadData}, and the pitching moment is taken about the reference point of the wind-tunnel study in the comparison with the measurements and about the origin of the geometry file in the parameter study.

The M600 is the 600~kW energy kite of Makani \cite{Echeverri2020}. Its main wing, horizontal tail and vertical tail are reconstructed from the vortex-lattice input of the open-source Makani release \cite{MakaniCode}, with all control surfaces at zero deflection. Each section has the lift curve of thin-aerofoil theory for its mean line, with slope $2\pi$, a constant moment coefficient and no profile drag. The tail lies in the wake of the wing, which makes the pitching moment a sensitive indicator of the wake solution, and the moment is taken about the origin of the geometry file. The tether, bridle, rotors and crosswind playbook of the flight cases are those of the open-source Makani code \cite{MakaniCode}, namely a 425.8~m line of 29.4~mm diameter, the bridle pivot and leg length of the M600, and the land playbook at 9~m/s, whose smoothed entry commands a loop of 135~m radius at an elevation of 0.556 rad. The declared tether strain that places the circle on the reachable sphere is the value at which the line alone gives, over a loop, the mean tension that the rigid-body balance of the prescribed motion requires, 0.0054.

\begin{table*}[!t]
\caption{\label{tab:2}The two kite models and the baseline settings of the fixed-kite study and of the flight computations.}
\small\setlength{\tabcolsep}{5pt}\renewcommand{\arraystretch}{1.12}
\begin{tabular}{@{}>{\raggedright\arraybackslash}p{0.26\textwidth}>{\raggedright\arraybackslash}p{0.27\textwidth}>{\raggedright\arraybackslash}p{0.41\textwidth}@{}}
\toprule
 & V3 kite & M600 wing and tails \\
\midrule
\multicolumn{3}{@{}l}{\emph{Geometry and discretisation}} \\
Geometry & public design geometry \cite{V3Mesh,Poland2026load} & vortex-lattice input of the Makani release \cite{MakaniCode} \\
Lifting surfaces, spanwise panels & wing, 36 & wing 34, horizontal tail 8, vertical tail 7 \\
Panel width, smallest / mean (m) & 0.20 / 0.32 & wing 0.30 / 0.76, horizontal tail 0.62, vertical tail 0.24 / 0.51 \\
Section data & two-dimensional RANS polars \cite{VSMCode}, $-10^\circ$ to $21.5^\circ$ & thin-aerofoil theory, no profile drag \\
Reference area $S$ (m$^2$), chord $c$ (m), span $b$ & 19.435, 2.5675, $3.2\,c$ & 32.9285, 1.2831, $20\,c$ \\
\addlinespace[2pt]
\multicolumn{3}{@{}l}{\emph{Fixed kite}} \\
Free stream $U_\infty$, angles of attack & 2.851~m/s; $3.08^\circ$, $5.41^\circ$, $7.35^\circ$ & 20~m/s; $-4^\circ$, $0^\circ$, $4^\circ$ \\
Time step $\Delta t$ ($U_\infty\Delta t/c$), ladder time & 0.2~s (0.22), 16~s ($17.8\,c$) & 0.02~s (0.31), 1.6~s ($24.9\,c$) \\
Core radius $r_c$ ($r_c/c$) & 0.15~m (0.058) & 0.15~m (0.117) \\
Reference data & wind tunnel, RANS, same-input VSM & same-input VSM \\
\addlinespace[2pt]
\multicolumn{3}{@{}l}{\emph{Crosswind flight (M600)}} \\
Wind, kite speed, incidence & & 5, 9 and 13~m/s; playbook, 45~m/s mean at 9~m/s; playbook, $1.7^\circ$ to $4.9^\circ$ at 9~m/s \\
Tether & & 425.8~m, 29.4~mm, 0.917~kg/m, $EA$ 18~MN, $C_{d,t}$ 0.7 \cite{MakaniCode}, 19 nodes, 10 sub-steps \\
Rotors & & 8 actuator discs of 2.3~m, Rev4 table \cite{MakaniCode}, 97.4~rad/s, rings of 16 filaments (tube T3), surface core 0.1~m, ring core 0.45~m, kept 2~s \\
Time step, steps per loop, far wake & & 0.02~s (0.01~s for one kite with rotors, 0.005~s for model D), 936 steps at 9~m/s; complete (Sec.~IV\,A) or frozen beyond one loop (Secs.~IV\,B, IV\,C) \\
\bottomrule
\end{tabular}
\end{table*}

\subsection{Verification}

The straight-wake VSM of OpenFAWE was compared with the open-source vortex step code \cite{VSMCode} on the same geometry, polars, reference quantities and flow conditions. On the M600, with the implementation retained in the test suite of that code, which returns no pitching moment, the two codes agree to $10^{-9}$ in $C_L$ and $4\times10^{-10}$ in $C_D$ at all three angles of attack, below the tolerance of the circulation iteration. On the V3, with the current implementation and its aerodynamic-centre load option, tabulated viscous polars and a strongly arched planform, the lift agrees within 0.25~\%, the drag within 0.88~\% and the pitching moment within 0.23~\%, below the 99~\% confidence half-widths of the measured lift and drag \cite{Poland2026load}, 0.7~\% to 2.0~\% and 6.2~\% to 16~\%. Figure~4~shows the coefficients of the V3 for all models together with the wind-tunnel data \cite{Poland2026load,V3LoadData}, the RANS results \cite{Vire2020,Poland2026load} and the same-input code, Table~III gives the differences to the measurements, and Fig.~5~shows the M600.

\begin{figure*}[!t]
\centering
\includegraphics[width=1\textwidth]{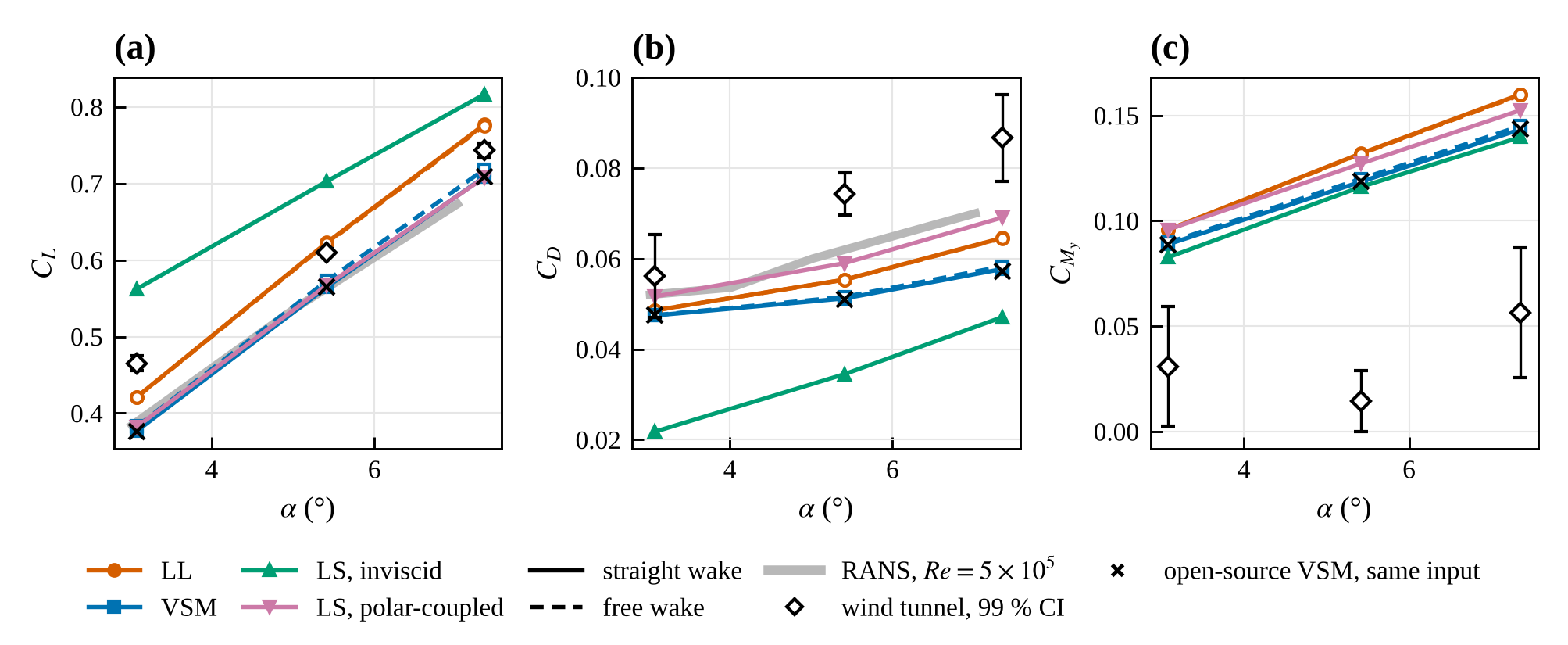}
\caption{\label{fig:4}The fixed V3 kite. (a) Lift, (b) drag and (c) pitching-moment coefficient against the angle of attack for the three bound-vortex models with a straight wake (solid) and a free wake of 40 chords (dashed), the RANS result without struts, the wind-tunnel measurements with their 99~\% confidence intervals and the open-source vortex step code on the same input. The pitching moment is taken about the reference point of the wind-tunnel study.}
\end{figure*}

\begin{table}[!t]
\caption{\label{tab:3}Relative difference of the computed lift and drag coefficients of the V3 to the wind-tunnel values at the three angles of attack. The 99~\% confidence half-widths of the measurements are 2.0~\%, 0.7~\% and 1.3~\% of $C_L$ and 16~\%, 6.2~\% and 11~\% of $C_D$, and the RANS values are interpolated linearly to the measured angles.}
\footnotesize\setlength{\tabcolsep}{3pt}\renewcommand{\arraystretch}{1.1}
\begin{tabular}{@{}llrrrrrr@{}}
\toprule
 & & \multicolumn{3}{c}{$C_L$ difference (\%)} & \multicolumn{3}{c}{$C_D$ difference (\%)} \\
\cmidrule(lr){3-5}\cmidrule(l){6-8}
Model & Wake & $3.08^\circ$ & $5.41^\circ$ & $7.35^\circ$ & $3.08^\circ$ & $5.41^\circ$ & $7.35^\circ$ \\
\midrule
VSM & straight & $-19.0$ & $-7.5$ & $-4.7$ & $-15.5$ & $-31.1$ & $-33.4$ \\
VSM & free & $-17.7$ & $-6.1$ & $-3.3$ & $-15.5$ & $-30.6$ & $-32.7$ \\
LL & straight & $-9.4$ & $+2.0$ & $+4.5$ & $-13.5$ & $-25.5$ & $-25.5$ \\
LL & free & $-9.5$ & $+1.9$ & $+4.3$ & $-13.5$ & $-25.6$ & $-25.6$ \\
LSi & straight & $+20.8$ & $+15.1$ & $+9.9$ & $-61.2$ & $-53.6$ & $-45.7$ \\
LSp & straight & $-17.8$ & $-7.1$ & $-4.8$ & $-8.1$ & $-20.6$ & $-20.3$ \\
RANS \cite{Vire2020,Poland2026load} & & $-16.4$ & $-7.7$ & n.a. & $-7.2$ & $-16.6$ & n.a. \\
\bottomrule
\end{tabular}
\end{table}

The polar-based models with a three-quarter-chord or lattice closure agree with RANS in lift. At 3.08$^\circ$ and 5.41$^\circ$ the VSM differs from the RANS values by $-$3.1~\% and +0.2~\% with the straight wake and by $-$1.6~\% and +1.7~\% with the free wake, and the polar-coupled lifting surface by $-$1.7~\% and +0.6~\%. These models and RANS all lie below the measured lift, by 16~\% to 19~\% at 3.08$^\circ$ and by 3~\% to 8~\% at the two higher angles, and below the measured drag at all three angles, and Poland et al. found that all numerical models predict a higher maximum lift-to-drag ratio than was measured, noting that the scale model has a thicker canopy \cite{Poland2026load}. The lifting line gives 8~\% to 12~\% more lift than the VSM, which places it closer to the measurement at 5.41$^\circ$ and further from RANS, and the inviscid lifting surface, without the viscous polar, overshoots the lift and undershoots the drag by far. The computed pitching moments lie between 0.08 and 0.16 and the measured ones between 0.01 and 0.06 with confidence half-widths of 0.015 to 0.03 [Fig.~4(c)], an offset that the same-input open-source code shows as well and that a mismatch of the moment reference point between measurement and simulation shifts directly \cite{Poland2026load}.

\begin{figure}[!htb]
\centering
\includegraphics[width=1\linewidth]{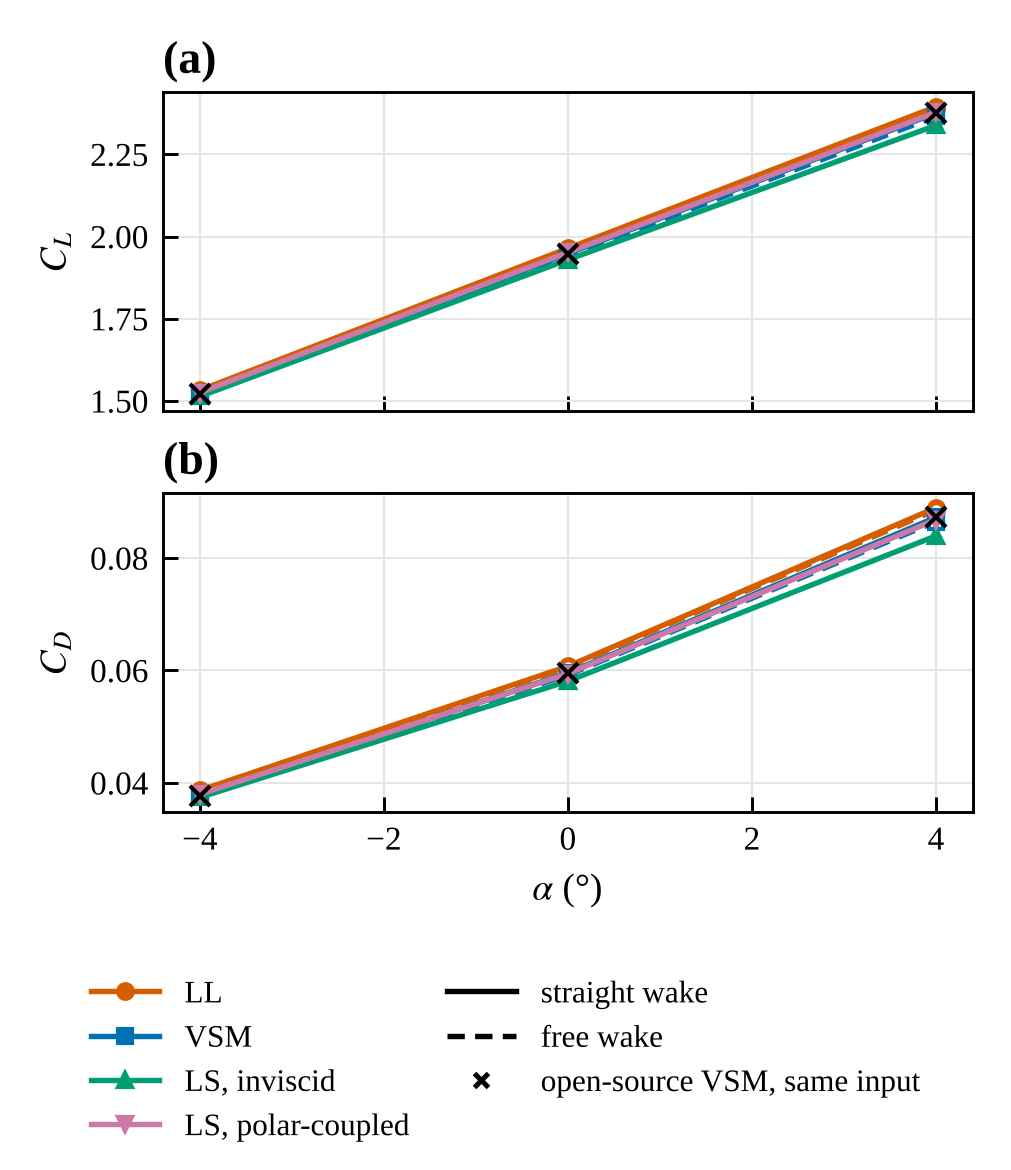}
\caption{\label{fig:5}Lift (a) and drag (b) of the fixed M600 against the angle of attack, with the symbols and lines of Fig.~4 and free wakes of 103 chords.}
\end{figure}

On the M600 (Fig.~5) the four bound-vortex models with a straight wake differ from the VSM by at most 1.6~\% in lift and 4~\% in drag at 4$^\circ$, and among themselves by 2.4~\% and 6.0~\%. With converged free wakes the two wake models agree within 0.5~\% in lift and 0.9~\% in drag for both closures, so that for a fixed wing in a uniform stream the deformation of the wake is a second-order effect, and on the V3 they agree within 0.2~\% for the lifting line, while the VSM gives a free-wake lift 1.4~\% to 1.6~\% above the straight-wake value, a core effect of the arched tip panels explained in Sec.~III D. With converged free wakes the lifting line gives 0.8~\% to 0.9~\% more lift than the VSM on the M600 and 8~\% to 10~\% more on the V3, so that the difference between the quarter-chord and the three-quarter-chord closure is small for the wing of high aspect ratio and large for the arched kite. The three-quarter-chord closure, which agrees with RANS on the arched kite and with the same-input code to the solver tolerance on the rigid wing, is retained for the rest of the paper.

Four checks apply to the moving kite, of invariance and consistency before the flight computations, of periodicity on every flight run and of superposition for two kites. A kite translating through still air gives the loads of the fixed kite in the opposite wind step by step, and over 40~steps of the free wake the largest relative difference of the force is $3.4\times10^{-13}$ for the V3 and $2.2\times10^{-14}$ for the M600, and of the moment about the body origin $5.9\times10^{-12}$ and $2.2\times10^{-12}$, against a limit of $10^{-8}$ set before the check. For every motion law the velocity and the acceleration are compared with fourth-order central differences of the sampled position and velocity, and the angular velocity with the same stencil applied to the rotation vectors of the sampled attitude, and the largest relative difference is $1.8\times10^{-7}$ for the Makani loop at 0.02~s, below the limit of $10^{-6}$ set before the runs. A run is accepted as periodic when the largest relative difference of the total load of the kite, aerodynamic, rotor and tether, between consecutive periods at equal phase stays below 1~\% over the last two periods in force and in moment. For the downstream unit of the two-unit runs the same comparison is also made for the aerodynamic load of wing and tails alone (Sec.~IV C 2). With two kites placed 20 km apart, the joint computation reproduces the two single-kite computations to $10^{-9}$ in the force and the moment of every step, with and without rotors, and with one kite it reproduces the single-kite computation bit for bit, rotor loads and rings included.

The frozen far wake and the evaluation on a graphics processor were checked on the M600 in flight. On a circle of 130~m radius at 47~m/s in a 9~m/s wind, the first check case of the motion interface, without tether, a run with the newest lap convected freely and the older laps drifting with the wind gives at equal phase of the third lap a force that differs from the complete free wake by at most $2.9\times10^{-5}$ in relative terms, and the runs of Secs.~IV B and IV C include cases with twice that free length. Part of the runs with rotors and two kites evaluate the tree code of the convection on a graphics processor in mixed precision, with the pairs close to a filament axis evaluated again in double precision, and the others use the double-precision tree code on the CPU. On five stored wake states of the M600 with up to 1872 rows the largest difference to the double-precision tree code was $3.9\times10^{-6}$ of the induced velocity, two orders of magnitude below the error of the tree code against the direct sum, $3.6\times10^{-4}$ to $2.7\times10^{-3}$. The first loop of the run without rotors on the graphics processor reproduces the lap-averaged loads of Sec.~IV A, computed on the CPU, to $1.3\times10^{-5}$ in $C_Z$ and $5.2\times10^{-5}$ in the aerodynamic power, relative, and with the far wake frozen beyond one loop the next three loops reproduce the complete free wake to $1.6\times10^{-5}$ and $9.5\times10^{-5}$, and $C_Z$ at every step to $1.7\times10^{-4}$.

The accelerated evaluations were compared with the reference sum over complete 40-step histories of both models with both closures, where the coefficients of the compiled direct sum differ from the NumPy reference by at most $5.1\times10^{-13}$ (VSM-F) and $4.9\times10^{-12}$ (LL-F) and are bitwise independent of the number of threads. The error of the tree code against the direct sum was required to stay within $|\Delta C_L|/|C_L| \le 5\times10^{-4}$, $|\Delta C_D|/|C_D| \le 5\times10^{-4}$ and $|\Delta C_{M_y}| \le 2\times10^{-4}$ at every step of a 200-step history and below one tenth of the smallest discretisation effect of the convergence ladders. With $\theta = 0.2$ and second-order moments the worst-step differences are $4.6\times10^{-6}$ in $C_L$, $2.5\times10^{-6}$ in $C_D$ and $9.0\times10^{-7}$ in $C_{M_y}$ for the V3 and $7.8\times10^{-6}$, $7.1\times10^{-5}$ and $3.4\times10^{-5}$ for the M600, whereas with $\theta = 0.3$ the M600 exceeds the moment budget while the starting vortex passes the tail. One wake-on-wake evaluation on the 400-step wake of the M600, with 20 852 targets and 40 501~segments, takes 38.7~s with the direct sum on one thread of an AMD Ryzen 7 255 and 4.2~s on 16 threads, and the tree code takes 2.2~s and 0.49~s.

\subsection{Wake length}

The convergence of the loads of the fixed configurations with the wake length, the time step, the spanwise panelling and the core radius was measured by refinement ladders in the manner of grid-refinement studies \cite{Roache1994,Celik2008}, with the levels and the tree error budget fixed before the runs. For the time step the observed order is one, the formal order of the half-explicit Euler scheme, and $U_\infty\Delta t/c$ of 0.22 (V3) and 0.31 (M600) keeps the lift within 0.02~\% and 0.004~\% of the finest level, so the wake length and the core radius are the parameters that matter. For an impulsively started wing, wake length and elapsed time coincide, because the starting vortex sits at the downstream end of the wake, and a wake that is too short produces the largest errors of the fixed-kite study. Table~IV and Fig.~6(a) follow the coefficients of the M600 as the starting vortex moves downstream. The lift at 20 chords is still 0.8~\% low and the drag 5~\% high, and drag and pitching moment come within 0.3~\% of their final values only at 80 chords, whereas the lift and drag of the V3 are within 0.07~\% of their 80-chord values at 20 chords. The deviations decay with the inverse square of the distance of the starting vortex, and the transient is the finite-wing counterpart of the lift build-up after an impulsive start described by Wagner \cite{Wagner1925} and Jones \cite{Jones1940}. Drag and pitching moment of the M600 are more sensitive than its lift because the tail, four to five chords behind the wing, responds to the downwash of the developing wake. In a cross-flow plane 1.6 chords behind the horizontal tail, a wake of 10 chords gives a downwash between the tip vortices that is stronger than in the developed wake by a nearly uniform $0.02\,U_\infty$, almost 30~\% more, and from 40 chords on the profiles coincide within $0.001\,U_\infty$. The same behaviour holds at a second angle of attack and for the lifting line. For the V3 at 7.35$^\circ$ the lift changes by 0.23~\%, 0.05~\% and 0.01~\% as the wake grows from 10 to 80 chords, and with the lifting line the M600 lift changes by 0.09~\% between 50 and 100 chords, so the wake-length requirement does not depend on the closure.

\begin{table}[!t]
\caption{\label{tab:4}Coefficients of the fixed M600 at 0$^\circ$ with VSM-F against the distance of the starting vortex behind the trailing edge, and their deviation from the value at 160 chords, for the pitching moment the deviation of its magnitude.}
\footnotesize\setlength{\tabcolsep}{3pt}\renewcommand{\arraystretch}{1.1}
\begin{tabular}{@{}lrrrrrr@{}}
\toprule
Wake & \multicolumn{2}{c}{$C_L$} & \multicolumn{2}{c}{$C_D$} & \multicolumn{2}{c}{$C_{M_y}$} \\
\cmidrule(lr){2-3}\cmidrule(lr){4-5}\cmidrule(l){6-7}
length & value & dev. (\%) & value & dev. (\%) & value & dev. (\%) \\
\midrule
$10\,c$ & 1.881342 & $-3.1$ & 0.068691 & $+16$ & $-0.038768$ & $-55$ \\
$20\,c$ & 1.926112 & $-0.83$ & 0.062206 & $+5.2$ & $-0.078333$ & $-9.9$ \\
$40\,c$ & 1.938629 & $-0.18$ & 0.059881 & $+1.3$ & $-0.085493$ & $-1.7$ \\
$80\,c$ & 1.941550 & $-0.03$ & 0.059276 & $+0.25$ & $-0.086732$ & $-0.29$ \\
$160\,c$ & 1.942216 & & 0.059129 & & $-0.086982$ & \\
\bottomrule
\end{tabular}
\end{table}

A wake of length $L$, closed by its starting vortex, differs from the steady semi-infinite wake by a horseshoe vortex of strength $-\Gamma_0$ whose bound segment is the starting vortex and whose legs run from there to infinity. At the wing centre, a distance $x$ upstream of the starting vortex, this horseshoe induces the downwash

\begin{equation}
\begin{aligned} \Delta w &= \Gamma_0\left[\frac{b'}{4\pi x\sqrt{x^2+b'^2/4}} - \frac{1}{\pi b'}\left(1-\frac{x}{\sqrt{x^2+b'^2/4}}\right)\right] \\ &\longrightarrow\; \frac{\Gamma_0\, b'}{8\pi x^2} \qquad (x \gg b'), \end{aligned}
\label{eq:12}
\end{equation}

\noindent with $b'$ the spacing of the legs. The starting vortex contributes $\Gamma_0 b'/(4\pi x^2)$ and the missing legs remove half of it. With $b'$ the Betz spacing, twice the lateral centroid $y_c = \int_0^{b/2}\Gamma\,\mathrm{d}y/\Gamma_0$ of the trailing vorticity of a half wing \cite{Betz1932}, the Kutta--Joukowski theorem \cite{Katz2001} gives for any spanwise loading

\begin{equation}
\Gamma_0\, b' = 2\int_0^{b/2}\Gamma\,\mathrm{d}y = \frac{F_L}{\rho U_\infty} = \tfrac12\,U_\infty S\, C_L 
\label{eq:13}
\end{equation}

\noindent with $F_L$ the lift, and for a configuration with several lifting surfaces the integral is the total over all of them. A uniform downwash $\Delta w$ acts like a change of incidence by $-\Delta w/U_\infty$, as the induced downwash does in lifting-line theory \cite{Prandtl1923,Katz2001}. With $C_w = -\partial C_L/\partial(w/U_\infty)$ the lift response of the wing to a uniform downwash, the relative lift deficit of a finite wake and the wake length for a deficit $\varepsilon$ follow as

\begin{equation}
\varepsilon_L = \frac{C_w\,\Delta w}{U_\infty C_L} = \frac{C_w S}{16\pi x^2}, \qquad L_\varepsilon = \left(\frac{C_w S}{16\pi\varepsilon}\right)^{1/2} - \frac{c}{4}
\label{eq:14}
\end{equation}

\noindent where $x = L + c/4$ is measured from the closure points. The deficit depends neither on $C_L$ nor on the spanwise loading, only on the reference area and on the response of the wing to a uniform downwash. A steady solution with a downwash of $10^{-3}\,U_\infty$ added at all closure and load points gives $C_w = 6.05$ for the M600 and 4.55 for the V3, within 1.2~\% and 3.2~\% of their lift slopes. In wing spans, $L_\varepsilon/b \approx (C_w/16\pi\varepsilon A)^{1/2}$ with the aspect ratio $A = b^2/S$, so a wing of high aspect ratio needs more chords but fewer spans than a kite of low aspect ratio. The prescribed straight wake, whose vortex rings all carry the current circulation and end at $L$, isolates the starting vortex, and Eq.~(14) reproduces its lift deficit within 5.1~\% from 7 to 320 chords for the M600 (Fig.~6) and within 3~\% from 5 to 320 chords for the V3, without a fitted constant. Below 7 chords the starting vortex of the M600 is near its tail. For $\varepsilon = 10^{-3}$, Eq.~(14) gives wakes of 49 chords (2.4~spans) for the M600 and 16 chords (5.0~spans) for the V3.

\begin{figure}[!htb]
\centering
\includegraphics[width=1\linewidth]{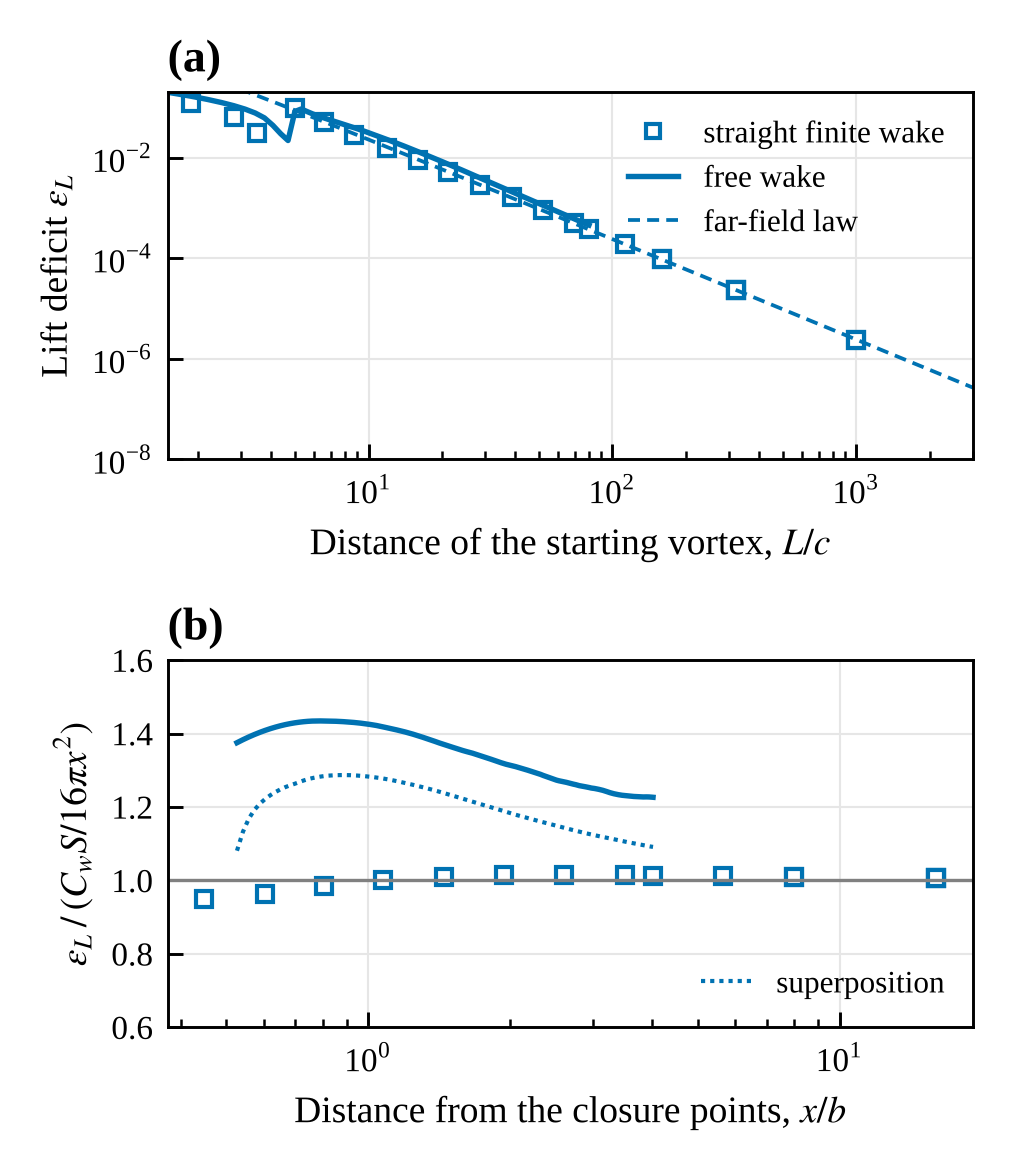}
\caption{\label{fig:6}Lift deficit of a finite wake on the fixed M600. (a) Against the distance of the starting vortex, (b) divided by Eq.~(14) against the distance from the closure points in spans.}
\end{figure}

In Fig.~6 the symbols give the prescribed straight wake that ends at $L$, the solid lines the free wake and the dashed line Eq.~(14), and the dotted line in Fig.~6(b) is the superposition of the straight-wake response over the shed history of the free wake introduced below. The free-wake deficits are taken relative to the value extrapolated from the two longest wakes of Table~IV with order two. The same downwash turns the force backward by $\Delta w/U_\infty$, the mechanism of induced drag \cite{Prandtl1923,Katz2001}, and changes the drag with the incidence, so that the relative drag error is $\kappa_D\,\varepsilon_L$ with

\begin{equation}
\kappa_D = \frac{\left(C_L - \partial C_D/\partial\alpha\right)C_L}{C_w\, C_D}.
\label{eq:15}
\end{equation}

For the inviscid sections of the M600, whose drag is induced drag \cite{Prandtl1923} and grows with $C_L^2$, $\partial C_D/\partial\alpha = 2C_D C_w/C_L$ and Eq.~(15) becomes $C_L^2/(C_w C_D) - 2 = 8.5$. The uniform-downwash solution gives 8.6 for the M600 and 1.0 for the V3, so a wing of high lift-to-drag ratio needs $\kappa_D^{1/2} \approx 2.9$ times the wake length of its lift to reach the same relative accuracy in drag. With Eqs.~(14) and (15), the M600 drag error is 0.32~\% at 80 chords and 0.08~\% at 160 chords, and the difference, 0.24~\%, is close to the 0.25~\% between the free-wake values of Table~IV. The pitching moment of the M600 converges slowest of all, because it is the small difference of the contributions of wing and tail and the tail sees the starting vortex at a shorter distance than the wing, and Eq.~(14) does not cover it.

In the free wake the lift deficit of the M600 is larger than Eq.~(14) by factors of 1.37, 1.43, 1.31 and 1.23 at 10, 20, 40 and 80 chords [Fig.~6(b)], and that of the V3 by 1.23, 1.10 and 1.04 at 10, 20 and 40 chords. Two mechanisms contribute. The circulation builds up after the impulsive start, so that the impulse $G = \sum_j \Gamma_j\,\mathrm{d}l_{y,j}$ of the released vortex rings, which is $F_L/(\rho U_\infty)$ by Eq.~(13), reaches 58~\% (M600) and 61~\% (V3) of its final value in the first step, 91~\% and 94~\% after ten steps, and 98~\% for the M600 only after 41~steps, 12.8 chords. The shed vorticity of this build-up lies between the wing and the starting vortex, and because the deficit is small it is the superposition $\varepsilon_F(t_n) = \sum_{m} (\Delta G_m/G_\infty)\,\varepsilon_S(x_m)$ of the straight-wake responses $\varepsilon_S$ to all impulse increments $\Delta G_m$ shed at the present distances $x_m$, the finite-wing counterpart of the superposition of indicial responses in unsteady aerofoil theory \cite{Leishman2006}. This superposition reproduces the free-wake deficit of the V3 within 1.3~\% from 7 to 40 chords, whereas the free-wake deficit of the M600~stays 11~\% to 13~\% above it from 14 to 80 chords. The second mechanism is the descent of the wake. Solving the closure on the free-wake geometry with all vortex rings at the current circulation removes the history and leaves the shape, and relative to the straight geometry of the same length it raises the deficit of the M600 by 12~\% at 40 chords and by 14~\% at 80 chords and changes that of the V3 by $-$1.5~\% and $-$0.5~\%. The impulse-weighted centroid of the shed vorticity of the free wake lies below the wing by 4.2~\% and 4.0~\% of its distance for the M600 at 40 and 80 chords. The field of this descended far end has a streamwise component $\Delta u$ at the closure points besides the downwash, and to first order

\begin{equation}
\varepsilon_L = \frac{C_w\,\langle\Delta w\rangle}{U_\infty C_L} - \frac{2\,\langle\Delta u\rangle}{U_\infty}
\label{eq:16}
\end{equation}

\noindent where the second term is the change of dynamic pressure and $\langle\cdot\rangle$ is the average over the closure points weighted by $\Gamma_j\,\mathrm{d}l_{y,j}$. Evaluated with the influence operators of Eq.~(6) and with the same reference for both sides, Eq.~(16) reproduces the deficits of both geometries within 3~\%. For the straight wake the dynamic-pressure term is at most 1.2~\% of the downwash term, and for the free wake of the M600 it is 13~\% and 12~\% at 40 and 80 chords, which accounts for the effect of the shape. For the V3 it is 3~\%, and a slightly smaller downwash term makes the net effect of the shape small and negative. The ratio of the two terms is $2(\Delta u/\Delta w)(C_L/C_w)$, which for comparable descent is larger for the M600 because its ratio $C_L/C_w$, an effective incidence, is 0.32 against 0.13 for the V3. Kites that fly at high lift coefficients therefore need free wakes longer than Eq.~(14) gives, which the rules of Sec.~III E cover with a factor of about 1.2.

The free wake behind the fixed kites rolls up as an inviscid sheet should, without the secondary vortices and the turbulence that enter the core of a real tip vortex at moderate Reynolds numbers \cite{Smith2021}. For the trailing filaments of the right half of the main wing, with strengths $\gamma_j$ and positions $(y_j, n_j)$ in a cross-flow plane, the circulation $\Gamma_h = \sum_j\gamma_j$ and the centroid $(y_c, n_c) = \sum_j \gamma_j (y_j, n_j)/\Gamma_h$ measure the roll-up. In an inviscid wake $y_c$ keeps the value $\int_0^{b/2}\Gamma\,\mathrm{d}y/\Gamma_0$ of the wing \cite{Betz1932}, and a pair of rolled-up vortices of strength $\Gamma_h$ at the spacing $b' = 2y_c$ descends with the velocity \cite{Lamb1932}

\begin{equation}
w_p = \frac{\Gamma_h}{2\pi b'} .
\label{eq:17}
\end{equation}

\begin{figure}[!htb]
\centering
\includegraphics[width=1\linewidth]{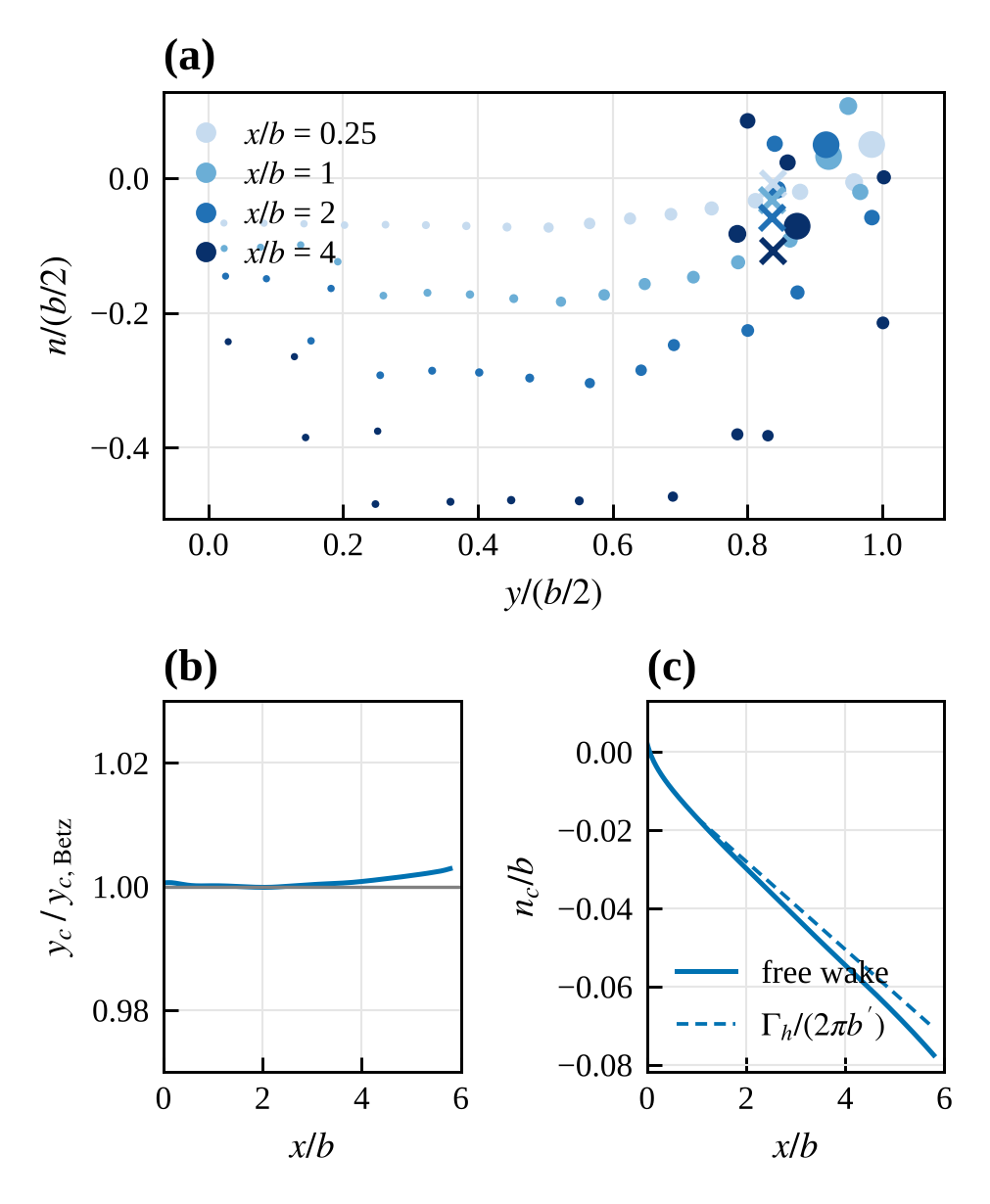}
\caption{\label{fig:7}Roll-up and descent behind the fixed M600. (a) Trailing filaments of the right half of the wing in cross-flow planes, (b) lateral and (c) vertical centroid of the trailing vorticity.}
\end{figure}

Figure~7 follows the 160-chord wake of the M600 at 0$^\circ$. In Fig.~7(a) the marker area is proportional to the filament circulation, $n$ is measured from the trailing-edge centroid of the main wing and the crosses are the centroids, and in Fig.~7(c) $n$ is measured from the trailing-edge centroid of all lifting surfaces and the dashed line is the descent with the velocity of Eq.~(17) from the centroid at $x = b$. The Betz spacing of the M600 wing is $b' = 0.836\,b$, above the value $\pi b/4$ that Eq.~(13) gives for elliptic loading, and the lateral centroid keeps the Betz value within 0.3~\% over six spans. The tip filament carries 34~\% of $\Gamma_h$, and from one span on about half of $\Gamma_h$ lies within one chord of it. The vertical centroid descends at 1.15 and 1.10 times $w_p$ between one and two and between two and four spans, and the excess decreases as the sheet rolls up and the bound vortex recedes. The weakly loaded inner sheet moves down faster than the centroid, as the fluid between a vortex pair does \cite{Lamb1932}. The arched wake of the V3, with a Betz spacing of $0.819\,b$, contracts by 0.9~\% within the first half span, then keeps its centroid within 1~\% over ten spans and descends at 1.10 and 1.04 times $w_p$ between one and three and between three and six spans. The free wake thus reproduces the inviscid invariants, which it does not impose, and Eq.~(17) gives the path of the wake of an upstream wing with an error of 4~\% to 15~\% over the first six spans.

\subsection{Vortex core}

The vortex core decides whether the closure is well posed. For wind turbines computed with a free vortex wake, an initial core radius of more than half the blade chord was found to be needed for stable convergence and accurate loads \cite{Xu2019}, whereas here the opposite bound appears, and with every panel divided in three at the baseline core of 0.15~m the Newton iteration stops converging on both models. Of the 24 free-wake runs of the spanwise and core ladders (V3 at 5.41$^\circ$ and 7.35$^\circ$, M600 at 0$^\circ$, with one to three sub-panels per panel and cores from 0.025 to 0.30~m), every run with $r_c \le 0.96\,w_{\min}$ but one completed, with $w_{\min}$ the smallest spanwise panel width, every run with $r_c \ge 1.9\,w_{\min}$ failed in the closure, and of the six runs between 1.1 and 1.5 five completed. The exception, the M600 with every panel divided in three and the core scaled down to 0.05~m ($r_c = 0.64\,w_{\min}$), failed by a reversed chordwise flow at a tail load point while the starting vortex passed the tail. The closure point of a VSM panel lies near mid-width, about half a panel width from the two trailing filaments that leave the ends of its own bound segment. At this point Eq.~(2) with a core radius equal to the panel width reduces the induction of these two filaments to 27~\% of its unregularised value, and a core of twice the width to 8~\%, while the two-dimensional term of Eq.~(7) is not regularised. The balance between three-dimensional induction and two-dimensional correction on which the closure rests is then lost on the narrowest panels. The lifting line keeps this balance, and its runs with every panel divided in three complete on both models with the 0.15~m core, so the OLAF guideline for the lifting line, a regularisation parameter of twice the average spanwise discretisation \cite{Shaler2020}, does not carry over to the VSM closure.

\begin{figure}[!htb]
\centering
\includegraphics[width=1\linewidth]{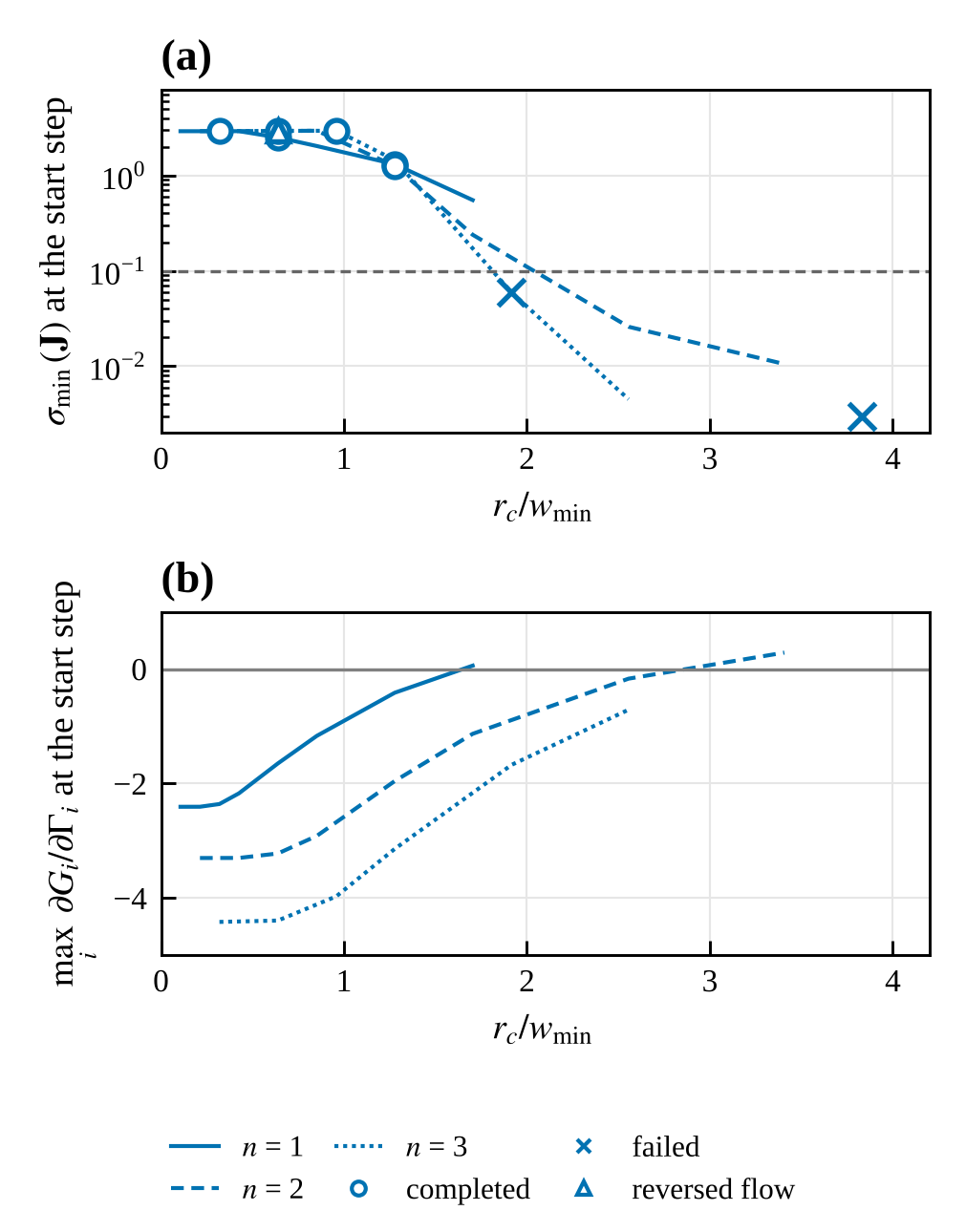}
\caption{\label{fig:8}Closure Jacobian of the M600 at the start step. (a) Smallest singular value and (b) largest self-coefficient against the ratio of the core radius to the smallest panel width.}
\end{figure}

The Jacobian of Eq.~(9) shows this before a run (Fig.~8, with $n$ sub-panels per panel and the outcome of every free-wake run of the M600, runs without a start-step solution drawn at the lower edge and the dashed line at the limit $\sigma_{\min} = 0.1$ used in Sec.~III E). Its smallest singular value $\sigma_{\min}$ at the start step, when the bound vortex rings alone carry the circulation, lies between 1.8 and 4.2 on both models up to $r_c/w_{\min} = 0.75$, falls by up to two orders of magnitude between $r_c/w_{\min} = 1$ and 2, by nearly three at 2.6, and at the largest ratios the start step has no solution. The self-coefficients show the mechanism. For small cores $\beta_i$ is strongly negative, $-$2.4 to $-$4.4 for the M600 and $-$10 to $-$16 for the V3, because the trailing filaments at the ends of a panel's own bound segment induce a downwash at its closure point that grows with its circulation and lowers its angle of attack. The core suppresses this induction while the two-dimensional term is not regularised, so the largest $\beta_i$ approaches zero and changes sign between $r_c/w_{\min} \approx 1$ and 3.4, and the panel loses its self-stabilising induction. $\sigma_{\min}$ collapses at smaller ratios than this change of sign, for the refined M600 while all $\beta_i$ are still negative, because narrow neighbouring panels share their trailing filaments and their closure equations become nearly dependent. Every completed free-wake run has $\sigma_{\min} \ge 0.106$ at the start step, and every run that failed in the closure has $\sigma_{\min} \le 0.080$ or no start-step solution. The start-step Jacobian, which costs one steady solution, thus identifies the usage boundary before a run, with an empirical criterion that rests on 19 completed runs and four closure failures of the two configurations.

With the core fixed at about one smallest panel width or less, the lift changes little under spanwise refinement. From two to three sub-panels it changes by 0.09~\% for the V3 at 5.41$^\circ$, by 0.17~\% at 7.35$^\circ$ and by 0.13~\% for the M600, the drag by 0.3~\% and 0.2~\% and the pitching moment by 0.9~\% and 0.8~\%. When the core is instead reduced in proportion to the panel width, the V3 lift changes by $-$1.0~\% and $-$0.34~\% between levels, mostly the effect of the changing core, so a spanwise refinement study of a VSM free-wake solution has to be made at a fixed core radius smaller than the finest panel width.

\begin{table}[!t]
\caption{\label{tab:5}Response of the fixed M600 at 0$^\circ$ to halving and doubling the baseline core radius of 0.15~m, with changes relative to 0.15~m, lift and drag with a wake of 25 chords and the pitching moment at 25 and 80 chords, the latter split into the parts of the wing and of the horizontal tail. For the pitching moments the change of the magnitude is given.}
\footnotesize\setlength{\tabcolsep}{2.5pt}\renewcommand{\arraystretch}{1.1}
\begin{tabular}{@{}lrrrrr@{}}
\toprule
 & \multicolumn{3}{c}{Core radius $r_c$} & \multicolumn{2}{c}{Change (\%)} \\
\cmidrule(lr){2-4}\cmidrule(l){5-6}
 & 0.075~m & 0.15~m & 0.30~m & 0.075~m & 0.30~m \\
\midrule
\multicolumn{6}{@{}l}{\emph{Wake of 25 chords}} \\
$C_L$ & 1.932251 & 1.932147 & 1.944221 & $+0.005$ & $+0.6$ \\
$C_D$ & 0.061143 & 0.061127 & 0.061444 & $+0.03$ & $+0.5$ \\
$C_{M_y}$ & $-0.082946$ & $-0.082099$ & $-0.047805$ & $+1.0$ & $-42$ \\
\addlinespace[2pt]
\multicolumn{6}{@{}l}{\emph{Wake of 80 chords}} \\
$C_{M_y}$ & $-0.087560$ & $-0.086733$ & $-0.053583$ & $+1.0$ & $-38$ \\
\quad wing part & $-0.268287$ & $-0.268286$ & $-0.268182$ & $+0.0006$ & $-0.04$ \\
\quad tail part & 0.180717 & 0.181543 & 0.214589 & $-0.5$ & $+18$ \\
\bottomrule
\end{tabular}
\end{table}

Table~V gives the response of VSM-F on the original panelling to halving and doubling the baseline radius. Between 0.075 and 0.15~m the M600 coefficients change by at most 1~\%, and the V3 lift and drag by 1.2~\% and 1.1~\%. Doubling the core to 0.30~m, larger than the smallest panel width of both models, reduces the magnitude of the M600 pitching moment by 42~\% at the ladder time and by 38~\% with an 80-chord wake, and raises the V3 lift by 2.7~\%. The sensitivity follows the ratio of core radius to panel width. The V3 panels are 0.20 to 0.45~m wide, so the baseline core already reduces the induction of a panel's trailing filaments at its own closure point and the lift depends on the core over the whole range, whereas the M600 wing panels are 0.76~m wide on average and its lift is insensitive up to 0.15~m. The contribution of the M600 wing to the pitching moment stays within 0.04~\%. The horizontal tail carries a download whose moment opposes that of the wing, and its contribution grows by 18~\% with the 0.30~m core and accounts for 99.7~\% of the change of the total, which is 38~\% because the total is the small difference of the two contributions. Directly behind the horizontal tail the 0.30~m core changes the downwash by up to $0.005\,U_\infty$, about 10~\% of the local value, and at the mid-width of a horizontal-tail panel, 0.62~m wide, Eq.~(2) leaves 74~\% of the induction of the panel's own trailing filaments with the 0.30~m core, against 99.5~\% with 0.15~m. The pitching moment of a configuration with a tail computed with a regularised free wake is therefore meaningful only together with its core radius, which should be small compared with the narrowest panels, those of the tail included.

\begin{table*}[!t]
\caption{\label{tab:6}Rules for the vortex-model parameters of free-wake VSM computations of kites, with their basis in this section and the values for the M600 at the baseline core of 0.15~m on the original panelling.}
\small\setlength{\tabcolsep}{5pt}\renewcommand{\arraystretch}{1.12}
\begin{tabular}{@{}>{\raggedright\arraybackslash}p{0.17\textwidth}>{\raggedright\arraybackslash}p{0.36\textwidth}>{\raggedright\arraybackslash}p{0.17\textwidth}>{\raggedright\arraybackslash}p{0.22\textwidth}@{}}
\toprule
Parameter & Rule & Basis & M600 \\
\midrule
Wake length, lift & $L \ge (C_w S/16\pi\varepsilon)^{1/2} - c/4$ & Eq.~(\ref{eq:14}), Fig.~\ref{fig:6} & $49\,c$ for $\varepsilon = 10^{-3}$ \\
Wake length, drag & lift value $\times\,\kappa_D^{1/2}$ & Eq.~(\ref{eq:15}) & $143\,c$ with $\kappa_D = 8.6$ \\
Free-wake margin & length $\times$ 1.2 & Eq.~(\ref{eq:16}) and superposition & $59\,c$ (lift) \\
Time step & $U_\infty\Delta t/c \approx 0.2$ to 0.3, first order & refinement ladder & 0.004~\% in $C_L$ at 0.31 \\
Core radius & $r_c \le 0.75\,w_{\min}$ on all surfaces, reported with the result & Table~\ref{tab:5}, Fig.~\ref{fig:8} & $0.64\,w_{\min}$ \\
Check of the closure & $\sigma_{\min}(\mathbf{J}) \ge 0.1$ at the start step & Eq.~(\ref{eq:9}), Fig.~\ref{fig:8} & 2.50 \\
Spanwise refinement & at fixed $r_c$ & refinement ladder & 0.13~\% in $C_L$ from 2 to 3 sub-panels \\
Core growth & not needed for the loads & $\delta = 10$ to 1000 & $< 0.01$~\% in $C_L$ \\
Tree code & $\theta = 0.2$, second order, 32 segments per leaf & error budget & within budget \\
\bottomrule
\end{tabular}
\end{table*}

A core that grows with the age of the filament represents viscous diffusion, and the documentation of OLAF \cite{Shaler2020} gives typical eddy-viscosity factors $\delta$ from 1 to 1000. Runs with $\delta = 10$, 100 and 1000 and the baseline initial core grow the cores of the oldest filaments to 0.16 to 0.63~m (M600, 80 chords) and 0.22 to 1.63~m (V3, 40 chords), and they change the loads by less than 0.05~\% in lift and drag and by at most 0.3~\% in pitching moment, one to two orders of magnitude less than doubling the core everywhere. The core sensitivity is therefore a property of the near field, the bound vortex rings and the newest wake vortex rings around the closure points, whose cores grow at most by the age of one time step, and it does not come from the viscous diffusion of the wake. The same argument applies in flight, where the older wake of earlier loops passes the kite at a distance of many chords.

\subsection{Rules for the numerical parameters}

These results give rules for choosing the numerical parameters of a free-wake vortex step or lifting-line computation of a kite before it is run (Table~VI). The wake length follows from Eq.~(14) for the lift accuracy that is required, multiplied by $\kappa_D^{1/2}$ of Eq.~(15) if the drag must be equally accurate and by about 1.2 for a free wake, whose deficit was up to 1.43 times Eq.~(14) in the cases computed here, and a configuration with a tail needs a separate check of its pitching moment. In crosswind flight the wake of the earlier loops takes the place of the starting vortex, and the rule turns into the number of loops that the wake must hold before the loads repeat, which Sec.~IV A measures. The time step sets the streamwise spacing of the vortex rings, and with the first-order scheme $U_\infty\Delta t/c$ of 0.2 to 0.3 kept the time-step error at or below 0.02~\% in lift for both fixed models. The core radius has to stay below the smallest spanwise panel width, tail panels included, should be about three quarters of it or less, is held fixed under spanwise refinement and is reported with every result. The smallest singular value of the start-step Jacobian tests a chosen pair of panelling and core radius before a run, and a core that grows with the age of the filaments is not needed for the loads.

\section{Results and discussion}\label{sec:iv}

\subsection{Single kite on the crosswind loop}

\subsubsection{Loads and periodicity}

The M600 flies the nominal loop of its controller at 9~m/s for four laps of 936~steps with the complete free wake and without its rotors, and for three laps at 5 and 13~m/s. At 9~m/s the path length is 848~m per lap, the period 18.7~s and the airspeed 40.6 to 56.5~m/s with a mean of 46.5~m/s. The tether, run alone for two laps before the coupled run, gives a lap-averaged tension whose relative change is $3\times10^{-7}$ from the first to the second lap of the coupled run and less than $10^{-8}$ afterwards. No step failed, and the Newton iteration needed five iterations at the first step, at most four in the next eight steps and at most three from then on. The 3744~steps took 16.1 h on 12 threads of a desktop processor, and the mean cost per step grew with the free wake from 1.0~s in the first lap to 39~s in the fourth.

\begin{figure*}[!t]
\centering
\includegraphics[width=0.75\textwidth]{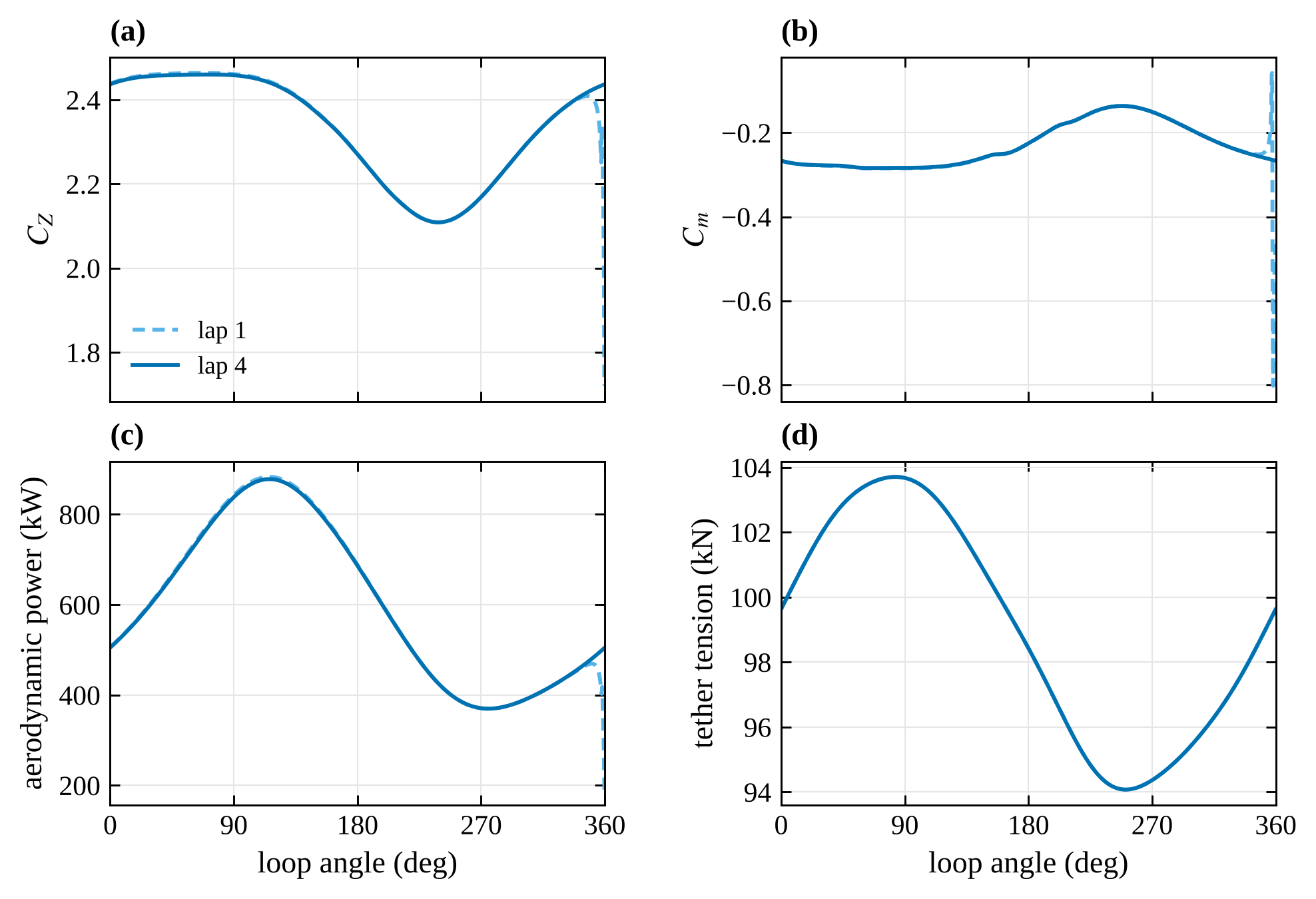}
\caption{\label{fig:9}\label{fig:loads}Loads of the M600 on the loop at 9~m/s in the first and the last of four laps. (a) Lift coefficient, (b) pitching-moment coefficient, (c) aerodynamic power and (d) tether tension.}
\end{figure*}

Figure~9 gives the lift coefficient $C_Z$ in body axes, the pitching-moment coefficient $C_m$ about the body origin, the aerodynamic power and the tether tension at the kite against the controller's loop angle, which is 0 at the side of the loop, 90$^\circ$ at the bottom and 270$^\circ$ at the top, with the kite flying in the direction of decreasing angle. The first lap differs from the later ones only during the first 3~\% of the lap, where the impulsive start produces the lift transient of a starting wing (Sec.~III C). Over the laps the largest relative difference of the total load at equal phase is $1.4\times10^{-3}$ in the force and $2.0\times10^{-3}$ in the moment between laps 2 and 3, and $4.6\times10^{-4}$ and $6.8\times10^{-4}$ between laps 3 and 4, well below the 1~\% set before the run. The lap-averaged lift of laps 1, 2 and 3 differs from that of lap 4 by $-$0.09~\%, +0.02~\% and +0.005~\%. Two laps of wake are thus enough for the loads to repeat to 0.1~\%, the counterpart in loops of the wake length of Eq.~(14). The wake of earlier laps lies $U_\infty T$ = 168~m, 0.62 loop diameters, and more downstream of the loop, where the far-field decay of Eq.~(12) makes its induction small.

Over the last lap the lift coefficient varies between 2.11 near the top of the loop and 2.46 near the bottom, and the pitching-moment coefficient between $-$0.28 and $-$0.14, following the airspeed, which gravity changes along the loop in the controller's law. The tether tension varies between 94.1 and 103.7~kN with a mean of 98.6~kN, and the aerodynamic power, the rate of work of the aerodynamic loads on the moving kite, varies between 370 and 878~kW with a mean of 574~kW. The force that the prescribed motion requires beyond the aerodynamic, tether and gravity loads of the model is 28~kN on average, and its component along the flight path averages $-$8.3~kN. The wing pushes the kite along its path, and in flight the drag of the rotors in generation acts against this push (Sec.~IV B).

\subsubsection{Wake structure}

\begin{figure*}[!t]
\centering
\includegraphics[width=0.75\textwidth]{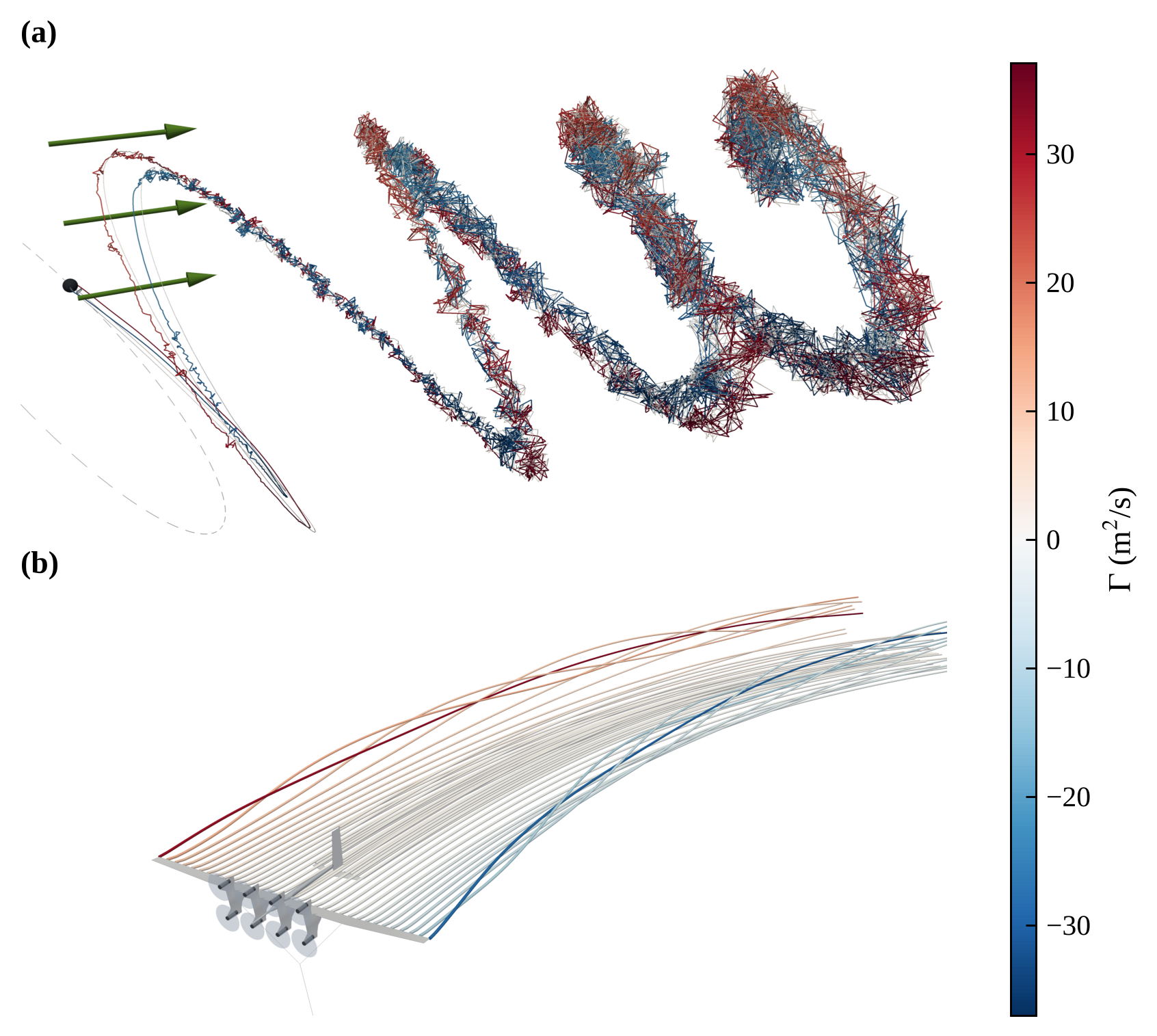}
\caption{\label{fig:10}\label{fig:wake3d}Wake of the M600 after four laps at 9~m/s. (a) The whole wake, (b) the kite and its near wake.}
\end{figure*}

Figure~10(a) shows the whole wake after four laps with the ground point, the tether drawn straight to the kite, the loop path of the last lap (dashed), the wind (arrows) and the trailing filaments of the main wing at the tips and quarter spans, and Fig.~10(b) the kite with the newest 120~steps of its wake and every trailing filament of wing and tails as a tube. The filaments of both panels are coloured by their circulation on one scale, with the fuselage, pylons and rotor discs drawn as in Fig.~3. The two edges of the sheet shed by the main wing roll up into the counter-rotating vortices of the inner and the outer wing tip. The wake of the loop is thus a helical vortex pair with one turn per lap, which the wind carries 0.62 loop diameters downstream per turn. The pair stays coherent for about two turns and then loses its order, as the tip lines become jagged and the sheet between them folds.

\begin{figure*}[!t]
\centering
\includegraphics[width=0.75\textwidth]{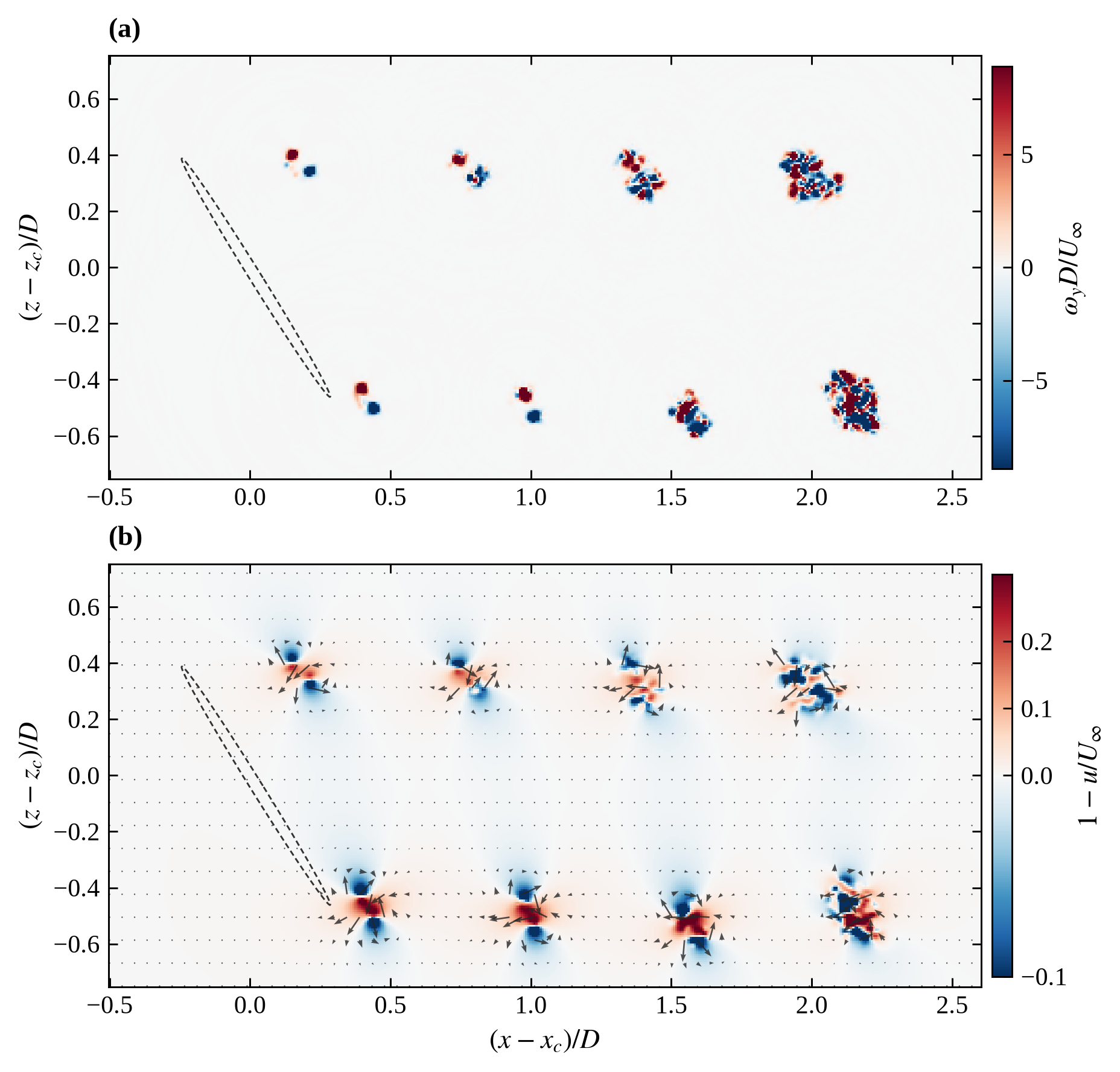}
\caption{\label{fig:11}\label{fig:meridional}Wake of the M600 at 9~m/s in the vertical plane through the path centre along the wind. (a) Out-of-plane vorticity, (b) axial velocity deficit.}
\end{figure*}

Figure~11 cuts the wake with the vertical plane through the path centre along the wind, where every turn crosses twice, at the top and at the bottom of the loop. It shows the out-of-plane vorticity $\omega_y D/U_\infty$ and the axial velocity deficit $1 - u/U_\infty$ with the in-plane induced velocity $(u - U_\infty, w)$ as arrows capped at $0.1\,U_\infty$, the velocity being induced by the stored filament network with a display core of 3~m. The dashed line is the projection of the last loop on the plane, the coordinates are measured from the path centre in loop diameters $D$ = 270~m and the colour scale of the deficit ends at 0.3. Each crossing shows the pair as two compact regions of opposite vorticity. Integrated over each region, the circulation is 108 and 111~m$^2$/s at the first two crossings at the bottom and 71 and 82~m$^2$/s at the top, following the lift of the wing, which is largest near the bottom of the loop. The distance between the two vortices is 21.0 to 21.4~m at the first crossings, close to the Betz spacing $0.836\,b$ = 21.5~m of the fixed wing (Sec.~III C), and 18.6~m at the second crossing at the top. The pair induces on itself a velocity against the wind, and the slow fluid between the two vortices forms the velocity deficit of the wake [Fig.~11(b)]. At the first two crossings the deficit reaches 0.41 and 0.37 $U_\infty$ at the bottom and 0.25 and 0.23 $U_\infty$ at the top, and it exceeds 0.1 $U_\infty$ over 550 to 640~m$^2$ at the bottom and 280 to 320~m$^2$ at the top, a region about one span across, while the interior of the loop is slowed by less than 0.05 $U_\infty$. From the third crossings on, 1.4 and 1.55 loop diameters downstream at the top and the bottom, the two vortices merge into a cloud of mixed vorticity in which the separation of the signs is lost.

\begin{figure*}[!t]
\centering
\includegraphics[width=1\textwidth]{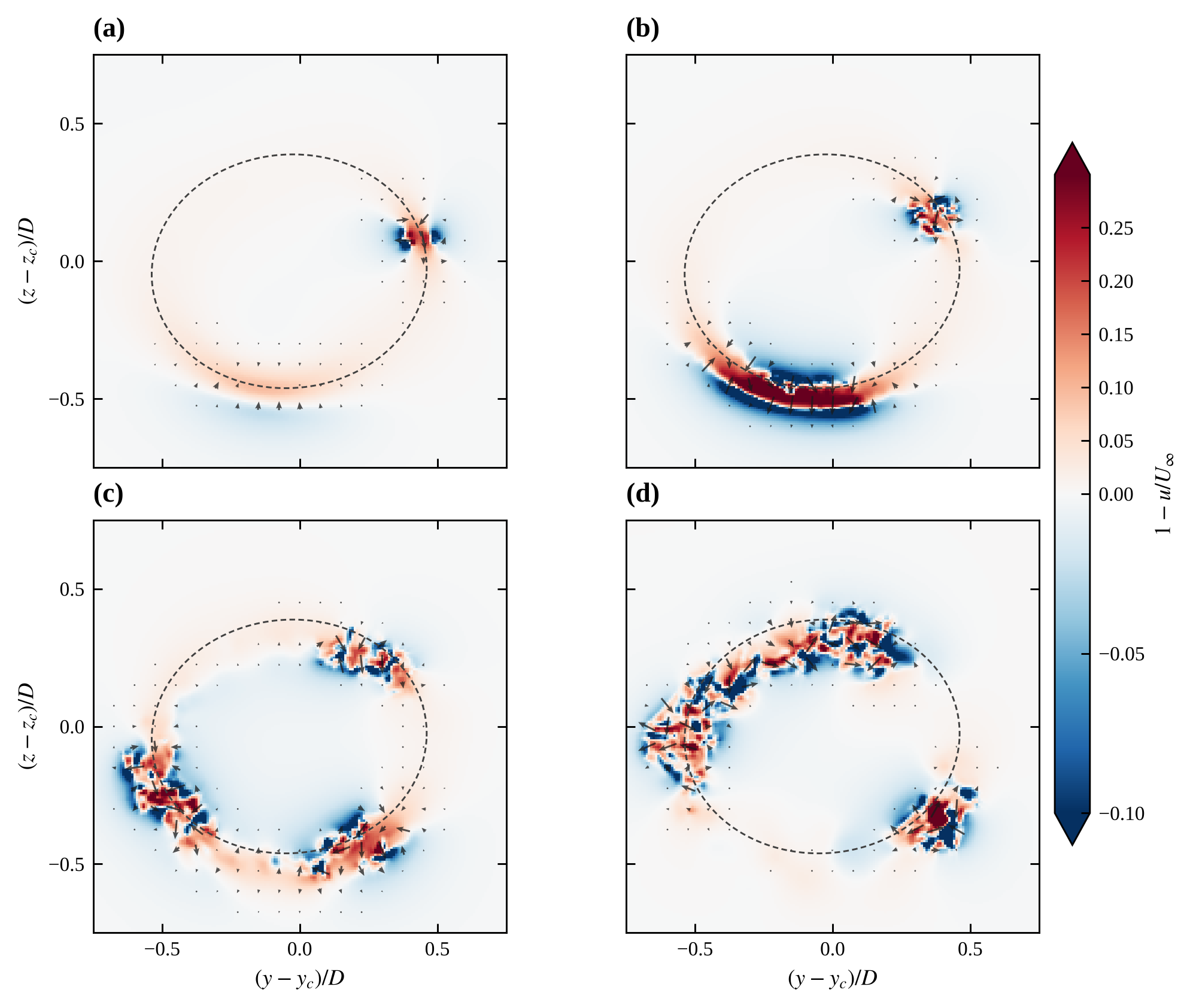}
\caption{\label{fig:12}\label{fig:planes}Cross-flow planes of the wake of the M600 at 9~m/s after four laps, normal to the wind at (a) 0.5, (b) 1, (c) 1.5 and (d) 2 loop diameters downstream of the path centre. Axial velocity deficit $1 - u/U_\infty$ and in-plane induced velocity (arrows), induced by the stored filament network.}
\end{figure*}

\begin{figure*}[!t]
\centering
\includegraphics[width=1\textwidth]{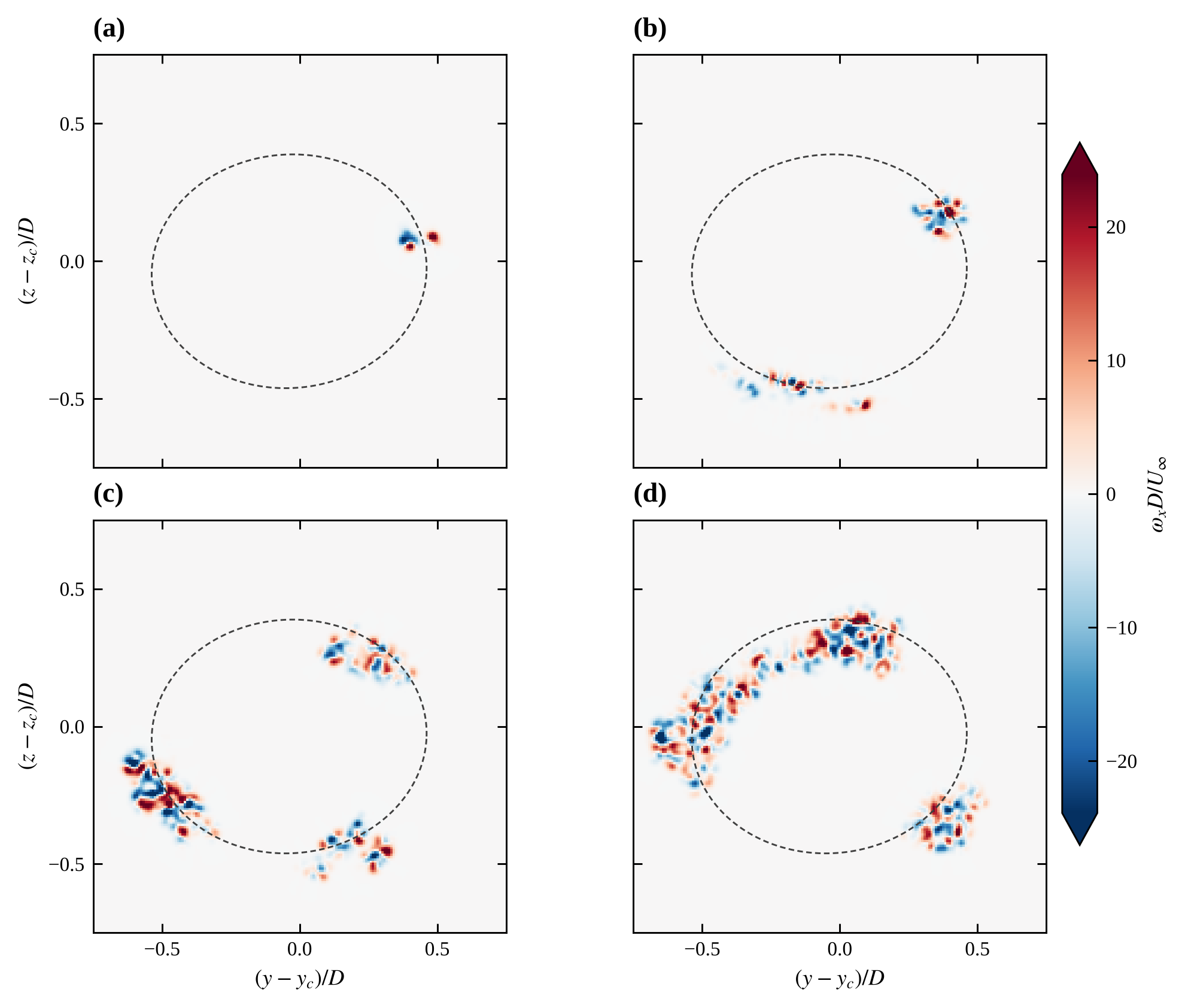}
\caption{\label{fig:13}\label{fig:vorticity}Streamwise vorticity of the wake of the M600 at 9~m/s on the cross-flow planes of Fig.~12, as the flux of the crossing filaments spread over 3~m.}
\end{figure*}

Figures~12 and 13 cut the same wake with planes normal to the wind at 0.5, 1, 1.5 and 2 loop diameters downstream of the path centre, where the wake is 0.8, 1.6, 2.4 and 3.2 laps old, the stored wake ending at 2.56 diameters. The velocity is again induced by the stored network with a display core of 3~m, the arrows are capped at $0.15\,U_\infty$, the dashed line is the projection of the last loop, and the streamwise vorticity $\omega_x D/U_\infty$ is the flux of the filaments crossing the plane spread with a Gaussian of 3~m, with the same limits in the four panels. The loop is inclined by 32$^\circ$ and extends 0.53 diameters along the wind, so a plane cuts the helix of pitch 0.62 diameters once or several times. At 0.5 diameters it cuts the newest turn once, at the side of the loop, where the sheet appears as a compact pair of counter-rotating vortices with the slow fluid between them [Figs.~12(a) and 13(a)]. At 1 diameter it cuts the sheet obliquely near the bottom of the loop, where the deficit forms a band of about one span in width along the path, and once more near the top, and the deficit exceeds $0.1\,U_\infty$ over 0.06 $D^2$ against 0.008 $D^2$ at 0.5 diameters [Fig.~12(b)]. The wake is an annulus. The azimuthal mean of the deficit about the axis of the loop peaks at 0.46 to 0.54 loop diameters from the axis with 0.030, 0.062, 0.041 and 0.028 $U_\infty$ at the four stations, and the interior of the loop flows slightly faster than the wind, by up to $0.01\,U_\infty$ in the azimuthal mean at 1.5 diameters. Outside the sheet the flow is also faster than the wind, so that the band of deficit is flanked by a speed-up on both sides, as the flow around a vortex sheet that is cut obliquely must be. At 1.5 and 2 diameters the crossings, 2.4 and 3.2 laps old, have become clouds of vortices of alternating sign [Figs.~12(c), 12(d), 13(c) and 13(d)], the deficit still exceeds $0.1\,U_\infty$ over 0.07 $D^2$, but the peak of its azimuthal mean falls and the in-plane velocity becomes irregular. The circulation crossing each plane in either sign grows from 640~m$^2$/s at 0.5 diameters to 12 700~m$^2$/s at 2 diameters as the plane cuts more turns, and the net flux is zero at every station.

\begin{figure}[!htb]
\centering
\includegraphics[width=1\linewidth]{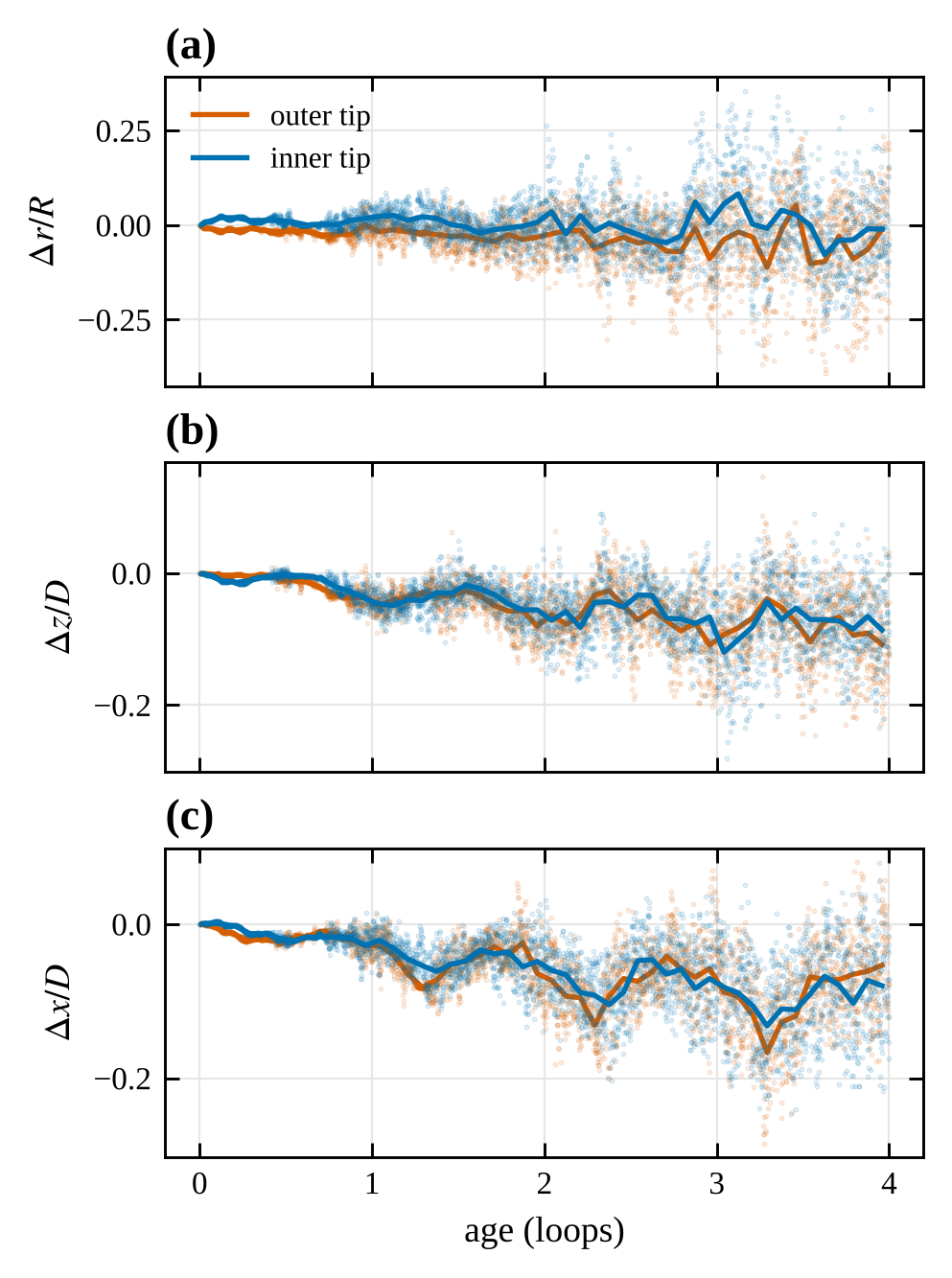}
\caption{\label{fig:14}\label{fig:tips}Displacement of the wing-tip nodes of the M600 wake from pure convection with the wind. (a) Radial, (b) vertical and (c) streamwise.}
\end{figure}

The displacement of the tip nodes from the position that the wind alone would give them measures the motion induced by the wake itself (Fig.~14, every wake row of the last stored state as dots and means over 1/12 loop as lines, the outer tip in vermilion and the inner tip in blue, against the age in loops, radially about the axis of the loop in loop radii and vertically and streamwise in loop diameters). Both tips descend by 0.02 to 0.026 loop diameters per loop, 0.3 to 0.4~m/s, and lag behind the wind by 0.018 loop diameters at an age of about one loop and 0.06 at three, an induced streamwise velocity of about 0.3~m/s or 3.4~\% of the wind speed. In the first loop the outer tip moves towards the axis of the loop by 0.018 $R$ and the inner tip away from it by 0.007 $R$, the roll-up of the sheet towards the Betz centroids, which moves each tip by 0.016 $R$. Later both tips move towards the axis, the outer one by 0.054 $R$ and the inner one by 0.012 $R$ after three loops, so that the helix contracts slowly as it ages. Beyond two loops the scatter of the displacements grows with the loss of coherence of Fig.~11. A straight counter-rotating pair of circulation $\Gamma$ and spacing $b'$ moves at $\Gamma/(2\pi b')$, Eq.~(17), 0.67~m/s for $\Gamma$ = 90~m$^2$/s and $b'$ = 21.3~m, 1.4 to 1.6 times the induced motion of the tips, 0.42 to 0.49~m/s, and is unstable to long-wave perturbations that lead to the reconnection of the two vortices \cite{Leweke2011}. Here the pair is bent into a helix whose neighbouring turns pass within 0.6 loop diameters of each other, and the computation does not separate a physical instability of this system from short-wave disturbances of the regularised filaments, whose growth depends on the time-marching scheme \cite{Bhagwat2001}. The loads repeat to $7\times10^{-4}$ from the third lap on, so the disordered far wake acts on the kite only through its integral induction. The centre line of the wake confirms this picture (Fig.~15, from the circulation-weighted centroids of the positive and the negative trailing vorticity of every wake row, averaged over one lap of rows). The two centroids descend together, from 0.015 loop diameters below the path centre at 0.28 diameters downstream to 0.073 diameters, about 20~m, at 2.1 diameters, their lateral positions stay within 0.01 diameters of the path centre and 0.01 diameters apart, and the rms radius of the trailing vorticity about its centroid grows from 0.045 loop diameters at 0.5 diameters downstream to 0.089 at 2 diameters, with a scatter from row to row that grows with the loss of order.

\begin{figure}[!htb]
\centering
\includegraphics[width=1\linewidth]{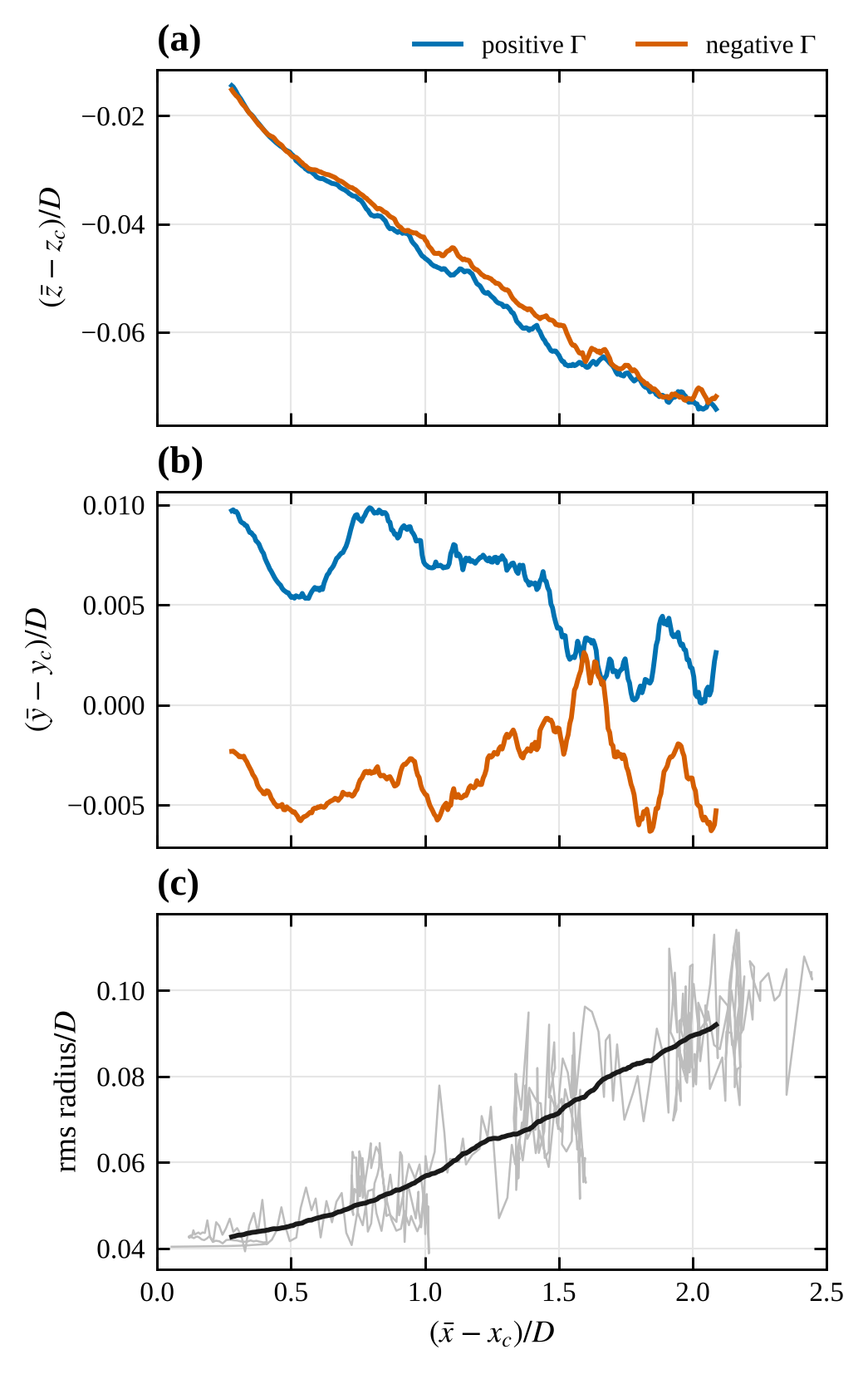}
\caption{\label{fig:15}\label{fig:centre}Centre line of the M600 wake at 9~m/s against the downstream distance. (a) Height and (b) lateral position of the circulation-weighted centroids of the positive and the negative trailing vorticity, lap means over the rows, (c) rms radius of the trailing vorticity about its centroid, every row (grey) and its lap mean.}
\end{figure}

\subsubsection{Loss by the kite's own wake}

The induction of the kite's own wake of the previous loops lowers its loads. With the same poses solved step by step with the straight wake of Sec.~II, which holds no history of the loop, the free wake lowers the lap-averaged lift and the aerodynamic power of the fourth lap by 0.55~\%. In the first lap, which still carries the starting vortex, the loss is 0.64~\%, and within the last lap it varies between 0.30~\% and 0.82~\% with the loop angle. An analytical vortex model of the wake of a kite flying circles estimates this loss \cite{Trevisi2023a,Trevisi2023b}. It replaces the wake of the earlier loops by a helical vortex sheet of radius $R_0$ and pitch $h_0$ and gives its contribution to the induced drag as

\begin{equation}
\begin{aligned} C_{D_i}^{\mathrm{f}} &\approx \frac{C_L^2}{4\pi^2 A}\,\kappa_0^{\pi/2}\,\lambda_0^{3/2}, \\ \kappa_0 &= \frac{b}{2R_0}, \qquad \lambda_0 = \frac{2\pi R_0}{h_0}, \end{aligned}
\label{eq:18}
\end{equation}

\noindent with the aspect ratio $A$ and the pitch $h_0$, the advance of the wake over one period at the axial wind speed reduced by the near-wake induction $u_0 C_L/(\pi A)$, where $u_0$ is the tangential speed \cite{Trevisi2023b}. For this loop, with $R_0$ = 135~m, $b$ = 25.7~m, $A$ = 20, the tangential speed of 45.3~m/s, the wind component of 7.6~m/s along the axis of the loop and the lap-averaged $C_L$ = 2.33, Eq.~(18) gives $\kappa_0$ = 0.095, $\lambda_0$ = 7.6 and $C_{D_i}^{\mathrm{f}}$ = 0.0036, 4~\% of the induced drag of the near wake, $C_L^2/(\pi A)$. Read as an induced incidence $C_{D_i}^{\mathrm{f}}/C_L$ = 0.088$^\circ$, it corresponds to a lift loss of 0.40~\% with the lift response $C_w$ = 6.05 of Sec.~III, against the 0.55~\% of the free wake. The model assumes a loop whose axis is along the wind and a helix that keeps its shape, and it shortens the pitch by the near-wake induction $u_0 C_L/(\pi A)$ = 1.7~m/s. In the computed wake the tips move against the wind along the axis of the loop at only 0.40 to 0.41~m/s (Fig.~14). With this advance, $h_0$ = 0.50 $D$, Eq.~(18) gives $\lambda_0$ = 6.3 and a loss of 0.30~\%, so the far wake of the model is tighter than the computed one. The rest of the difference lies in what the model leaves out, the inclination of the loop by 32$^\circ$ and the curvature of the near wake, which follows the path in the free solution and the straight line in the reference.

\begin{figure*}[!t]
\centering
\includegraphics[width=1\textwidth]{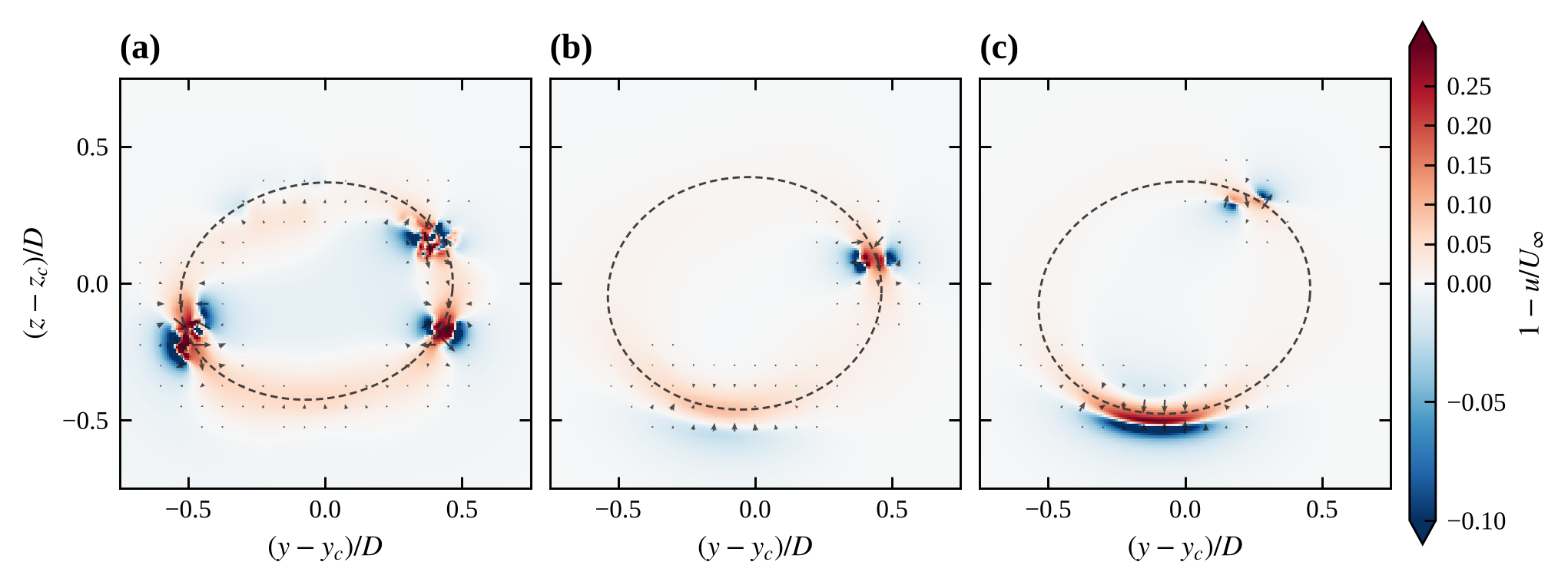}
\caption{\label{fig:16}\label{fig:wind}Cross-flow plane at 0.5 loop diameters downstream of the path centre at the end of the runs at (a) 5, (b) 9 and (c) 13~m/s, axial velocity deficit and in-plane velocity, each run in its own loop diameter.}
\end{figure*}

At 5 and 13~m/s the playbook sets the loop radius to 164 and 143~m and the elevation to 37.5$^\circ$ and 31.9$^\circ$, against 135~m and 31.9$^\circ$ at 9~m/s, and the wake advances 0.35 and 0.88 loop diameters per lap against 0.62. The torsional parameter $\lambda_0$ of Eq.~(18), with the pitch shortened by the near-wake induction \cite{Trevisi2023b} and the lap-averaged $C_L$ of each run, is then 18.5 and 5.0 against 7.6 at 9~m/s. At 5~m/s Eq.~(18) gives $C_{D_i}^{\mathrm{f}}$ = 0.0089 with $C_L$ = 2.20, a lift loss of 1.1~\%, nearly three times the 0.40~\% at 9~m/s. The free wake at 5~m/s lowers the lift of the third lap by 0.54~\% against the straight wake, as much as at 9~m/s, and within the lap by 0.26~\% to 0.87~\%. The loss of aerodynamic power is larger, 0.95~\%, and grows from lap to lap, 0.59~\%, 0.71~\% and 0.95~\%, while the lap-averaged lift loss stays between 0.51~\% and 0.56~\% and the loads of the second and third laps agree to 0.3~\% at every step. The tips of the free wake at 5~m/s move against the wind along the axis of the loop at 0.36~m/s, against 0.40 to 0.41~m/s at 9~m/s, whereas the near-wake shortening of the model is 1.6~m/s. With the computed advance the pitch is 84~m, $\lambda_0$ = 12.3, and Eq.~(18) gives a loss of 0.61~\%, close to the 0.54~\% of the free wake, while the same procedure gives 0.30~\% against 0.55~\% at 9~m/s. The near-wake shortening is 39~\% of the axial wind at 5~m/s and 22~\% at 9~m/s.

In the meridional plane at 5~m/s the pair crosses as two compact vortices at the first two crossings and as a cloud at the third, 0.6 and 1.0 loop diameters downstream at the top and the bottom, so it loses its order after two turns as at 9~m/s, but closer to the kite. The cross-flow plane at 0.5 loop diameters shows the tighter helix directly (Fig.~16). At 5~m/s the plane cuts the sheet four times, 1.4 laps behind the kite, and the deficit exceeds $0.1\,U_\infty$ over 0.040 $D^2$, against one cut 0.8 laps old and 0.008 $D^2$ at 9~m/s and two cuts 0.6 laps old and 0.025 $D^2$ at 13~m/s, while the azimuthal mean of the deficit peaks at about half a loop diameter from the axis at all three wind speeds, with 0.033, 0.030 and 0.024 $U_\infty$.

At 13~m/s Eq.~(18) gives $C_{D_i}^{\mathrm{f}}$ = 0.0016 with $C_L$ = 2.26, a lift loss of 0.19~\%, against 0.52~\% for the free wake, which lowers the aerodynamic power by 0.48~\% and the lift within the lap by 0.24~\% to 0.83~\%. The tips advance against the wind at 0.35 to 0.38~m/s, the pitch is then 0.72 loop diameters, and Eq.~(18) with this pitch gives 0.16~\%. In the meridional plane at 13~m/s the pair crosses as two compact vortices of equal and opposite circulation at the first two crossings of each half, 0.3 to 1.3 loop diameters downstream, and as a cloud at the third, 2.1 diameters downstream, so the faster wind carries it farther before it loses its order. Over the three wind speeds the model gives 1.1~\%, 0.40~\% and 0.19~\%, falling as the wind rises, whereas the loss of the free wake stays between 0.52~\% and 0.55~\%.

\subsection{Rotor slipstream}

\subsubsection{Free rings and the slipstream tube}

The M600 now flies the loop of Sec.~IV A with its eight rotors at a constant 97.4~rad/s. Along the loop the axial inflow at the hubs is 36 to 62~m/s, the thrust coefficient in wind-turbine notation 0.26 to 0.34 and the axial induction of the discs 0.07 to 0.09, so momentum theory has a solution at every step and the rotor table is never left. A run without rotors at the same time step is the reference, and the difference of the two runs is the effect of the rotors at equal numerical settings. The treatment of the slipstream was chosen on the first second of flight, which every variant below shares, through the effect of the rotors on the body-axis coefficients averaged over 0.5 to 1.0~s, and the whole loops then give the effect along the loop.

\begin{figure}[!htb]
\centering
\includegraphics[width=1\linewidth]{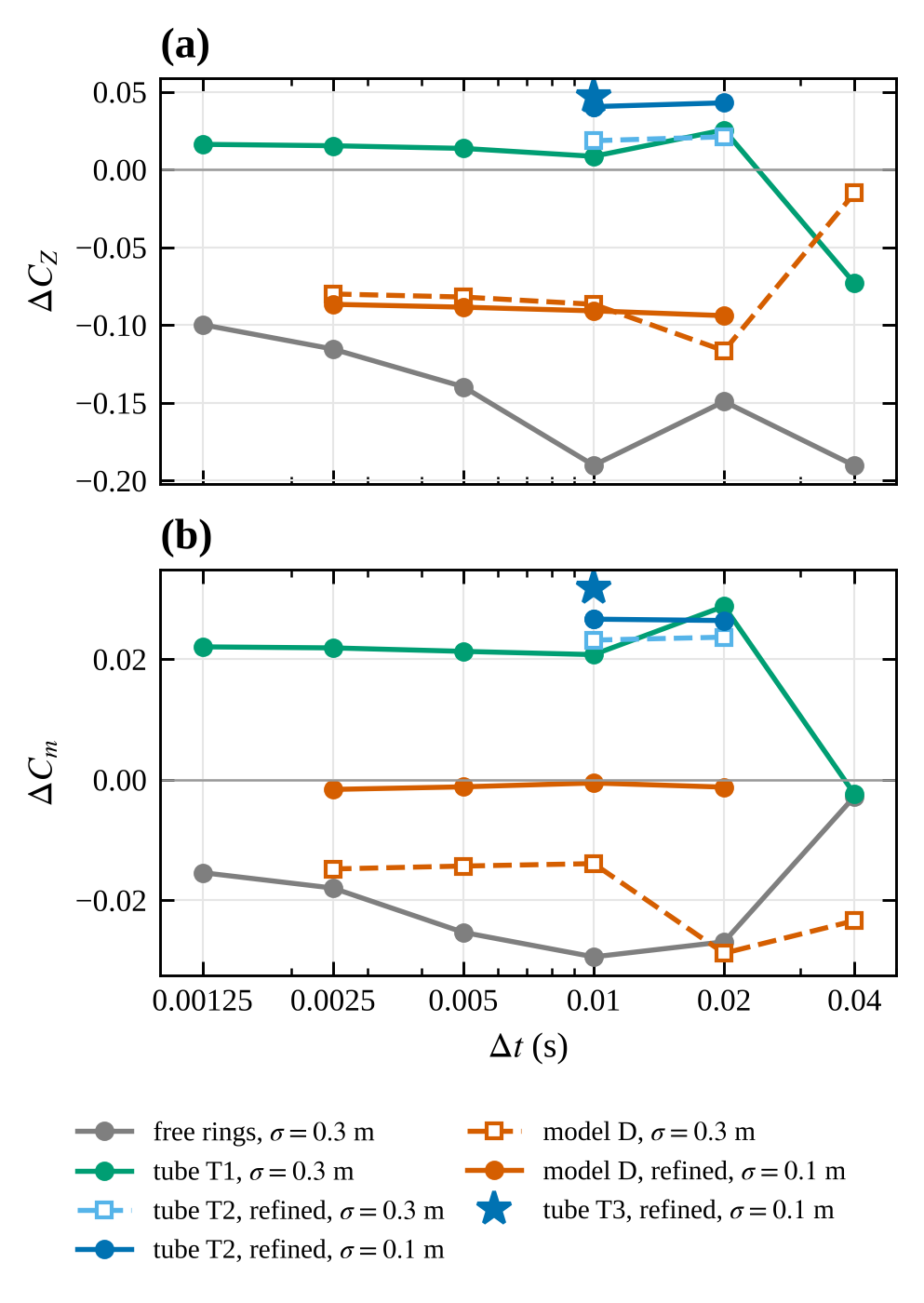}
\caption{\label{fig:17}\label{fig:dt}Effect of the eight rotors on (a) the lift coefficient and (b) the pitching-moment coefficient of wing and tails over the first second against the time step, for the treatments of the slipstream.}
\end{figure}

When the rings convect freely with the induced velocity, as the wake of the wing does, the effect of the rotors on $C_Z$ over the first second moves between $-$0.19 and $-$0.10 for time steps from 0.04 to 0.00125~s without settling, and on $C_m$ between $-$0.003 and $-$0.029 (Fig.~17, free rings, $\sigma$ the surface core). Down to one eighth of the production step each halving changes the effect on $C_Z$ by 0.016 to 0.05, so the free rings give no converged answer. The rings of the four lower discs pass the bound vortex of the wing at about 0.1~m, inside its core of 0.15~m, and move 0.2 to 1.7~m during one step there. Where a ring lies after this passage depends on the phase of the steps relative to the wing, the ring is torn apart within a few steps, and its remains cross the lifting surfaces. Over the first lap, too, halving the step from 0.02 to 0.01~s changed the effect of the rotors on the lap-averaged $C_Z$ by a third. Vortex particle methods meet the same difficulty where a wake reaches a lifting surface, and they keep particles from crossing the panels \cite{Tugnoli2021} or spread the circulation of the wing over its chord with an actuator surface \cite{AlvarezNing2024}.

The effect converges when each ring convects freely only during its first step and then drifts with the wind (tube T1 in Fig.~17). From 0.02 to 0.00125~s it is +0.0256, +0.0085, +0.0133, +0.0153 and +0.0162 on $C_Z$ and +0.029, +0.021, +0.022, +0.022 and +0.022 on $C_m$, and the differences between neighbouring steps fall from 0.017 to 0.0009. The tube is the straight vortex tube with which lifting-surface methods represent a propeller slipstream that the wing does not deform \cite{Nederlof2023,Trevisi2026}. Its rings are those of the free vortex-ring actuator disc, whose far wake is a vortex cylinder of strength $\Gamma_r/\Delta s$ for rings $\Delta s$ apart \cite{Yu2017}. Two features of tube T1~set its error at a finite step. During its first step a ring deforms with the local induction, and at $\Delta t$ = 0.02~s the rings of the four upper discs, whose tubes pass less than 0.1~m from the plane of the wing, grow from 1.15 to 1.19~m in radius and keep that size. The released circulation of Eq.~(11), that of the free vortex-ring actuator disc \cite{Yu2017}, belongs to a wall that advances relative to the disc at $V + \gamma/2$, the mean of the axial velocities outside and inside the tube. A ring that drifts with the wind recedes at $V$ instead, so that the wall of tube T1 is weaker than momentum theory by the axial induction, about 9~\%. Rings that drift with the wind from their release (tube T2) remove the first error. Rings that drift with the wind plus $\gamma/2$ along the axis of their disc at release, with the circulation released over this drift spacing, $\Gamma_r = \gamma(V_\infty + \gamma/2)\Delta t$ with the ambient inflow $V_\infty$ along the axis, give a wall of exactly the strength $\gamma$ of momentum theory (tube T3, Sec.~II D). At $\Delta t$ = 0.01~s the effect on $C_Z$ over the first second is +0.0405 for T2, +0.0446 for the same tube advancing at $\gamma/2$, 10~\% more, close to the ratio $1/(1-a)$ of the two wall strengths for the induction $a$, and +0.0467 with the circulation released over the ambient spacing, another 4.7~\%, all three with the refinement described next.

\begin{figure}[!htb]
\centering
\includegraphics[width=1\linewidth]{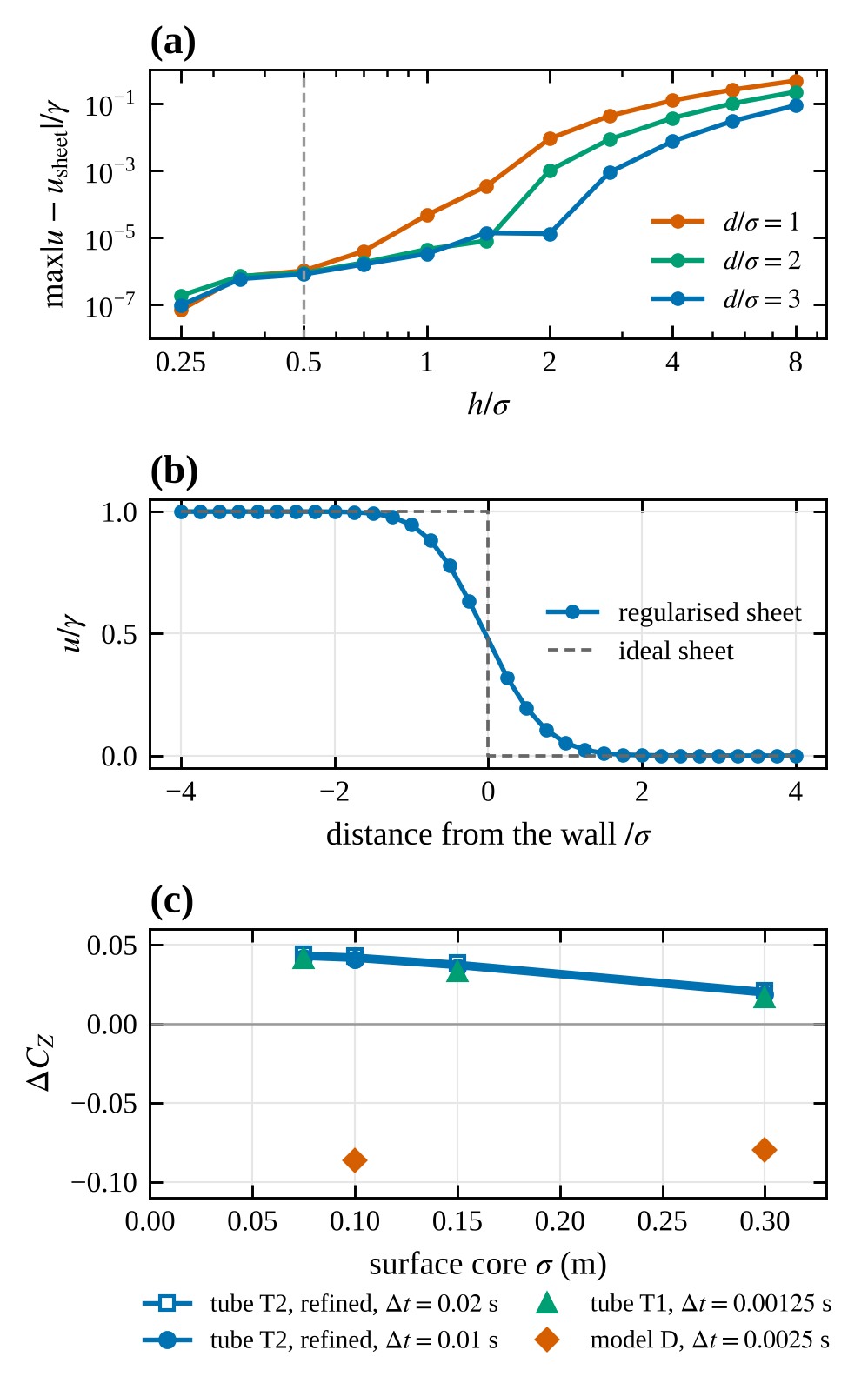}
\caption{\label{fig:18}\label{fig:tube}The slipstream tube near a lifting surface. (a) Error of a row of discrete rings, (b) axial velocity across the regularised wall, (c) effect of the rotors on the lift against the surface core.}
\end{figure}

Three lengths decide how the tube acts at a closure or load point of the wing (Fig.~18), the spacing $h$ of the rings along the tube, about $(V_\infty + \gamma/2)\Delta t$, 0.36 to 0.54~m at $\Delta t$ = 0.01~s, the core radius $\sigma$ of the ring filaments at the surface points, and the distance $d$ from the tube wall to the point. A row of discrete rings induces at a point outside the wall a velocity that oscillates about that of the continuous sheet with the phase of the point between two rings. Computed with the kernel of the code for a tube of radius $11.5\sigma$ and sheet strength $\gamma$, the largest deviation is below $10^{-4}\gamma$ while $h \le \sigma$, and it grows to $9\times10^{-3}\gamma$ at $h = 2\sigma$ and $0.12\gamma$ at $h = 4\sigma$ for $d = \sigma$ [Fig.~18(a), where the dashed line marks the spacing reached with the refinement of Eq.~(20)]. The regularised sheet in turn spreads the jump of the axial velocity from $\gamma$ inside to zero outside over about two core radii, so that a point at $d = \sigma$ outside the wall sees $0.05\gamma$ and a point at $d = 2\sigma$ sees $1.5\times10^{-3}\gamma$ [Fig.~18(b), negative distances inside the tube]. Both errors stay below $2\times10^{-3}\gamma$ at a point for which

\begin{equation}
h \le \sigma/2, \qquad \sigma \le d/2.
\label{eq:19}
\end{equation}

With the production step the first condition would require a core of about 1~m, far more than the distance of the tubes from the surfaces. At the end of the fourth lap of the single unit with the undeformed tube, the closure points of the main wing lie at least 0.12~m from the nearest ring filament, next to the pylons, those of the vertical tail 0.09~m and those of the horizontal tail 0.23~m, against a median of 0.53~m over all 49 points. The rings are therefore refined at the surface points only. For the induction at the closure and load points every ring $k$ of a disc stands for the stretch of tube between the midpoints to its neighbours, of largest node displacement $s_k$, and where this stretch reaches within $2s_k + 4\sigma + \ell_k$ of a point, with $\ell_k$ the longest edge of the ring, it is replaced by

\begin{equation}
m_k = \left\lceil \lambda\, s_k/\sigma \right\rceil, \qquad \lambda = 2
\label{eq:20}
\end{equation}

\noindent sub-rings interpolated linearly between the neighbouring rings that share the circulation of ring $k$ in proportion to their stretches. The sub-ring spacing at the surface points is then at most $\sigma/2$. Vortex particle methods likewise need overlapping particle cores to converge \cite{Beale1986} and use a core of about twice the particle spacing in propeller and wing computations \cite{AlvarezNing2024}. The convection of the rings and the inflow at the hubs are not changed. In the pairs at 0.02~s the refinement replaces 190 to 248 rings of the two units by up to 22~sub-rings each, and it costs 22~\% more time per step at 0.02~s and 13~\% to 20~\% at 0.01~s.

With the refinement the effect converges as the surface core shrinks [Fig.~18(c)]. For $\sigma$ = 0.3, 0.15, 0.1 and 0.075~m it is +0.0211, +0.0384, +0.0429 and +0.0442 on $C_Z$ at $\Delta t$ = 0.02~s and +0.0186, +0.0360, +0.0405 and +0.0417 at 0.01~s, and from 0.1 to 0.075~m it changes by 0.0012 on $C_Z$ and by 1.0~\% to 1.3~\% on $C_m$. A core of 0.3~m, as large as the distance of the walls from the closest points, smears the wall over the load points and halves the effect on $C_Z$, so a core of the order of the wall distance acts as a model parameter. Independently of the refinement, tube T1 without it at 0.00125~s, where the rings are 0.05~m apart, gives +0.0328 at $\sigma$ = 0.15~m and +0.0412 at 0.075~m, within 0.002 of the refined tube extrapolated to a vanishing step. The production core is 0.1~m, half the wall distance of all but the few closest points, and the remaining sensitivity of 3~\% in $C_Z$ from 0.1 to 0.075~m comes from those points.

With tube T3, the refinement and $\sigma$ = 0.1~m, the effect over whole laps of the single unit changes by +0.0026 to +0.0027 on $C_Z$ and by 1.4~\% to 1.6~\% on $C_m$ from $\Delta t$ = 0.01 to 0.02~s in each of the four laps, as over the first second, where the refined tube T2 changes by 0.0024 between the same steps. The single unit is computed at 0.01~s and the pairs of Sec.~IV C at 0.02~s, whose error in the effect on $C_Z$ is thus about +0.0026, 6~\% of the effect. Over the first second, 32 nodes per ring instead of 16 change the effect on $C_Z$ by $2\times10^{-4}$, a core of the ring filaments away from the surfaces of 0.15 or 0.9~m instead of the node spacing of 0.45~m by at most $10^{-4}$ and 1.3~\% in $C_m$, and keeping the rings for 0.5~s instead of 2~s by 0.0022 (5~\%). Over two whole laps, rings kept for 4~s change the lap-averaged effect on $C_Z$ by $-1.4\times10^{-4}$ and on $C_m$ by less than 0.1~\%, and in the pair of Sec.~IV C rings kept for two laps change it by $-2.4\times10^{-4}$ at the downstream unit. Table~VII collects the rules, with the effects of the rotors on the coefficients of wing and tails over the first second at $\Delta t$ = 0.01~s unless whole laps are named.

\begin{table*}[!t]
\caption{\label{tab:7}Rules for the rotor slipstream of a crosswind kite computed with a vortex step method and a free vortex wake, with their basis and the values and sensitivities for the M600.}
\small\setlength{\tabcolsep}{5pt}\renewcommand{\arraystretch}{1.12}
\begin{tabular}{@{}>{\raggedright\arraybackslash}p{0.15\textwidth}>{\raggedright\arraybackslash}p{0.37\textwidth}>{\raggedright\arraybackslash}p{0.13\textwidth}>{\raggedright\arraybackslash}p{0.27\textwidth}@{}}
\toprule
Item & Rule & Basis & M600 value and sensitivity \\
\midrule
Slipstream & tube that the wing does not deform (T3), with the deflected slipstream (model D) as the second treatment & Figs.~\ref{fig:dt}, \ref{fig:loop} & $\Delta C_Z$ $+0.045$ and $-0.078$ over a lap \\
Release & one ring per disc and step at the rim, $\Gamma_r = \gamma(V_\infty+\gamma/2)\Delta t$, drift with the wind plus $\gamma/2$ along the disc axis & Eq.~(\ref{eq:11}) & $\Delta C_Z$ $+10$~\% against T2 for the drift, a further $+4.7$~\% for the ambient spacing \\
Spacing at the surface points & $h/m_k \le \sigma/2$ by Eq.~(\ref{eq:20}) & Fig.~\ref{fig:tube}(a) & up to 22 sub-rings, 13~\% to 22~\% more time per step \\
Surface core & $\sigma \le d/2$, $d$ the smallest wall distance & Fig.~\ref{fig:tube}(b, c) & 0.1~m; 0.0012 from 0.1 to 0.075~m \\
Time step & 0.01~s single unit, 0.02~s pairs with the single-unit error & whole laps & $+0.0026$ from 0.01 to 0.02~s \\
Nodes per ring & 16 & first second & $2\times10^{-4}$ with 32 \\
Ring core away from surfaces & node spacing, 0.45~m & first second & $\le 10^{-4}$ for 0.15 and 0.9~m \\
Rings kept & 2~s & whole laps, Sec.~IV\,C & $-1.4\times10^{-4}$ with 4~s, $-2.4\times10^{-4}$ at unit B with two laps \\
Free rings & not used & Fig.~\ref{fig:dt} & no convergence down to 0.00125~s \\
\bottomrule
\end{tabular}
\end{table*}

\subsubsection{Effect of the rotors on the wing along the loop}

\begin{figure}[!htb]
\centering
\includegraphics[width=1\linewidth]{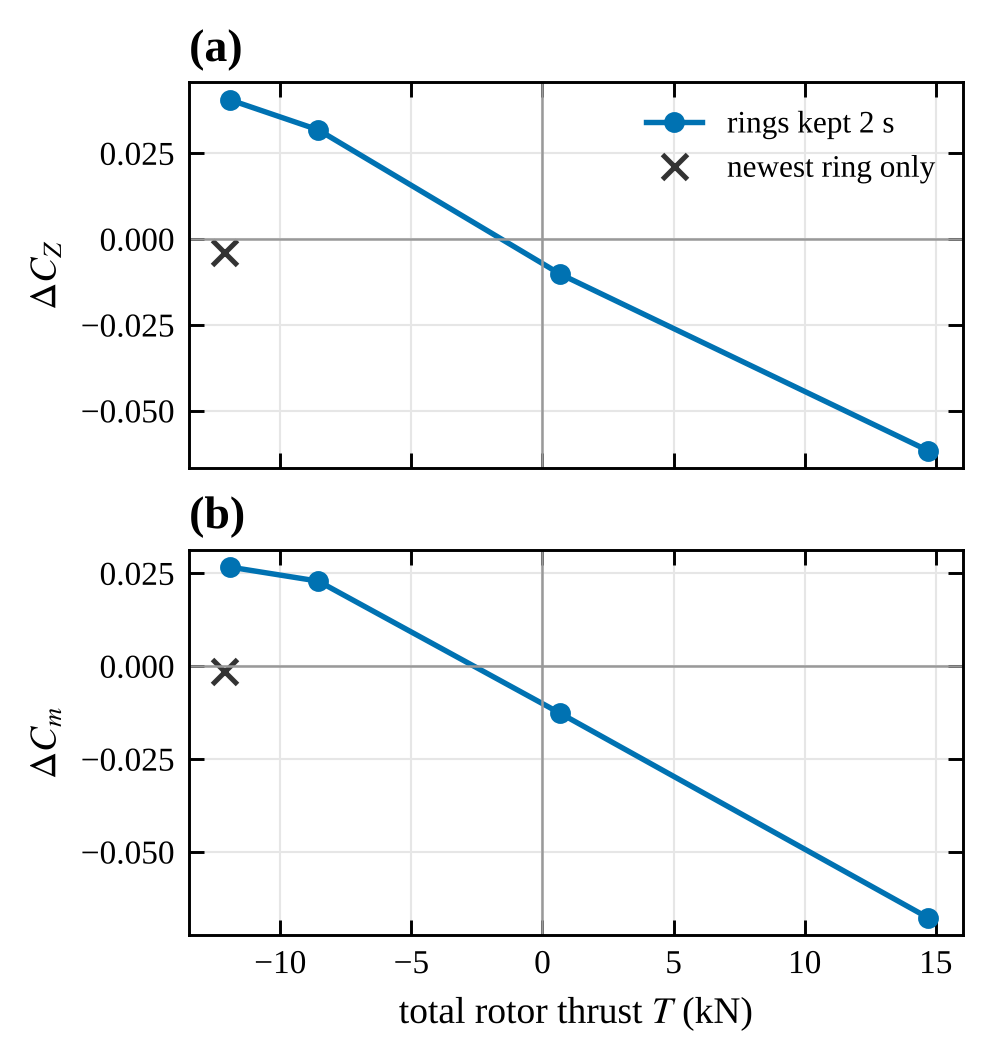}
\caption{\label{fig:19}\label{fig:thrust}Effect of the rotors on (a) the lift coefficient and (b) the pitching-moment coefficient over the first second against the total thrust of the eight rotors.}
\end{figure}

Over the first second the effect follows the total thrust of the rotors almost linearly and changes sign with it (Fig.~19, tube T2 with refinement, $\sigma$ = 0.1~m, $\Delta t$ = 0.01~s, rings kept 2~s, negative thrust being generation). At 60, 97.4, 160 and 200~rad/s, with the total thrust $-$8.6, $-$11.9, +0.7 and +14.7~kN and the shaft power 262, 417, $-$77 and $-$816~kW, it is +0.0318, +0.0405, $-$0.0102 and $-$0.0616 on $C_Z$ and +0.023, +0.027, $-$0.013 and $-$0.068 on $C_m$. Between the generating point and the strongest motoring point

\begin{equation}
\Delta C_Z \approx -0.0052 - 0.0038\,T/\mathrm{kN} \approx -0.005 - 0.17\,T/(q S)
\label{eq:21}
\end{equation}

\noindent with $qS \approx 44$ kN, so the effect is smallest where the rotors carry no thrust and grows with the thrust in either direction. With only the newest ring of each disc kept (the cross in Fig.~19), so that no tube reaches the wing, the effect is $-$0.0037 on $C_Z$ and $-$0.0012 on $C_m$, a tenth and a twentieth of the full effect and of opposite sign. The effect of the rotors on the wing is thus carried by the walls of the slipstream tubes where they pass the wing and the tails, and its sign is the sign of the wall strength $\gamma$, set by the thrust. The rotors of the M600~sit on pylons above and below the wing, which stays outside the tubes. A wing immersed in a slipstream loses lift when the slipstream carries a velocity deficit \cite{Nederlof2023}, whereas the undeformed tubes passing the M600 wing in generation raise its lift.

\begin{figure*}[!t]
\centering
\includegraphics[width=0.85\textwidth]{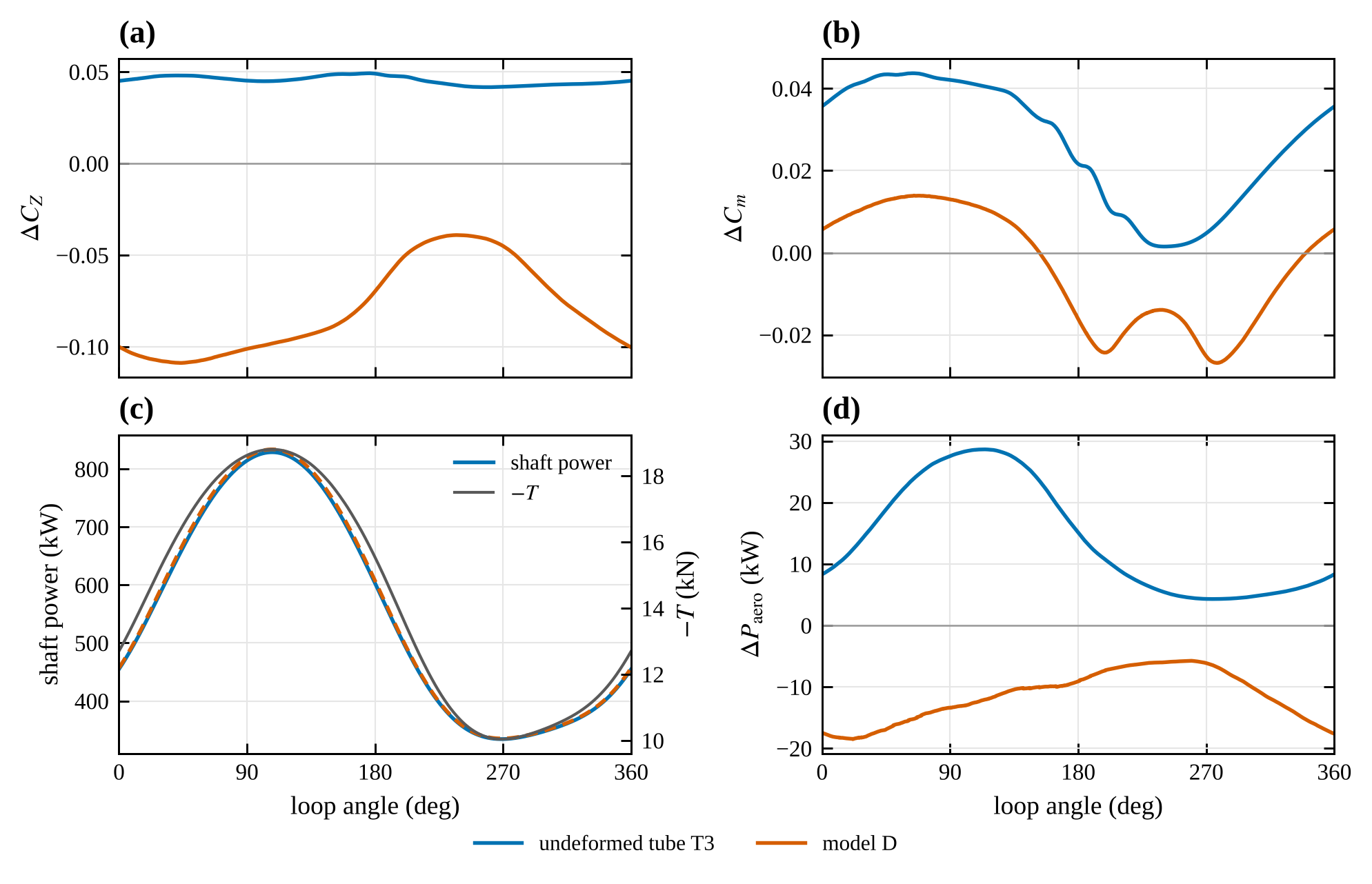}
\caption{\label{fig:20}\label{fig:loop}Effect of the eight rotors along the last lap at 9~m/s for tube T3 and model D. (a) Lift coefficient, (b) pitching-moment coefficient, (c) shaft power and thrust, (d) aerodynamic power.}
\end{figure*}

With tube T3, the refinement, $\sigma$ = 0.1~m and $\Delta t$ = 0.01~s, the unit with rotors repeats its loads from lap to lap to $4.4\times10^{-4}$ in force and $1.4\times10^{-3}$ in moment between the third and fourth laps, and its step-to-step scatter of $C_Z$, $1.2\times10^{-5}$ in the root mean square of the second difference, equals that without rotors, against $1.9\times10^{-3}$ with free rings at $\Delta t$ = 0.02~s. Ring nodes enter the surface core of a load point at 217 of 1872~steps, 0.13 nodes per step on average. Over the fourth lap the eight rotors give a shaft power of 527~kW, from 334~kW at the top of the loop, at a loop angle of 269$^\circ$, to 829~kW near the bottom, at 108$^\circ$, and a thrust of $-$13.8~kN, from $-$10.0 to $-$18.8~kN [Fig.~20(c)], and they take 637~kW from the motion of the kite. They raise the lap-averaged $C_Z$ of wing and tails by +0.045, from 2.328 to 2.374, or the lift by 1.9~\%, between 1.8~\% and 2.2~\% along the loop, and the pitching-moment coefficient by +0.025, from $-$0.227 to $-$0.202, by +0.002 at 240$^\circ$, after the top of the loop, to +0.044 at 64$^\circ$, after the bottom [Fig.~20(a), 20(b)]. The aerodynamic power of wing and tails rises by 12.9~kW, from 574.3 to 587.2~kW [Fig.~20(d)]. The effect is the same in every lap to $6\times10^{-4}$ in $C_Z$ and close to the +0.047 of the first second.

\subsubsection{Slipstream deflected by the wing}

The deformation of the slipstream by the wing is significant at high lift \cite{Duivenvoorden2026} and lies outside a prescribed rotor wake \cite{Erhard2024}, so model D lets the rings convect freely and regularises, for their convection only, the bound vortex system of the lifting surfaces over $r_b$ = 0.64~m, half the mean chord. It stands for a wing whose circulation is spread over its chord, as an actuator-surface wing spreads it in a vortex particle method \cite{AlvarezNing2024}, so that a ring no longer passes through the core of a lifting line. The loads, the wake of the wing and the inflow at the hubs are computed as before. Over the first second model D converges with the time step (Fig.~17). With $\sigma$ = 0.1~m and the refinement the effect on $C_Z$ is $-$0.094, $-$0.091, $-$0.088 and $-$0.087 from 0.02 to 0.0025~s, and with $\sigma$ = 0.3~m $-$0.116, $-$0.087, $-$0.082 and $-$0.080. With $\sigma$ = 0.1~m the differences between neighbouring steps fall by a factor of only 0.75 to 0.80 per halving, an observed order of about 0.4. The effect on $C_m$ is $-$0.0016 with $\sigma$ = 0.1~m and $-$0.015 with 0.3~m and changes by at most $5\times10^{-4}$ between the two finest steps. The criterion set before these runs required the differences between neighbouring steps to fall for both coefficients and the last one to stay within 0.003 in $C_Z$ and 10~\% in $C_m$. It is met in $C_Z$ for both cores. For $\sigma$ = 0.1~m it is met in the absolute change of $C_m$ and missed in the relative one, because the effect on $C_m$ is close to zero, and for $\sigma$ = 0.3~m the changes of $C_m$, at most $5\times10^{-4}$, do not fall.

Over whole laps at $\Delta t$ = 0.005~s model D lowers the lap-averaged $C_Z$ by 0.078, the lift by 3.5~\%, between 1.9~\% near 237$^\circ$ and 4.4~\% near 44$^\circ$ along the loop, and it changes $C_m$ by $-$0.005, between $-$0.027 at 277$^\circ$ and +0.014 at 64$^\circ$ along the loop (Fig.~20, model D against its run without rotors at the same step, third lap, dashed in Fig.~20(c)). Its loads are as smooth as without rotors, $1.1\times10^{-5}$ in the scatter of $C_Z$, and repeat between the second and third laps to $1.7\times10^{-3}$ in force and $3.2\times10^{-3}$ in moment. At 0.01~s the lap-averaged effect on $C_Z$ is $-$0.082. Extrapolated at first order the effect at a vanishing step would be $-$0.075, and with the observed order $-$0.064 to $-$0.068, so the value at 0.005~s overstates the loss of lift by at most about 0.014 in $C_Z$. At $\Delta t$ = 0.01~s a surface core of 0.3~m instead of 0.1~m changes the lap-averaged effect on $C_Z$ by +0.003 and on $C_m$ by $-$0.008. The rotors give the same shaft power as with the undeformed tube within 0.6~\%, 530~kW, and the aerodynamic power of wing and tails falls by 11.5~kW. A vortex particle computation of an M600-like kite with actuator-line rotors and wing, whose slipstreams the wing deforms, was made on a straight path at one operating point \cite{Mehr2025}. In every configuration studied it found that the rotors lower the lift of the wing, by up to about 7~\%, and more so when they sit closer to the wing and farther ahead of it \cite{Mehr2025}. Model D has the sign of that result.

\begin{figure*}[!t]
\centering
\includegraphics[width=0.85\textwidth]{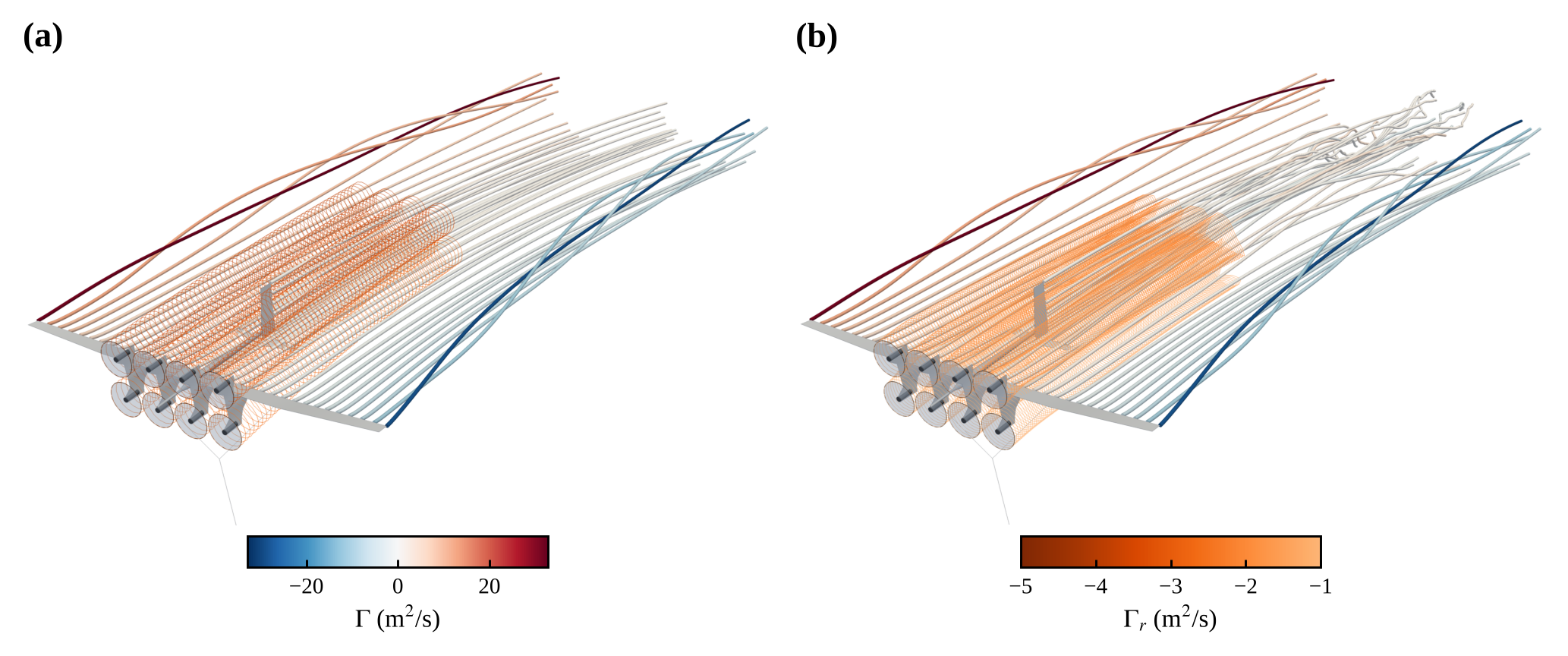}
\caption{\label{fig:21}\label{fig:renders}The slipstream of the eight rotors at the end of a lap at 9~m/s. (a) Tube T3, (b) model D.}
\end{figure*}

\begin{figure*}[!t]
\centering
\includegraphics[width=1\textwidth]{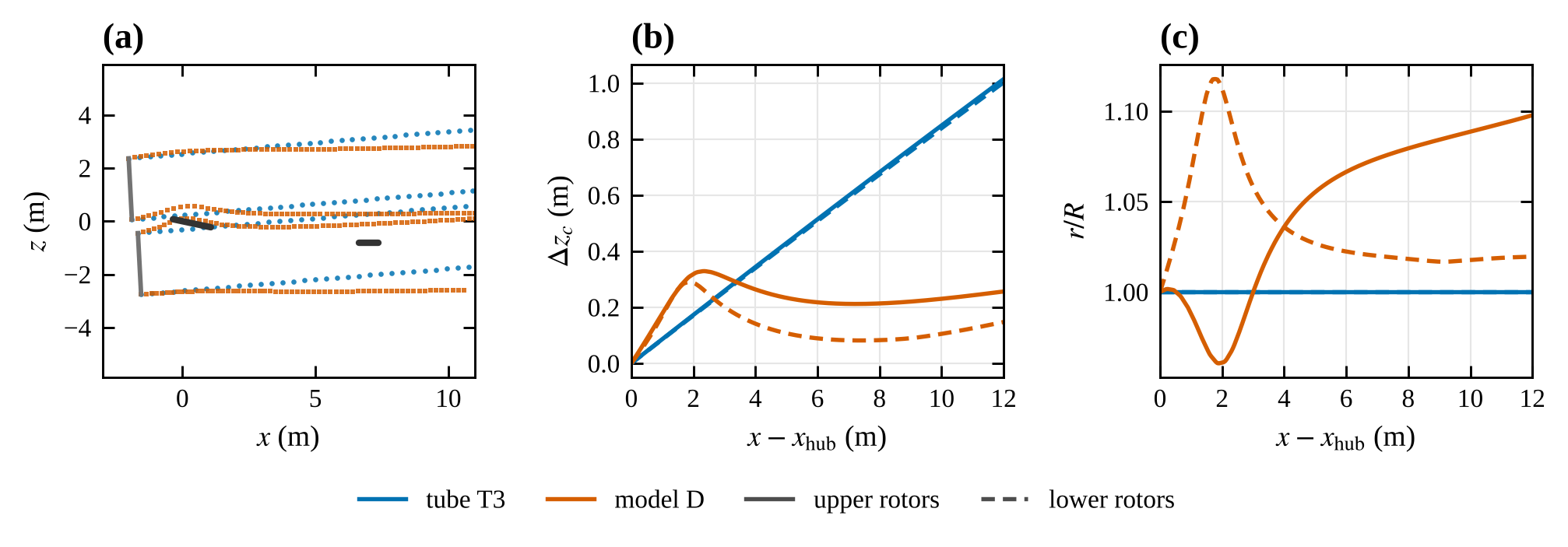}
\caption{\label{fig:22}\label{fig:shape}Shape of the slipstream behind the rotors. (a) Ring extremes in the body $x$-$z$ plane through the starboard inboard discs, (b) height of the ring centres and (c) mean ring radius against the distance behind the hub.}
\end{figure*}

Figure~21~shows the slipstream of the two treatments at the end of a lap, at the same pose of the kite at a loop angle of 0$^\circ$, for the single unit of Fig.~20, tube T3 at $\Delta t$ = 0.01~s after four laps and model D at 0.005~s after three laps. The renders draw the trailing filaments of wing and tails up to 1.2~s old coloured by their circulation $\Gamma$, the rotor rings up to about 0.6~s old coloured by theirs, $\Gamma_r$, and the complete airframe as in Fig.~3. Figure~22(a) gives the highest and lowest node of every ring of the upper and lower inboard discs on the starboard side in the body $x$-$z$ plane through their axes, with the sections of the main wing and the horizontal tail there, and Figs.~22(b) and 22(c) the means over the four upper (solid) and the four lower (dashed) discs against the distance behind the hub along the body $x$ axis. In the undeformed tube the rings of every disc keep the disc radius and lie on a straight line along the apparent wind, which rises by 0.085~m per metre along the body $x$ axis at this point of the loop. In model D the rings move with the velocity induced by the wing and its wake. Where they pass the wing the bound vortex lifts the ring centres of the upper and lower discs to 0.33 and 0.29~m above their hubs, about 0.13~m more than in the undeformed tube, and behind the wing the downwash turns the slipstream towards the body $x$ axis. At the leading edge of the horizontal tail, 8.3 to 8.6~m behind the discs, the ring centres lie 0.22~m above the hubs of the upper discs and 0.08~m above those of the lower ones, against 0.73 and 0.70~m in the undeformed tube. The rings of the upper discs, whose tubes pass the suction side, contract to 0.96 of the disc radius over the wing and grow to 1.08 at the tail, and those of the lower discs, below the pressure side, expand to 1.12 over the wing and return to 1.02. The deflected walls come as close to the lifting surfaces as the undeformed ones, within 0.04~m of a closure point of the main wing against 0.12~m, and within 0.2~m of 14 of the 49 closure points against 7. The two treatments therefore differ in where the walls of the tubes lie relative to the wing and the tails, by up to 0.6~m at the tail, and in the contraction and growth of the tubes, while the strength of the walls and the shaft power are the same.

Because the rings near the wing move with its induced velocity in model D alone, the two treatments give the effect of the rotors on the wing for a slipstream that the wing does not deform and for one that it deflects. The undeformed tube raises the lift by 1.9~\% and model D lowers it by 3.5~\%, a spread of 0.12 in $C_Z$, larger than either effect, while the shaft power of the rotors and their thrust agree to 0.6~\%. The undeformed tube is a common model of slipstreams in lifting-surface methods \cite{Nederlof2023}, and a recent design study of an airborne wind energy windplane switched the interaction off for rotors on pylons \cite{Trevisi2026}. For the M600 the present computations give an interaction several times larger than the loss by the kite's own wake (Sec.~IV A), whose sign depends on how the slipstream is modelled near the wing.

The power taken by the rotors also enters the energy balance of the prescribed loop. Without rotors the aerodynamic loads deliver 574~kW over a lap and the tether takes 139~kW, so the loads give 435~kW more than the prescribed motion needs, and the kite would accelerate. With the rotors at 97.4~rad/s the balance is $587 - 139 - 637 = -189$ kW with the undeformed tube and $563 - 139 - 640 = -216$ kW with model D, and the kite would slow down. On the prescribed loop the balance closes for a power taken by the rotors from the motion of about 425 to 450~kW, which a lower rotor speed would give. A model of the land-based unit with on-board generation therefore needs the rotors for the power that sets the speed of the loop.

\subsection{Two units}

\subsubsection{Wakes of the two units}

Two M600 units fly the same nominal loop at 9~m/s, each on its own tether, in one wake (Sec.~II E). The ground point of the downstream unit B is one or two loop diameters downwind of unit A, in one case one diameter downwind and half a diameter to the side. Every configuration is computed without rotors and with the rotors of both units treated as the undeformed tube T3 of Sec.~IV B at $\Delta t$ = 0.02~s, whose error in the effect of the rotors on $C_Z$ is about +0.0026. The rings are kept for 2~s. At one diameter a second run keeps them for two laps, 37~s, so that the slipstream tubes of A, which the wind carries over the 270~m to the loop of B in about 30~s, reach B, and older rings lie more than 60~m behind the loop of B, a distance of more than fifty tube radii. The far wake drifts with the wind beyond one lap. The runs cover five laps at one diameter and seven with the offset and at two diameters, so that the wake of A reaches B after 1.6 and 3.2 laps and B then flies 2.4 to 5.4 laps in it, and a run without rotors with two free laps at one diameter tests the frozen far wake.

\begin{figure*}[!t]
\centering
\includegraphics[width=0.85\textwidth]{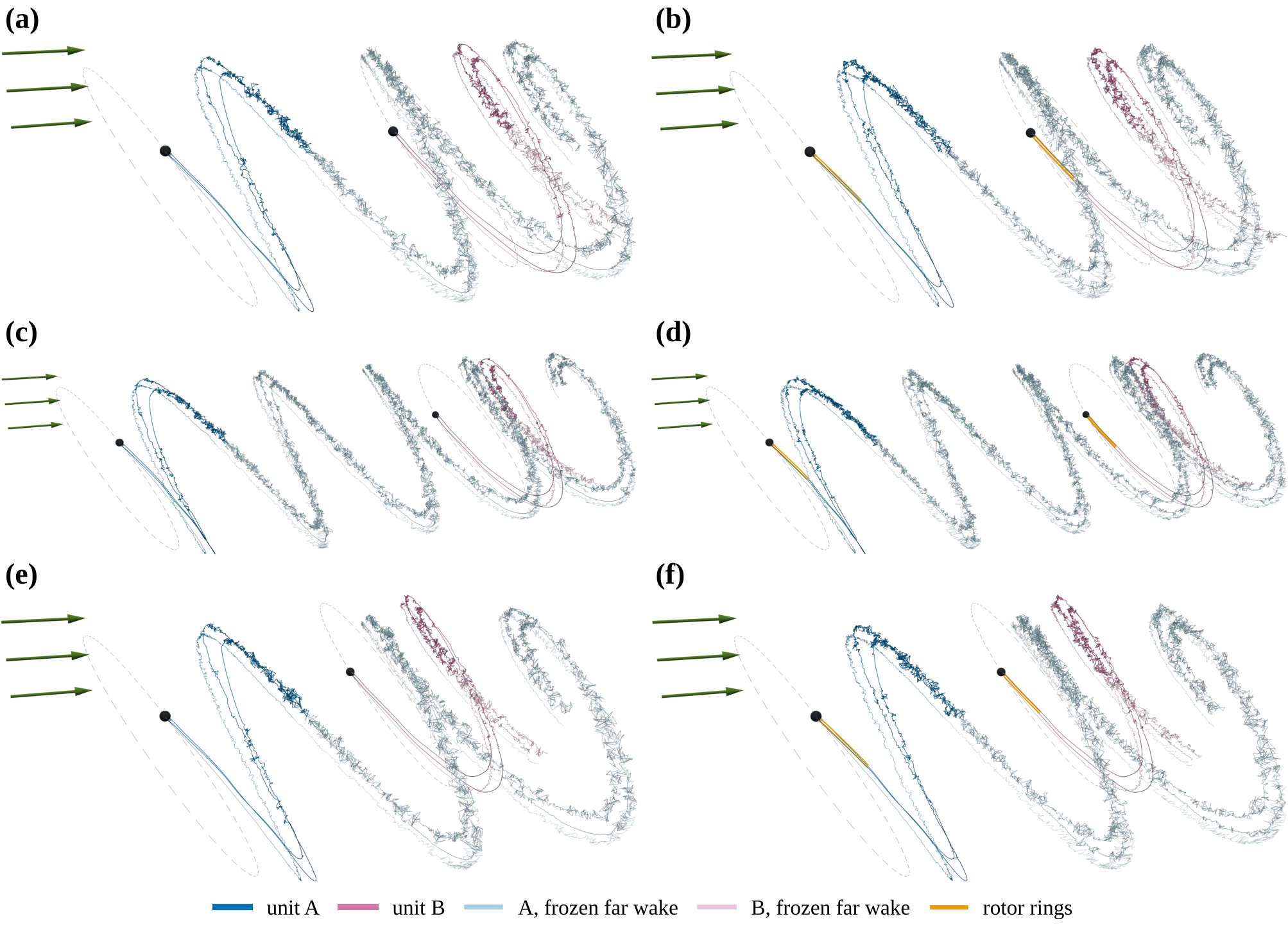}
\caption{\label{fig:23}\label{fig:pairs}Wakes of two M600 units at the end of their runs. (a), (b) Tandem at one loop diameter, (c), (d) tandem at two diameters, (e), (f) offset, each without and with rotors.}
\end{figure*}

Figure~23~shows the two wakes at the last stored step, a loop angle of 0$^\circ$, with the ground points, the tethers drawn straight to the kites, the loop paths of the last lap (dashed) and the wind (arrows). The trailing filaments of the main wing of each unit are drawn at the tips and at mid-span, in blue for A and pink for B, lighter where the wake is older than its free part and drifts frozen with the wind, and up to one loop diameter downwind of the centre of the loop of B. In tandem the helical wake of A drifts along the wind into the loop of B. It reaches B 1.6 laps old at one diameter [Figs.~23(a) and 23(b)] and 3.2 laps old at two diameters [Figs.~23(c) and 23(d)], after drifting frozen for about 0.6 and 2.2 laps in the helical form it had when it left the free part. With the offset of half a diameter [Figs.~23(e) and 23(f)] the wake of A passes beside the loop of B and meets the path of B near the bottom of the loop. The slipstream tubes of the rotors, drawn as orange lines through the ring centres, trail up to 82~m behind their kite along its path, because the kite flies at about 45~m/s while its rings drift with the slower flow of Sec.~II D and are kept for 2~s. In tandem at two diameters the tip vortices of A pass B on its pressure side, whereas with the offset no trailing filament of the main wing of A comes within 75~m of B at this step. When the rings are kept for two laps, the slipstream tubes of A trail up to 415~m behind A and reach the loop of B, and at this step nine rings of A lie within 75~m of B.

\subsubsection{Losses of the downstream unit}

\begin{table*}[!t]
\caption{\label{tab:8}Lap-averaged lift and aerodynamic power of the downstream unit B relative to the upstream unit A, the shaft power $P_s$ of the rotors of B relative to A over the last lap, and the periodicity of B between the last two laps as the largest change at any step of the total load (aerodynamic, rotor and tether) and of the aerodynamic load of wing and tails alone, in force and in moment.}
\small\setlength{\tabcolsep}{4.5pt}\renewcommand{\arraystretch}{1.12}
\begin{tabular}{@{}llllcccccccc@{}}
\toprule
\multicolumn{2}{@{}l}{Configuration} & & & & \multicolumn{3}{c}{Ratios of B to A} & \multicolumn{4}{c}{Periodicity of B over the last two laps (\%)} \\
\cmidrule(r){1-2}\cmidrule(lr){6-8}\cmidrule(l){9-12}
 & & & & & & & & \multicolumn{2}{c}{total load} & \multicolumn{2}{c}{wing and tails} \\
\cmidrule(lr){9-10}\cmidrule(l){11-12}
Spacing & Offset & Rotors & Rings kept & Laps & $L_B/L_A$ & $P_B/P_A$ & $P_{s,B}/P_{s,A}$ & force & moment & force & moment \\
\midrule
$1\,D$ & 0 & no & & 5 & 0.9949 & 0.9785 & & 0.40 & 0.62 & 0.13 & 0.38 \\
$1\,D$ & 0 & yes & 2~s & 5 & 0.9950 & 0.9790 & 1.0002 & 0.42 & 1.06 & 0.12 & 0.43 \\
$1\,D$ & 0 & yes & 2 laps & 4 & 0.9950 & 0.9790 & 1.0002 & 0.34 & 1.07 & 0.12 & 0.89 \\
$1\,D$, two free laps & 0 & no & & 4 & 0.9946 & 0.9774 & & 3.4 & 20 & 1.0 & 14 \\
$2\,D$ & 0 & no & & 7 & 0.9920 & 0.9682 & & 0.31 & 0.98 & 0.09 & 0.74 \\
$2\,D$ & 0 & yes & 2~s & 7 & 0.9923 & 0.9694 & 1.0002 & 0.37 & 0.67 & 0.10 & 0.36 \\
$1\,D$ & $0.5\,D$ & no & & 7 & 0.9991 & 0.9957 & & 0.20 & 2.8 & 0.08 & 0.50 \\
$1\,D$ & $0.5\,D$ & yes & 2~s & 7 & 0.9991 & 0.9957 & 1.0002 & 0.25 & 2.4 & 0.11 & 0.37 \\
\bottomrule
\end{tabular}
\end{table*}

The tether tension is set by the prescribed motion and is the same for both units. In tandem at one diameter without rotors B loses 0.51~\% of its lift and 2.2~\% of its aerodynamic power against A over the fifth lap (Table~VIII, Fig.~24). The loss of power is four times the loss of lift because the wake of A also lowers the forward force of B along its body $x$ axis, by 6.6~\% or 0.22~kN against 0.53~kN of lift, and this component delivers about a quarter of the aerodynamic power but 10 of the 12~kW that B loses. Along the loop the lift of B is lower than that of A by 0.01~\% to 1.3~\%, by more than 1.2~\% between loop angles of 354$^\circ$ and 13$^\circ$, on the side where the kite climbs, again by 0.6~\% near 153$^\circ$ and almost not at all near 262$^\circ$ at the top of the loop. Unit A loses 0.03~\% of its lift and 0.1~\% of its power against the single unit through the induction of B and its wake. At two diameters B loses more, 0.80~\% of its lift and 3.2~\% of its power, and with the offset of half a diameter it loses only 0.09~\% and 0.43~\%. The wake of A reaches B at two diameters after it has drifted frozen for about two laps, and it then keeps the helical arrangement and the compact cross-section of its tip-vortex pair [Fig.~25(a)], although its filament lines are jagged (Fig.~23), whereas the complete free wake loses this order after two turns (Sec.~IV A), so that the loss at two diameters belongs to an ordered wake that a free wake would not keep. At one diameter, where the wake reaching B is younger than two laps, two free laps change the lift ratio from 0.9951 to 0.9946 and the power ratio from 0.9791 to 0.9774 in the fourth lap. In that run B has not yet settled, its total load changing by up to 20~\% between the third and fourth laps, so the difference of 0.05~\% in lift and 0.17~\% in power is an estimate of the effect of the frozen far wake at one diameter.

\begin{figure}[!htb]
\centering
\includegraphics[width=1\linewidth]{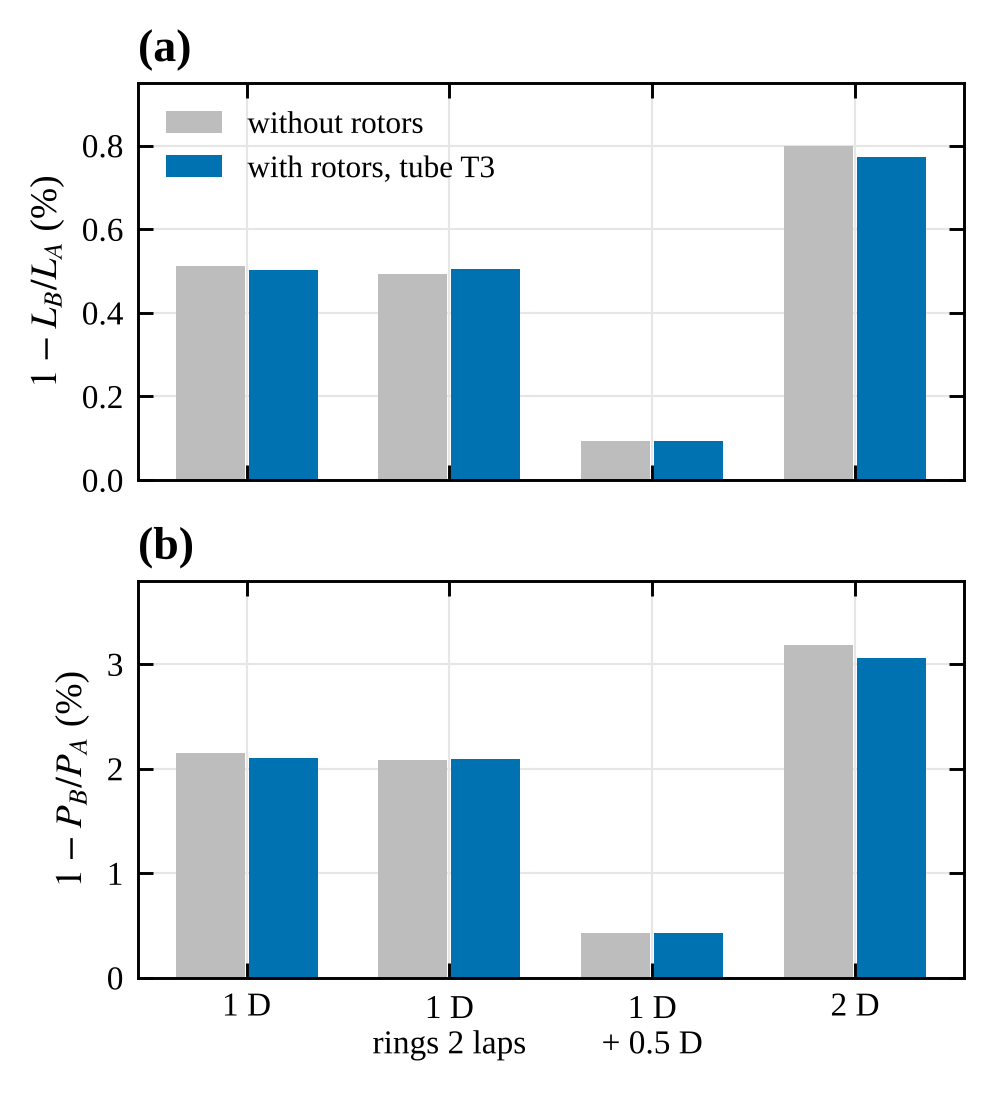}
\caption{\label{fig:24}\label{fig:ratios}Losses of the downstream unit relative to the upstream unit over the last lap, without rotors and with the rotors of both units as tube T3. (a) Lift, (b) aerodynamic power.}
\end{figure}

With the rotors of both units the ratios of B to A change by at most 0.03~\% in lift and 0.12~\% in power (Table~VIII, Fig.~24, where the bars compare each configuration without rotors and with the rotors of both units as tube T3, for the tandem at one diameter with the rings kept for 2~s and for two laps, the offset and the tandem at two diameters). The rotors of B give 0.02~\% more shaft power than those of A in every configuration, so the slower flow of the wake of A does not lower the power of the rotors of B measurably. The effect of the rotors on $C_Z$ of B is that on A within $5\times10^{-4}$, +0.047 to +0.048 in every configuration. When the rings are kept for two laps, so that the slipstream tubes of A reach B, the lap-averaged $C_Z$ of B changes by $-2.4\times10^{-4}$ against the run in which they are removed after 2~s, its aerodynamic power by less than 0.1~kW and the shaft power of its rotors by $-$0.1~kW, and the lift and power ratios change by $5\times10^{-5}$ and $1\times10^{-4}$. The slipstream tubes of the upstream rotors thus do not reach the downstream unit with a measurable effect. In this treatment they are thin tubes that the wind carries along the loop of A, and B passes through the region they occupy only briefly. The interaction of the two units is carried by the wakes of their wings, with or without rotors.

The total load of B, aerodynamic, rotor and tether together, changes between the last two laps by up to 1.1~\% in moment at one diameter with rotors and by up to 2.8~\% with the offset (Table~VIII), above the 1~\% set before the runs. The tether balances most of the aerodynamic moment. With the offset the steps above 1~\% lie between loop angles of 90$^\circ$ and 127$^\circ$, where B passes the tip vortices of A (Fig.~25), and there the total moment is 13 to 15~kN m against an aerodynamic moment of about 86~kN m, so that a change of the aerodynamic moment of B by 0.37~\% to 0.50~\% appears as 2.4~\% to 2.8~\% of the total. The aerodynamic load of wing and tails of B repeats to 0.9~\% in moment and 0.13~\% in force at every step in every run with one free lap, and its lap-averaged $C_Z$ and $C_m$ change by at most $8.4\times10^{-4}$ and 0.1~\% between the last two laps. The ratios of Table~VIII are therefore periodic, and B carries a local, slightly irregular load where its path crosses the wake of A.

\subsubsection{Downwash of the upstream tip vortices}

\begin{figure}[!htb]
\centering
\includegraphics[width=1\linewidth]{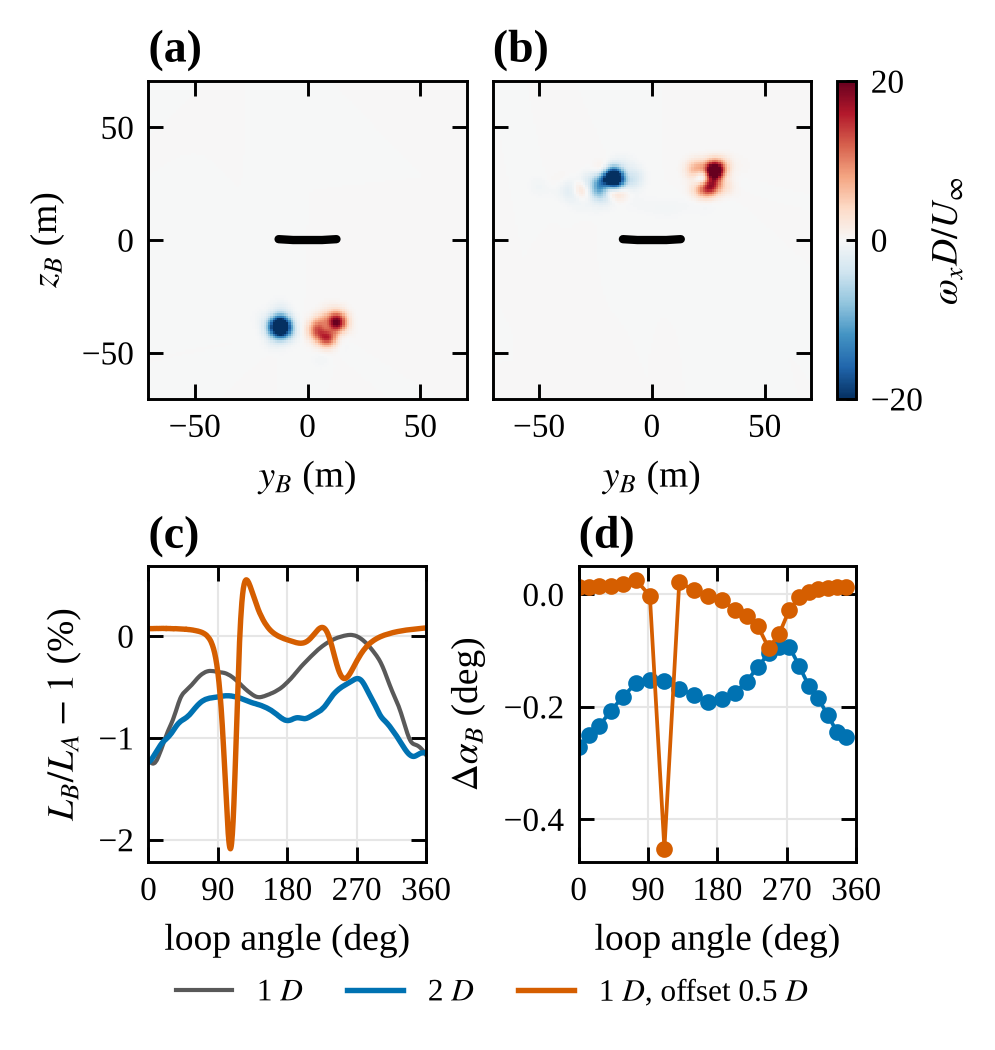}
\caption{\label{fig:25}\label{fig:wakeB}The wake of unit A at unit B over the last lap. (a), (b) Streamwise vorticity of the wake of A in the plane of B, (c) change of the lift of B relative to A, (d) change of the angle of attack of B induced by the wake of A.}
\end{figure}

\begin{figure*}[!t]
\centering
\includegraphics[width=1\textwidth]{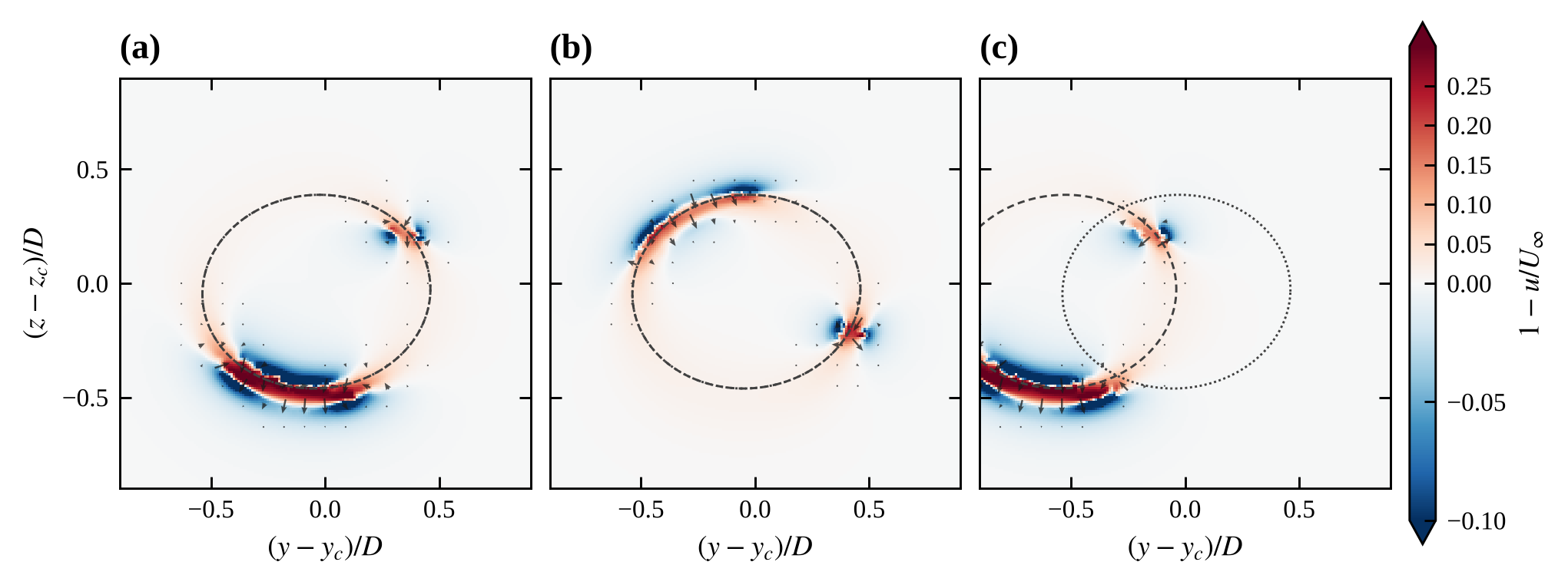}
\caption{\label{fig:26}\label{fig:pairplanes}Axial velocity deficit induced by the wake of the lifting surfaces of unit A alone in the plane normal to the wind through the path centre of unit B at the end of the runs without rotors, with the loop of A (dashed) and of B (dotted). (a) Tandem at one loop diameter, (b) tandem at two diameters, (c) offset.}
\end{figure*}

Figure~25~shows the wake that B crosses and the induced incidence it receives in the runs with rotors. Figures~25(a) and 25(b) give the streamwise vorticity of the wake of the lifting surfaces of A in the plane normal to the body $x$ axis of B through its body origin, in the body axes of B, as the flux of the crossing filaments spread over 3~m, with the bound line of the main wing of B in black, at the stored step at which the lift of B is lowest relative to A, for the tandem at two loop diameters at a loop angle of 0$^\circ$ and for the offset at 111$^\circ$. Figure~25(c) gives the change of the lift of B relative to A along the last lap for the tandem at one and two diameters and for the offset, and Fig.~25(d) the mean over the span of the main wing of B of the change of its angle of attack induced by the wake of A at 24~stored steps of the last lap. The velocity induced by the wake of the lifting surfaces of A along the main wing of B changes the airspeed of B by less than 0.3~\% and its angle of attack by up to $-$0.27$^\circ$ at two diameters and $-$0.45$^\circ$ with the offset. The lift of B relative to A follows this change of the angle of attack, with a correlation of 0.96 at two diameters and 0.90 with the offset over the 24~steps and a slope of 4.4~\% and 3.5~\% of lift per degree. At two diameters the wake of A reaches B as a helical pair of counter-rotating vortices, which lies about 40~m from B on its pressure side when B is halfway up the climbing side of its loop, at a loop angle of 0$^\circ$ [Fig.~25(a)]. B flies in the downwash of the pair all along the loop, by 0.09$^\circ$ near 270$^\circ$ at the top to 0.27$^\circ$ at 0$^\circ$, and its lift is lower than that of A by 0.42~\% at 271$^\circ$ to 1.24~\% at 0$^\circ$. In tandem at one diameter the loss has the same form along the loop, largest near 0$^\circ$ and smallest near the top. With the offset B flies beside the wake of A, and over about two thirds of the loop its lift is lower than that of A by less than 0.05~\%. Between loop angles of 92$^\circ$ and 116$^\circ$, near the bottom of the loop, B passes the tip vortices of A at 25 to 30~m on its suction side [Fig.~25(b)]. Its angle of attack falls by up to 0.45$^\circ$ and its lift by up to 2.1~\% at 106$^\circ$, and the lift then rises 0.55~\% above that of A at 127$^\circ$ as B leaves the vortices. A second, weaker passage near 254$^\circ$ lowers the lift by 0.42~\%. These passages are the local irregularity of the total moment of B with the offset in Table~VIII. The loss of the downstream unit is thus the downwash of the tip-vortex pair of the upstream unit, set by where the downstream unit flies relative to the pair.

Figure~26~shows the deficit that the wake of A alone induces in the plane through the path centre of B for the three configurations without rotors. In tandem at one diameter the sheet of A crosses the plane as the band seen at one diameter behind the single kite (Fig.~12), near the bottom of the loop of B and once near its top, with a deficit above $0.1\,U_\infty$ over 0.062 $D^2$ and above $0.3\,U_\infty$ over 0.013 $D^2$. At two diameters the band has moved with the frozen wake and its peak has fallen to about $0.3\,U_\infty$, the deficit above $0.1\,U_\infty$ covering 0.040 $D^2$ and that above $0.3\,U_\infty$ only 0.0002 $D^2$, yet B loses more there than at one diameter, because the loss is set by the downwash of the ordered pair on the path of B rather than by the deficit in its loop. With the offset the band of A lies beside and below the loop of B, which meets it only near the bottom of its path, where the passages of Fig.~25(b) occur.

\FloatBarrier

\section{Conclusions}\label{sec:v}

A vortex step method coupled to a time-marching free vortex wake was formulated for airborne wind energy kites on prescribed crosswind motions with a lumped-mass tether, actuator-disc rotors that release vortex rings and several kites in one wake, implemented in the open-source code OpenFAWE, verified on the TU Delft V3 kite and the Makani M600 with its tails, and applied to the M600 on its crosswind loop alone, with its rotors and as a pair of units.

\begin{enumerate}
\item The implementation reproduces an independent open-source vortex step code to $10^{-9}$ on the M600 and within 0.3~\% in lift and 0.9~\% in drag on the V3, whose lift agrees with RANS within 3.1~\% while both lie below the wind-tunnel values. A kite translating through still air reproduces the fixed kite in the opposite wind to $10^{-12}$, two kites 20 km apart reproduce two single computations to $10^{-9}$, and the tree code stays within the error budget fixed before the runs while its evaluation on a graphics processor reproduces it to $4\times10^{-6}$.
\item The lift deficit of a finite wake is the far field of its starting vortex for any spanwise loading, so that the wake length for a required accuracy follows before a run from the reference area and the lift response of the kite, 49 chords for the M600 and 16 chords for the V3 for 0.1~\% in lift, about three times longer for the drag of a wing of high lift-to-drag ratio and about 1.2 times longer for a free wake, whose shed history and descent add to the deficit. The free wake reproduces the inviscid invariants of roll-up, the Betz centroid within 1~\% and the descent of a vortex pair within 4~\% to 15~\%.
\item The vortex step closure with regularised filaments is well posed while the core radius stays below the smallest spanwise panel width, tail panels included, and beyond it the smallest singular value of the closure Jacobian falls by two orders of magnitude and the runs fail, which the Jacobian at the start step tests before a run. Doubling the core changes the M600 pitching moment by 38~\% through the load of the tail and the V3 lift by 2.7~\%, whereas viscous core growth in the wake changes the loads by at most 0.3~\%, so the core sensitivity is a near-field property of the closure. A time step of 0.2 to 0.3 chords of travel keeps the lift within 0.02~\% of the finest level.
\item On its crosswind loop the M600 reaches loads that repeat from lap to lap within 0.1~\% after two laps. Its wake is a helical pair of counter-rotating tip vortices at close to the Betz spacing, with a velocity deficit of up to 0.4 of the wind speed between the vortices, which descends by about 0.02 loop diameters per loop, contracts slowly and loses its order after about two turns. The induction of this wake lowers the lift by about 0.5~\% at 5, 9 and 13~m/s, whereas an analytical helical-wake model, whose helix is shortened by the induction of the near wake, predicts a loss that falls from 1.1~\% to 0.19~\% over these wind speeds.
\item Free vortex rings from the actuator discs of the rotors pass the bound vortex of the wing within its core, and the effect of the rotors on the lift coefficient does not converge with the time step. A slipstream tube that the wing does not deform converges when two conditions hold at the lifting-surface points, a ring spacing of at most half the surface core, reached by subdividing the rings at these points only, and a surface core of at most half the distance of the points from the tube walls. With this tube the eight rotors draw 527~kW of shaft power and raise the lift by 1.9~\%, an effect that follows the thrust almost linearly, changes sign with it and is carried by the tube walls that pass the wing and the tails. When the wing deflects the slipstream, the same rotors give the same shaft power and lower the lift by 3.5~\%, so the treatment of the slipstream near the wing sets the sign of the rotor effect, which is several times the loss by the kite's own wake. The rotors take about 640~kW from the motion, more than the 435~kW that the wing has to spare on the prescribed loop.
\item A downstream unit loses lift and power in the downwash of the tip-vortex pair of the upstream unit, 0.51~\% of its lift and 2.2~\% of its aerodynamic power one loop diameter downstream, 0.80~\% and 3.2~\% at two diameters in a far wake that stays ordered, and 0.09~\% and 0.43~\% when it is also offset sideways by half a diameter, where the loss is concentrated in the passages of the downstream unit through the tip vortices of the upstream one. The rotors change these losses by at most 0.03~\% in lift and 0.12~\% in power, the rotors of the downstream unit give the shaft power of the upstream ones, and the slipstream tubes of the upstream unit do not reach the downstream unit with a measurable effect, so that the interaction is carried by the wakes of the wings.
\end{enumerate}

The next steps are the closed-loop flight of the complete land-based unit with its rotors and ground station and the wake interaction of several units in a turbulent wind.

\section*{Author Declarations}

\subsection*{Conflict of Interest}

The authors have no conflicts to disclose.

\subsection*{Author Contributions}

\textbf{Yinong Tian:} Conceptualization (lead); Data curation (lead); Formal analysis (lead); Investigation (lead); Methodology (lead); Software (lead); Validation (lead); Visualization (lead); Writing -- original draft (lead); Writing -- review \& editing (equal). \textbf{Shan Wang:} Investigation (supporting); Methodology (supporting); Supervision (lead); Writing -- review \& editing (equal).

\subsection*{Use of generative AI}

The source code of OpenFAWE and the analysis scripts were written with the AI coding assistants Claude (Anthropic) and Codex (OpenAI) under the direction of the first author. As the authors are not native speakers of English, Claude (Anthropic) was also used to edit the language of the manuscript. The authors reviewed all content and take full responsibility for it.

\section*{Data Availability}

The source code of the aerodynamic module of OpenFAWE is openly available at \url{https://github.com/robytian/OpenFAWE} \cite{OpenFAWE}. The kite models, the run scripts, the result files and the analysis scripts of the runs reported here are released in the same repository with a DOI upon publication. Every run is identified by the commit of the source code that produced it. The reference data are available from their original sources, namely the wind-tunnel loads \cite{V3LoadData}, the V3 geometry \cite{V3Mesh}, the RANS results \cite{Vire2020}, the open-source Makani code \cite{MakaniCode} and the open-source vortex step code \cite{VSMCode}.

\bibliography{references}

\end{document}